%% file: main.tex
\documentclass[11pt, a4paper]{book}

\usepackage[textwidth=15.5cm]{geometry}
\usepackage[utf8]{inputenc}
\usepackage[english]{babel}
\usepackage{fancyhdr}
\usepackage{indentfirst}
\usepackage{graphicx, fancyhdr}
\usepackage{amsmath,amssymb,amsfonts}
\usepackage{setspace}
\usepackage{booktabs}
\usepackage{braket}
\usepackage{nicefrac}
\usepackage{bbm}                    %fonte \mathbbm{} do símbolo 1 da matriz identidade
\usepackage{subfigure}              %opção para figuras múltiplas
\usepackage{fancyhdr}               %cabeçalhos e rodapés decentes
\usepackage{slashed}
\usepackage{hyperref}
\usepackage{cite}
	\usepackage[font=small]{caption}
\newcommand{\eq}[1]{(\ref{#1})}     %coloca as referências de equações entre parênteses
\fancypagestyle{plain}{%            %
\fancyhead{}                        %sem cabeçalhos em páginas limpas
 }

\begin{document}

	\input{tituloportugues.tex}                  %folha de rosto
                                     %
%\hspace{-0.8cm}
\thispagestyle{empty}
\includegraphics[scale=0.78]{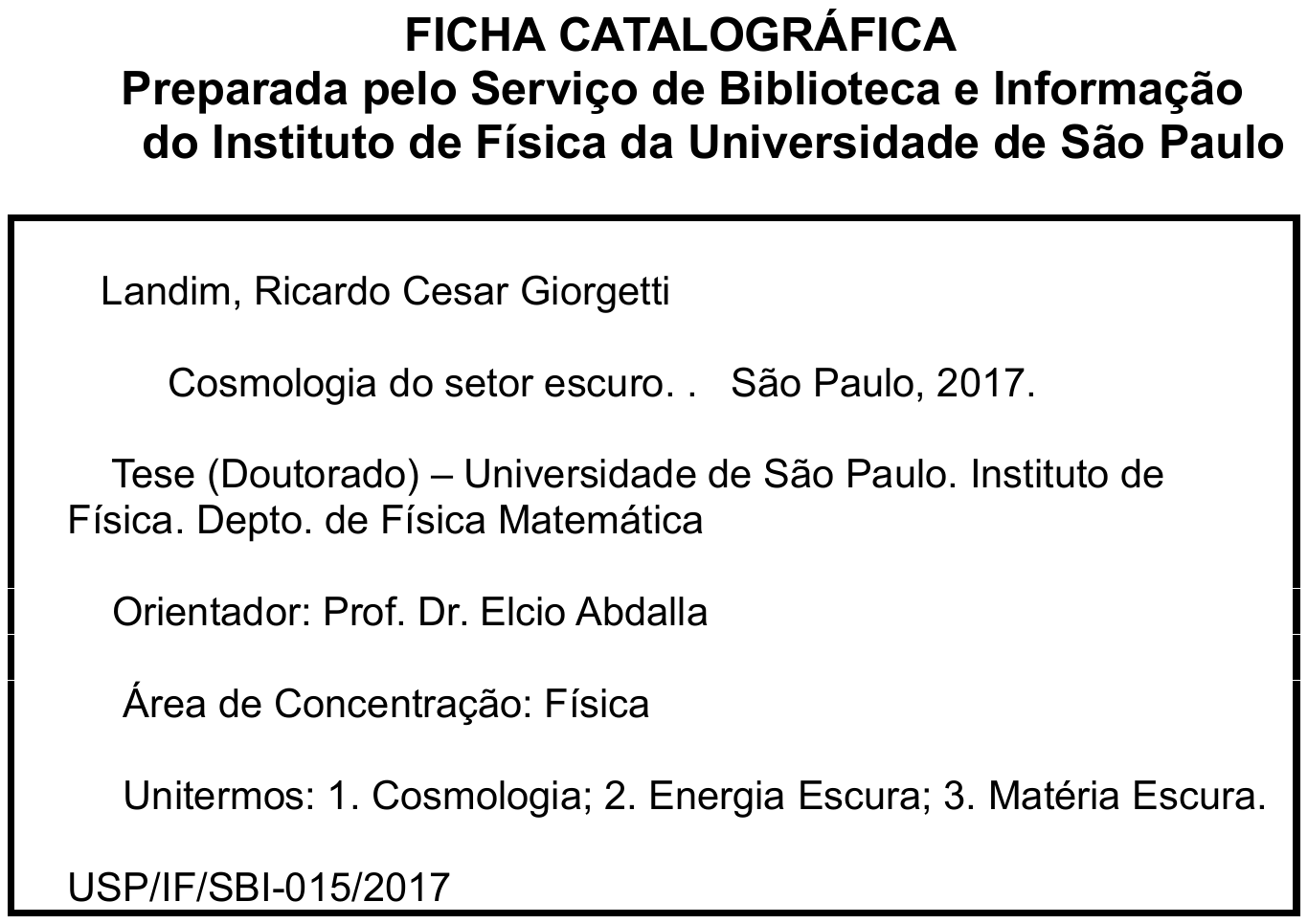}
 %Aqui vai a ficha catalográfica!    %ficha catalográfica 
	                       %
%\newpage                            %
%\baselineskip 8.75 mm               %
                                    %
%\thispagestyle{empty}          %
%\vspace*{-0.7cm}                    %

\input{titulo.tex}                  %folha de rosto
              \clearpage{\pagestyle{empty}\cleardoublepage}                      %
%\newpage                            %
%\baselineskip 8.75 mm               %
                                    %
           %
                                    %
%\hspace{-0.8cm}
%\includegraphics[scale=0.78]{fc_fabricio}
 %Aqui vai a ficha catalográfica!    %ficha catalográfica
                                    %
                                    %
\thispagestyle{empty}               %
                                    %
\input{preliminares.tex}            %dedicatórias, versos, agradecimentos
        \clearpage{\pagestyle{empty}\cleardoublepage}   
																	\thispagestyle{empty}%

\input{resumo.tex}  
  %\clearpage{\pagestyle{empty}\cleardoublepage}    
				\thispagestyle{empty}
\input{abstract.tex}                  %resumo / abstract
     
        %resumo / abstract
     % \clearpage{\pagestyle{empty}\cleardoublepage}                                %
%\renewcommand{\contentsname}{Índice}%traduz o nome "Contents" para "Índice"
\frontmatter
\tableofcontents             \clearpage{\pagestyle{empty}\cleardoublepage}              %cria o índice automaticamente
\mainmatter                         %
\input{introduction.tex}     \clearpage{\pagestyle{empty}\cleardoublepage}       
   \input{cosmoBasics.tex}     \clearpage{\pagestyle{empty}\cleardoublepage}                           %
\input{dynanalysis.tex}  \clearpage{\pagestyle{empty}\cleardoublepage}

\input{sugra.tex} \clearpage{\pagestyle{empty}\cleardoublepage}

\input{MDE.tex} \clearpage{\pagestyle{empty}\cleardoublepage}

 \input{conclusions.tex}    \clearpage{\pagestyle{empty}\cleardoublepage}                             %
\appendix
\input{notations.tex}\clearpage{\pagestyle{empty}\cleardoublepage}

	\newpage
	\addcontentsline{toc}{chapter}{Bibliography}
	
\bibliographystyle{unsrt}
\bibliography{trab1}

\end{document}

%% file: tituloportugues.tex
%%%%%%%%%%%%%%%%%%%%%%%%%%%%%%%%%%%%%%%%%%%%%%%%%%%%%%%%%%%%%%%%%%%%%%%%%%%%%%%%%%%%

\begin{titlepage}

%\frontmatter

\thispagestyle{empty}

\newcommand{\lyxline}[1][1pt]{
  \par\noindent
    \rule[.5ex]{\linewidth}{#1}\par}
\vspace*{-2cm}

\parskip = .4\baselineskip

\begin{center}
UNIVERSIDADE DE S\~AO PAULO \\
INSTITUTO DE F\'ISICA
\par\end{center}

\vspace{5cm}

%==================================================================================%
\begin{doublespace}
\begin{center}
\textbf{\huge{Cosmologia do setor escuro  }}\\

\par\end{center}{\Large \par}
\end{doublespace}

\vspace{0.8cm}

%==================================================================================%
\begin{center}
\textbf{\Large Ricardo Cesar Giorgetti Landim}
\par\end{center}{\large \par}

\vspace{2.2cm}

%==================================================================================%
\begin{flushright}

\begin{minipage}[c][1\totalheight][t]{10.5cm}%
\textbf{Orientador: Prof. Dr. Elcio Abdalla}

%\lyxline{\normalsize}

\vspace{0.2cm}

Tese de doutorado apresentada ao Instituto de F\'isica para a obtenção do t\'itulo de Doutor em Ci\^encias.

%Monografia apresentada ao Instituto de F\'isica como requisito parcial para a qualifica\c{c}\~ao no Curso de Doutorado em Ci\^encias.
\vspace{0.2cm}

%\lyxline{\normalsize}

\vspace{0.5cm}

\end{minipage}
\par\end{flushright}

%==================================================================================%
\vspace{1.6cm}

\noindent \textbf{\small Banca Examinadora: }\\
Prof. Dr. Diego Trancanelli (IF-USP)\\
Prof. Dr. Elcio Abdalla (IF-USP)\\
Prof. Dr. Enrico Bertuzzo (IF-USP) \\
Prof. Dr. Julio Cesar Fabris (UFES) \\
Prof. Dr. Rogério Rosenfeld (IFT-UNESP)

\vspace{1.6cm}%1.6cm original

%==================================================================================%
\begin{center}
S\~ao Paulo\\
2017
\par
\end{center}

\end{titlepage}

%==================================================================================%
%%%%%%%%%%%%%%%%%%%%%%%%%%%%%%%%%%%%%%%%%%%%%%%%%%%%%%%%%%%%%%%%%%%%%%%%%%%%%%%%%%%%

%% file: titulo.tex
%%%%%%%%%%%%%%%%%%%%%%%%%%%%%%%%%%%%%%%%%%%%%%%%%%%%%%%%%%%%%%%%%%%%%%%%%%%%%%%%%%%%

\begin{titlepage}

%\frontmatter

\thispagestyle{empty}

\newcommand{\lyxline}[1][1pt]{
  \par\noindent
    \rule[.5ex]{\linewidth}{#1}\par}
\vspace*{-2cm}

\parskip = .4\baselineskip

\begin{center}
UNIVERSIDADE DE S\~AO PAULO \\
INSTITUTO DE F\'ISICA
\par\end{center}

\vspace{5cm}

%==================================================================================%
\begin{doublespace}
\begin{center}
\textbf{\huge{Dark sector cosmology}}\\

\par\end{center}{\Large \par}
\end{doublespace}

\vspace{0.8cm}

%==================================================================================%
\begin{center}
\textbf{\Large Ricardo Cesar Giorgetti Landim}
\par\end{center}{\large \par}

\vspace{2.2cm}

%==================================================================================%
\begin{flushright}

\begin{minipage}[c][1\totalheight][t]{10.5cm}%
\textbf{Advisor: Prof. Dr. Elcio Abdalla}

%\lyxline{\normalsize}

\vspace{0.2cm}
A thesis submitted to the Institute of Physics in partial fulfillment of the requirements for the Degree of Doctor of Philosophy in Science

%A monograph submitted to the Institute of Physics in partial fulfillment of the requirements for the Degree of Doctor of Philosophy in Science

\vspace{0.2cm}

%\lyxline{\normalsize}

\vspace{0.5cm}

\end{minipage}
\par\end{flushright}

%==================================================================================%
\vspace{1.6cm}

\noindent \textbf{\small Examination Committee: }\\
Prof. Dr. Diego Trancanelli (IF-USP)\\
Prof. Dr. Elcio Abdalla (IF-USP)\\
Prof. Dr. Enrico Bertuzzo (IF-USP) \\
Prof. Dr. Julio Cesar Fabris (UFES) \\
Prof. Dr. Rogério Rosenfeld (IFT-UNESP)

\vspace{1.6cm}%1.6cm original

%==================================================================================%
\begin{center}
S\~ao Paulo\\
2017
\par
\end{center}

\end{titlepage}

%==================================================================================%
%%%%%%%%%%%%%%%%%%%%%%%%%%%%%%%%%%%%%%%%%%%%%%%%%%%%%%%%%%%%%%%%%%%%%%%%%%%%%%%%%%%%

%% file: preliminares.tex
%%%%%%%%%%%%%%%%%%%%%%%%%%%%%%%%%%%%%%%%%%%%%%%%%%%%%%%%%%%%%%%%%%%%%%%%%%%%%%%%%%%%
%%%%%%%%%%%%%%%%%%%%%%%%%%%%%%%%%%%%%%%%%%%%%%%%%%%%%%%%%%%%%%%%%%%%%%%%%%%%%%%%%%%%

%==================================================================================%

% Agradecimentos ==================================================================%
%\newpage
%\thispagestyle{empty}
%\frontmatter

%\bigskip
%\bigskip
%\begin{center}
%\end{center}
%\vspace{1cm}
\centerline{\Large{\bf Acknowledgments}}

\vspace{3.0cm}
\begin{quote}

The road from medical physics, which was the area I got my Master's, to cosmology was long and arduous. Besides my personal effort, some people were undoubtedly important to let the journey feasible. 

Firstly, my advisor had an essential role in the trip, because like a customs officer allows we cross the border, Elcio allowed me to enter in the realm of theoretical cosmology with no restriction on what I could study or do. The freedom he gave me pushed myself beyond my limits and such experience accelerated my maturation as a researcher. 

I am also grateful to my family (parents, brother and mother-in-law (!)) for their support. They were the umbrella in a rainy day. My profound gratitude  is to my near-future wife Mayara, who has been as essential as  the air we breathe in a long journey. As a bonus, I would like to thank our several dogs for the love they gave. They made the road greener due to the love that only former homeless dogs know how to give.

Some friends helped me to overcome obstacles in this trip. They were the ax or the hand saw needed to cut the trunks in the path. I thank specially Gian Camilo, Rafael Marcelino, Rafael Marcondes, Fabrizio Bernardi, Cedrick Miranda and Hugo (Camacho) Chavez.

Finally, I thank the financial support by CAPES, CNPq and specially FAPESP. They furnished the vital supplies needed in this long travel.

\end{quote}
 
\newpage

%==================================================================================%
%%%%%%%%%%%%%%%%%%%%%%%%%%%%%%%%%%%%%%%%%%%%%%%%%%%%%%%%%%

%% file: resumo.tex
%==================================================================================%
\centerline{\large \textbf{RESUMO}} 

\vspace{3.0cm}

\begin{quote}

O lado escuro do universo \'e misterioso e sua natureza \'e ainda desconhecida. De fato, isto talvez constitua o maior desafio da cosmologia moderna. As duas componentes do setor escuro (mat\'eria escura e energia escura)	correspondem hoje a cerca de noventa e cinco por cento do universo. O  candidato mais simples para a energia energia \'e uma constante cosmol\'ogica. Contudo, esta tentativa apresenta uma enorme discrep\^ancia de 120 ordens de magnitude entre a predi\c{c}\~ao te\'orica e os dados observados. Tal disparidade motiva os f\'isicos a investigar modelos mais sofisticados. Isto pode ser feito tanto buscando um entendimento mais profundo de onde a constante cosmol\'ogica vem, se deseja-se deriv\'a-la de primeiros princ\'ipios, quanto considerando outras possibilidades para a expans\~ao acelerada, tais como modifica\c{c}\~oes da relatividade geral, campos de mat\'eria adicionais e assim por diante. Ainda considerando uma energia escura din\^amica, pode existir a possibilidade de intera\c{c}\~ao entre   energia e mat\'eria escuras, uma vez que suas densidades s\~ao compar\'aveis e, dependendo do acoplamento usado, a intera\c{c}\~ao pode tamb\'em aliviar a quest\~ao de porqu\^e as densidades de mat\'eria e energia escura s\~ao da mesma ordem hoje. Modelos fenomenol\'ogicos tem sido amplamente estudados na literatura. Por outro lado, modelos de teoria de campos que visam uma descri\c{c}\~ao consistente da intera\c{c}\~ao energia escura/mat\'eria escura ainda s\~ao  poucos. Nesta tese, n\'os exploramos como candidato \`a energia escura um campo escalar ou vetorial em v\'arias abordagens diferentes, levando em conta uma poss\'ivel intera\c{c}\~ao entre as duas componentes do setor escuro. A tese \'e dividida em tr\^es partes, que podem ser lidas independentemente. Na primeira parte, n\'os analisamos o comportamento asint\'otico de alguns modelos cosmol\'ogicos usando campos escalares ou vetorial como candidatos para a energia escura, \`a luz 	da teoria de sistemas din\^amicos. Na segunda parte, n\'os  usamos um campo escalar em supergravidade para construir um modelo de energia escura din\^amico e tamb\'em para incorporar um modelo de energia escura hologr\'afica em supergravidade m\'inima. Finalmente, na terceira parte, n\'os propomos um modelo de energia escura metaest\'avel, no qual a energia escura \'e um campo escalar com um potencial dado pela soma de auto-intera\c{c}\~oes pares at\'e  ordem seis. N\'os inserimos a energia escura metaest\'avel em um modelo $SU(2)_R$ escuro, onde o dubleto de energia escura e o dubleto de mat\'eria escura interagem naturamente. Tal intera\c{c}\~ao abre uma nova janela para investigar o setor escuro do ponto-de-vista de f\'isica de part\'iculas. Esta tese \'e baseada nos seguintes artigos, dispon\'iveis tamb\'em no arXiv: 1611.00428, 1605.03550, 1509.04980, 1508.07248, 1507.00902 e 1505.03243. O autor tamb\'em colaborou nos trabalhos: 1607.03506  e 1605.05264. 

Palavras-chave: energia escura, matéria escura, teoria de sistemas dinâmicos, física de partículas, supergravidade

\end{quote}

\newpage

%==================================================================================%

%% file: abstract.tex
\centerline{\large \textbf{ABSTRACT}}

\vspace{3.0cm}

\begin{quote}
The dark side of the universe is mysterious and its nature is still unknown. In fact, this poses perhaps as the biggest challenge in the modern cosmology. The two components of the dark sector (dark matter and dark energy) correspond today to around ninety five percent of the universe. The simplest dark energy candidate is a cosmological constant. However, this attempt presents a huge discrepancy of 120 orders of magnitude between the theoretical prediction and the observed data. Such a huge disparity motivates physicists to look into a more sophisticated models. This can be done either looking for a deeper understanding of where the cosmological constant comes from, if one wants to derive it from first principles, or considering other possibilities for accelerated expansion, such as  modifications of general relativity, additional matter fields and so on. Still regarding a dynamical dark energy, there may exist a possibility of interaction between dark energy and dark matter, since their densities are comparable and, depending on the coupling used, the interaction can also alleviate the issue of  why dark energy and matter densities are of the same order today.  Phenomenological models have been widely explored in the literature. On the other hand, field theory models that aim a consistent description of the dark energy/dark matter interaction are still few. In this thesis, we explore either a scalar or a vector field as a dark energy candidate in several different approaches, taking into account a possible interaction between the two components of the dark sector. The thesis is divided in three parts, which can be read independently of each other. In the first part, we analyze the asymptotic behavior of some cosmological models using either scalar or vector fields as dark energy candidates, in the light of the dynamical system theory. In the second part, we  use a scalar field in the supergravity framework to build a model of dynamical dark energy and also to embed a holographic dark energy model into  minimal supergravity. Finally, in the third part, we propose a model of metastable dark energy, in which the dark energy is a scalar field with a potential given by the sum of even self-interactions up to order six. We insert the metastable dark energy into a dark $SU(2)_R$ model, where the dark energy doublet and the dark matter doublet naturally interact with each other. Such an interaction opens a new window to investigate the dark sector from the point-of-view of particle physics. This thesis is based on the following papers, available also in the arXiv: 1611.00428, 1605.03550, 1509.04980, 1508.07248, 1507.00902 and 1505.03243. The author also collaborated in the works 1607.03506  and 1605.05264. 

Keywords: dark energy, dark matter, dynamical system theory, particle physics, supergravity

\end{quote}

\newpage

%% file: introduction.tex
\chapter{Introduction}

Our universe is approximately 13.8 billion years old. Human beings have been interested in astronomy and cosmology since the ancient times, however,  modern cosmology started around a century ago. According to the modern classification, all components of the universe can be classified in three main categories: dark energy, matter and radiation. However, we do not know yet what dark energy and the majority of matter are, thus the full understanding of the universe is still not achieved.  Observations of Type IA Supernovae indicate that the universe undergoes an accelerated expansion \cite{reiss1998, perlmutter1999}, which is dominant at present times and the driving force sustaining the acceleration arises from some kind of dark energy corresponding to $\sim$ 68\% of all components of the universe \cite{Planck2013cosmological}. 

In addition to baryonic matter,\footnote{The term `baryonic' is deceptive because it includes all non-relativistic particles of the standard model, such as baryons, which are made up three quarks, mesons, which are composed of one quark and one anti-quark and electrons, for instance, which are leptons.} which forms galaxies and other astrophysical objects,   the remaining $27\%$ of the universe is an unknown form of matter that interacts in principle only gravitationally and it is known as dark matter. The nature of the dark sector is still mysterious and it is one of the biggest challenges in the modern cosmology. The simplest dark-energy candidate is the cosmological constant. This attempt, however,  suffers from the so-called cosmological constant problem \cite{Weinberg:1988cp}, a huge discrepancy of 120 orders of magnitude between the theoretical prediction and the observed data. Such a huge disparity motivates physicists to look into a more sophisticated models. This can be done either looking for a deeper understanding of where the cosmological constant comes from, if one wants to derive it from first principles, or considering other possibilities for accelerated expansion, such as  modifications of general relativity (GR), additional matter fields and so on (see \cite{copeland2006dynamics, dvali2000, yin2005,Dymnikova:2001ga,Dymnikova:2001jy,Mukhopadhyay:2007ed}). 

Among a wide range of alternatives, a scalar field is a viable candidate to be used  with a broad range of forms of the potentials. Its usage includes the canonical scalar field, called `quintessence' \cite{peebles1988,ratra1988,Frieman1992,Frieman1995,Caldwell:1997ii},  and the scalar field with the opposite-sign in the kinetic term, known as `phantom' \cite{Caldwell:1999ew,Caldwell:2003vq}. One of the first proposals for scalar potential was the inverse power-law. Although non-renormalizable, such a potential has the remarkable property that it leads to the attractor-like behavior for the equation of state and density parameter of the  dark energy, which are nearly constants for a wide range of initial conditions, thus alleviating the so-called `coincidence problem' \cite{Zlatev:1998tr,Steinhardt:1999nw}.\footnote{ This problem arises because there is no apparent reason for the dark energy and matter densities to be of the same order today.} Beyond the real scalar field case, a complex quintessence was also used in \cite{Gu2001} to account for the acceleration of the universe. The $U(1)$ symmetry associated with this complex scalar leads to a more sophisticated structure for the dark sector, and unless the Standard Model of particle physics is a very special case in the nature, there is no reason (apart from simplicity) not to consider a richer physics of the dark sector.\footnote{ A current example of a vector field that perhaps interacts with dark matter is the so-called `dark photon' (see \cite{Essig:2013lka} for a quick review).}

Scalar fields also appear in  supersymmetric theories, such as the supersymmetric version of GR, known as supergravity. Since supergravity is also the low-energy limit of the superstring theory, it is a natural option to investigate if it can furnish a model that describes the accelerated expansion of the universe, where the scalar field plays the role of the dark energy. In the framework of minimal supergravity, Refs. \cite{Brax1999,Copeland2000} were first the attempts to describe dark energy through quintessence.

Another dark energy candidate is a spin-1 particle, described by a vector field. To be consistent with the homogeneity and isotropy of the universe there should be three identical copies of the vector field, each one with the same magnitude and pointing  in mutually orthogonal directions. They are called cosmic triad  and were proposed in  \cite{ArmendarizPicon:2004pm}. Other possibilities of vector dark energy are shown in \cite{Koivisto:2008xf,Bamba:2008ja,Emelyanov:2011ze,Emelyanov:2011wn,Emelyanov:2011kn,Kouwn:2015cdw}.

 Still regarding a dynamical dark energy, there may exist a possibility of interaction between dark energy and dark matter \cite{Wetterich:1994bg,Amendola:1999er}, since their densities are comparable and, depending on the coupling used, the interaction can also alleviate the coincidence problem \cite{Zimdahl:2001ar,Chimento:2003iea}.  Phenomenological models have been widely explored in the literature \cite{Amendola:1999er,Guo:2004vg,Cai:2004dk,Guo:2004xx,Bi:2004ns,Gumjudpai:2005ry,yin2005,Wang:2005jx,Wang:2005pk,Wang:2005ph,Wang:2007ak,Costa:2013sva,Abdalla:2014cla,Costa:2014pba,Costa:2016tpb,Marcondes:2016reb}. On the other hand, field theory models that aim a consistent description of the dark energy/dark matter interaction are still few \cite{Farrar:2003uw,Abdalla:2012ug,D'Amico:2016kqm,micheletti2009}. 

In order to have a first idea whether a specific dark energy candidate is good enough to describe the current accelerated expansion of the universe or not, the dark energy candidate  can be considered in the presence of a barotropic fluid. Thus, the relevant evolution equations are converted   into an autonomous system and the asymptotic states of the cosmological models are analyzed. It has been done for uncoupled dark energy (quintessence, tachyon field and phantom field  for instance \cite{copeland1998,ng2001,Copeland:2004hq,Zhai2005,DeSantiago:2012nk,Dutta:2016bbs}) and coupled dark energy \cite{Amendola:1999er,Gumjudpai:2005ry,TsujikawaGeneral,amendola2006challenges,ChenPhantom,Landim:2015poa,Landim:2015uda,Mahata:2015lja}, but it remained to be done for a vector-like dark energy, whose interesting properties were explored in  \cite{ArmendarizPicon:2004pm}.   Furthermore, since a complex scalar field and coupled dark energy are generalizations of the real field and the uncoupled case, respectively,  we aimed to extend the analysis  for both possibilities together. We will explore these models in chapter \ref{dynanaly}, in the light of the linear dynamical systems theory.

In chapter \ref{sugrachapter} we also use  a complex scalar field as the starting point for another  dark energy model, based on supergravity. In the same chapter we explore as well the relation between scalar fields and an attempt to explain the acceleration from the holographic principle.   In this latter model we embedded a holographic dark energy model into minimal supergravity.

In chapter \ref{MDE} we explore another possibility of scalar field as a dark energy candidate. We propose a model of metastable dark energy, in which the dark energy is a scalar field with a potential given by the sum of even self-interactions up to order six. The parameters of the model can be adjusted in such a way that the difference between the energy of the true vacuum and the energy of the false one is the observed vacuum energy  ($10^{-47}$ GeV$^4$). Different models of false vacuum decay are found in \cite{Stojkovic:2007dw,Greenwood:2008qp,Abdalla:2012ug}. A different mechanism of metastable dark energy (although with same name) appeared recently in the literature \cite{Shafieloo:2016bpk}. In the same chapter, a dark $SU(2)_R$ model is presented, where the dark energy doublet and the dark matter doublet naturally interact with each other. Such an interaction opens a new window to investigate the dark sector from the point-of-view of particle physics. Models with $SU(2)_R$ symmetry are well-known in the literature as extensions of the standard model  introducing the so-called left-right symmetric models \cite{Aulakh:1998nn,Duka:1999uc,Dobrescu:2015qna,Dobrescu:2015jvn,Ko:2015uma}. Recently,  dark matter has also been taken into account  \cite{Bezrukov:2009th,Esteves:2011gk,An:2011uq,Nemevsek:2012cd,Bhattacharya:2013nya,Heeck:2015qra,Garcia-Cely:2015quu,Berlin:2016eem}.  However, there is no similar effort to insert dark energy in a model of particle physics. We begin to attack this issue in chapter \ref{MDE}, with the dark $SU(2)_R$ model.

Conclusions are presented in chapter \ref{conclusions}. This thesis is based on our papers \cite{Landim:2015upa,Landim:2016dxh,Landim:2015hqa,Landim:2015uda,Landim:2015poa,Landim:2016isc}, combined in chapters \ref{dynanaly}, \ref{sugrachapter} and \ref{MDE}. The author also collaborated in the works \cite{Marcondes:2016reb,Landim:2016gpz}. The former in presented in further detail in  \cite{Marcondes:2016zte} while the latter is a work in progress. 
We use Planck units $\hbar=c=8\pi G=k_B=1$ throughout the thesis, unless where is explicitly written.

%% file: cosmoBasics.tex
	 \chapter{Modern Cosmology}

In this chapter we review the necessary content on cosmology that will be used in the rest of the thesis. It is based on \cite{weinberg2008cosmology,Dodelson-Cosmology-2003,Kolb:1990vq}, excepting the scattering amplitude in Sect. \ref{DMboltz}, which we have calculated.

The universe seems to be homogeneous and isotropic (one part in about $10^{-5}$) when viewed at sufficient large scales. These observational evidences have been stated through the so-called `cosmological principle'. 

The assumption of homogeneity and isotropy leads to the unique metric (up to a coordinate transformation) that represents a spherically symmetric and isotropic universe to a set of freely falling observers \cite{Weinberg:100595}. This is now known as Friedmann--Lemaître--Robertson--Walker (FLRW) metric \cite{friedmann1,friedmann2,Lemaitre,robertson1,robertson2,walker}  and  is given by

\begin{equation}\label{FLRWmetric}
ds^2=-dt^2+a(t)^2\left[\frac{dr^2}{1-Kr^2}+r^2d\theta^2+r^2\sin^2\theta d\phi^2\right],
\end{equation}

\noindent where $a(t)$ is the scale factor, set to be one today ($a_0=1$). $K$ is the curvature and describes a spherical ($K=+1$), hyperspherical ($K=-1$)  or Euclidean ($K=0$) universe.

%%%%%%%%%%%%%%%%%%%%%%%%%%%%%%%%%%%%%%%%%%%%%%%%%%%%%%%%%%%%%%%%%%%%%%%%%%%%%%%%%%%%%%%%%%%%%%%%%%%%%%%%%%%%%%%%%%%%%%%%%%
\section{ Einstein equations from the action} We start with the Einstein-Hilbert action to get the Einstein equations. The action is

\begin{equation}
S_{EH}=\int{\sqrt{-g}R}\,d^4x\quad.
\label{eq:4.1}
\end{equation}

\noindent Variation with respect to the metric yields
\begin{equation}
\delta S_{EH}=\int{\delta(\sqrt{-g}R)}\,d^4x = \int{(R\delta\sqrt{-g}+\sqrt{-g}R_{\mu\nu}\delta g^{\mu\nu}+\sqrt{-g}g^{\mu\nu}\delta R_{\mu\nu})}\,d^4x
\label{eq:4.2}
\end{equation}

\noindent where we  evaluate separately each one of the three terms above. The Ricci tensor is given by
\begin{equation}
R_{\mu\lambda\nu}^{\lambda}= \frac{\partial \Gamma_{\mu\lambda}^{\lambda}}{\partial x^{\nu}} - \frac{\partial \Gamma_{\mu\nu}^{\lambda}}{\partial x^{\lambda}} + \Gamma_{\mu\lambda}^{\eta} \Gamma_{\nu\eta}^{\lambda} - \Gamma_{\mu\nu}^{\eta} \Gamma_{\lambda\eta}^{\lambda}\quad,
\label{eq:4.3}
\end{equation}

\noindent where 
\begin{equation}
\Gamma^{\lambda}_{\mu\nu}= \frac{1}{2}g^{\lambda\rho}\left(\frac{\partial g_{\rho\mu}}{\partial x^{\nu}} +\frac{\partial g_{\rho\nu}}{\partial x^{\mu}}-\frac{\partial g_{\mu\nu}}{\partial x^{\rho}}\right )\quad
\label{eq:1.2}
\end{equation}

\noindent is the Levi-Civita connection.\footnote{Connections are objects used in forming covariant derivatives. The torsion (absent in GR) and the Levi-Civita connection are parts of the affine connection.} The change in the Ricci tensor is
\begin{equation}
\delta R_{\mu\lambda\nu}^{\lambda}= \frac{\partial \delta\Gamma_{\mu\lambda}^{\lambda}}{\partial x^{\nu}} - \frac{\partial \delta\Gamma_{\mu\nu}^{\lambda}}{\partial x^{\lambda}} + \delta\Gamma_{\mu\lambda}^{\eta} \Gamma_{\nu\eta}^{\lambda} +\Gamma_{\mu\lambda}^{\eta} \delta\Gamma_{\nu\eta}^{\lambda}- \delta\Gamma_{\mu\nu}^{\eta} \Gamma_{\lambda\eta}^{\lambda}-\Gamma_{\mu\nu}^{\eta} \delta\Gamma_{\lambda\eta}^{\lambda}\quad.
\label{eq:4.4}
\end{equation}

 Once $\delta\Gamma_{\nu\mu}^{\rho}$ is the difference of two connections, it is a tensor and we can thus calculate its covariant derivative
\begin{equation}
\nabla_{\lambda}(\delta\Gamma_{\nu\mu}^{\rho})= \frac{\partial \delta\Gamma_{\nu\mu}^{\rho}}{\partial x^{\lambda}}+ \Gamma_{\lambda\sigma}^{\rho}\delta\Gamma_{\nu\mu}^{\sigma}-\Gamma_{\lambda\nu}^{\sigma}\delta\Gamma_{\sigma\mu}^{\rho}-\Gamma_{\lambda\mu}^{\sigma}\delta\Gamma_{\nu\sigma}^{\rho}\quad.
\label{eq:4.5}
\end{equation}

\noindent Using Eq. (\ref{eq:4.5}) and doing the following two sets of substitutions in the equation above: 1) $\nu \rightarrow \mu$, $\mu \rightarrow \lambda$, $\rho \rightarrow \lambda$, $\lambda \rightarrow \nu$; 2) $\nu \rightarrow \mu$, $\mu \rightarrow \nu$, $\lambda \rightarrow \lambda$, $\rho \rightarrow \lambda$, we have 

\begin{equation}
\delta R_{\mu\lambda\nu}^{\lambda}= \nabla_{\nu}\delta\Gamma_{\mu\lambda}^{\lambda}-\nabla_{\lambda}\delta\Gamma_{\mu\nu}^{\lambda}\quad.
\label{eq:4.6}
\end{equation}

\noindent Therefore, the contribution of the third term in Eq. (\ref{eq:4.2}) is 
\begin{align}
(\delta S_{EH})_3 =& \int{\sqrt{-g}g^{\mu\nu}\delta R_{\mu\nu}}\,d^4x = \int{\sqrt{-g}g^{\mu\nu}\left[\nabla_{\nu}\delta\Gamma_{\mu\lambda}^{\lambda}-\nabla_{\lambda}\delta\Gamma_{\mu\nu}^{\lambda}\right]}\,d^4x\quad,\notag\\
(\delta S_{EH})_3 = & \int{\sqrt{-g}\left[\nabla_{\nu}(g^{\mu\nu}\delta\Gamma_{\mu\lambda}^{\lambda})-\nabla_{\lambda}(g^{\mu\nu}\delta\Gamma_{\mu\nu}^{\lambda})\right]}\,d^4x\quad,
\label{eq:4.82}
\end{align}

\noindent where we used $\nabla_{\sigma}g^{\mu\nu}=0$. We have
\begin{equation}
\Gamma^{\lambda}_{\mu\lambda}= \frac{1}{2}g^{\lambda\rho}\left(\frac{\partial g_{\rho\mu}}{\partial x^{\lambda}} +\frac{\partial g_{\rho\lambda}}{\partial x^{\mu}}-\frac{\partial g_{\mu\lambda}}{\partial x^{\rho}}\right )= \frac{1}{2}g^{\lambda\rho}\frac{\partial g_{\rho\mu}}{\partial x^{\lambda}}\quad,
\label{eq:4.9}
\end{equation}

\noindent since the two last terms cancel each other (just change $\mu \leftrightarrow\rho$). Now, using the following relation for an arbitrary matrix M,

\begin{equation}
\frac{1}{\text{Det }M}\frac{\partial}{\partial x^{\lambda}} (\text{Det }M) = \text{Tr}\left(M^{-1}\frac{\partial}{\partial x^{\lambda}} M\right)\quad,
\label{eq:4.10}
\end{equation}

\noindent and taking $M$ to be $g_{\mu\nu}$, so Det $ M = $ Det $g_{\mu\nu} = g$, we have

\begin{equation}
g^{\lambda\rho}\frac{\partial g_{\rho\mu}}{\partial x^{\lambda}}= \frac{1}{g}\frac{\partial g}{\partial x^{\lambda}}\quad.
\label{eq:4.10.1}
\end{equation}

\noindent Then, using Eq. (\ref{eq:4.10.1}) in Eq. (\ref{eq:1.2}) we get

\begin{equation}
\Gamma^{\lambda}_{\mu\lambda}= \frac{1}{2g}\frac{\partial g}{\partial x^{\lambda}}= \frac{1}{\sqrt{-g}}\frac{\partial \sqrt{-g}}{\partial x^{\lambda}}\quad.
\label{eq:4.10.2}
\end{equation}
\noindent Furthermore, since

\begin{equation}
\nabla_{\lambda} V^{\lambda} = \frac{\partial V^{\lambda}}{\partial x^{\lambda}} + \Gamma^{\lambda}_{\mu\lambda}V^{\mu}\quad,
\label{eq:4.10.3}
\end{equation}

\noindent we use Eq. (\ref{eq:4.10.2}) in Eq. (\ref{eq:4.10.3}) and we get

\begin{equation}
\nabla_{\lambda} V^{\lambda} =  \frac{\partial V^{\lambda}}{\partial x^{\lambda}} + \frac{1}{\sqrt{-g}}\frac{\partial \sqrt{-g}}{\partial x^{\lambda}}V^{\mu} =\frac{1}{\sqrt{-g}}\frac{\partial }{\partial x^{\lambda}}(\sqrt{-g}V^{\mu})\quad.
\label{eq:4.10.4}
\end{equation}

\noindent Inserting Eq. (\ref{eq:4.10.4}) in Eq. (\ref{eq:4.82}) results

\begin{equation}
(\delta S_{EH})_3 = \int{\left[\frac{\partial }{\partial x^{\nu}}(\sqrt{-g}g^{\mu\nu}\delta\Gamma_{\mu\lambda}^{\lambda})-\frac{\partial }{\partial x^{\lambda}}(\sqrt{-g}g^{\mu\nu}\delta\Gamma_{\mu\nu}^{\lambda})\right]}\,d^4x=0\quad,
\label{eq:4.10.5}
\end{equation}

\noindent since  each term in the integrand vanishes at infinity.

To evaluate the first term in Eq. (\ref{eq:4.2}) we consider again Eq. (\ref{eq:4.10}). We have
\begin{equation}
\delta g=g(g^{\mu\nu}\delta g_{\mu\nu})\quad.
\label{eq:4.11}
\end{equation}

\noindent Furthermore, since
\begin{equation}
\begin{split}
\delta (g_{\mu\nu}g^{\nu\sigma}) =  \delta g_{\mu\nu}g^{\nu\sigma} +g_{\mu\nu}\delta g^{\nu\sigma}=\delta (\delta_{\mu}^{\sigma})=0\quad,
\end{split}
\label{eq:4.12}
\end{equation}

\noindent multiplying the equation above by $g^{\mu\rho}$ yields
\begin{equation}
\begin{split}
g^{\mu\rho} \delta g_{\mu\nu}g^{\nu\sigma} +g^{\mu\rho}g_{\mu\nu}\delta g^{\nu\sigma}=0\quad,\\
g^{\mu\rho}g^{\nu\sigma} \delta g_{\mu\nu} +g^{\rho}_{\nu}\delta (g^{\nu}_{\rho}g^{\rho\sigma})=0\quad,\\
g^{\mu\rho}g^{\nu\sigma} \delta g_{\mu\nu} +g^{\rho}_{\nu}\delta g^{\nu}_{\rho}g^{\rho\sigma}+g^{\rho}_{\nu} g^{\nu}_{\rho}\delta g^{\rho\sigma}=0\quad,\\
g^{\mu\rho}g^{\nu\sigma} \delta g_{\mu\nu} +g^{\rho}_{\nu}\delta (\delta^{\nu}_{\rho})g^{\rho\sigma}+g^{\nu}_{\nu}\delta g^{\rho\sigma}=0\quad,\\
\delta g^{\rho\sigma}=-g^{\mu\rho}g^{\nu\sigma} \delta g_{\mu\nu}\quad.
\end{split}
\label{eq:4.13}
\end{equation}

\noindent Using the equation above with $\rho=\nu$ and $\sigma=\mu$ we can write Eq. (\ref{eq:4.11}) as
\begin{equation}
\delta g=g(g^{\mu\nu}\delta g_{\mu\nu})=-g(g_{\mu\nu}\delta g^{\mu\nu})\quad.
\label{eq:4.14}
\end{equation}

 Using Eq. (\ref{eq:4.14}) we write

\begin{equation}
\delta \sqrt{-g}= -\frac{1}{2\sqrt{-g}}\delta g=\frac{g}{2\sqrt{-g}}g_{\mu\nu}\delta g^{\mu\nu}=-\frac{1}{2}\sqrt{-g}g_{\mu\nu}\delta g^{\mu\nu}\quad.
\label{eq:4.151}
\end{equation}

\noindent Putting Eq. (\ref{eq:4.151}) in Eq. (\ref{eq:4.2}) yields
\begin{align}
\delta S_{EH}= &\int{\left(\sqrt{-g}R_{\mu\nu}\delta g^{\mu\nu}-R\frac{1}{2}\sqrt{-g}g_{\mu\nu}\delta g^{\mu\nu}\right)}\,d^4x\quad,\notag\\
\delta S_{EH}= & \int{\sqrt{-g}\left(R_{\mu\nu}-\frac{1}{2}g_{\mu\nu}R\right)\delta g^{\mu\nu}}\,d^4x=0\quad.
\label{eq:4.17}
\end{align}

\noindent Finally, as $\delta S_{EH} = 0$ we should have
\begin{equation}
R_{\mu\nu}-\frac{1}{2}g_{\mu\nu}R=0\quad,
\label{eq:4.18}
\end{equation}

\noindent which are the Einstein equations in vacuum. If matter fields are also present, the variation of the correspondent action $S_m$ due to these fields with respect to the metric gives the energy-momentum tensor
\begin{equation}\label{eq:3}
  T_{\mu\nu}=-\frac{2}{\sqrt{-g}}\frac{\delta S}{\delta g^{\mu\nu}}\quad,
\end{equation}

\noindent therefore, the Einstein equations are 
\begin{equation}
R_{\mu\nu}-\frac{1}{2}g_{\mu\nu}R=8\pi G T_{\mu\nu}\quad.
\label{eq:4.19}
\end{equation}

We are going to obtain now an important result from the Bianchi identities, namely, the conservation of the energy-momentum tensor.

%%%%%%%%%%%%%%%%%%%%%%%%%%%%%%%%%%%%%%%%%%%%%%%%%%%%%%%%%%%%%%%%%%%%%%%%%%%%%%%%%%%%%%%%%%%%%%%%%%%%%%%
\subsection{The Bianchi identities} In order to get the Bianchi identities we use the Riemann tensor

\begin{equation}
R_{\mu\nu\kappa}^{\lambda}= \frac{\partial \Gamma_{\mu\nu}^{\lambda}}{\partial x^{\kappa}} - \frac{\partial \Gamma_{\mu\kappa}^{\lambda}}{\partial x^{\nu}} + \Gamma_{\mu\nu}^{\eta} \Gamma_{\kappa\eta}^{\lambda} - \Gamma_{\mu\kappa}^{\eta} \Gamma_{\nu\eta}^{\lambda}\quad,
\label{eq:4,1}
\end{equation}

\noindent and the Levi-Civita connection (\ref{eq:1.2}). Adopting a locally inertial frame in which $\Gamma_{\mu\nu}^{\lambda}$ (but not its derivatives) is zero, the covariant derivative of the Riemann tensor is
 \begin{equation}
R_{\mu\nu\kappa;\sigma}^{\lambda}= \frac{\partial}{\partial x^\sigma}\left(\frac{\partial \Gamma_{\mu\nu}^{\lambda}}{\partial x^{\kappa}} - \frac{\partial \Gamma_{\mu\kappa}^{\lambda}}{\partial x^{\nu}} \right)\quad.
\label{eq:4,2}
\end{equation}

Using Eq. (\ref{eq:1.2}) and deriving it we have

\begin{equation}
\frac{\partial \Gamma^{\lambda}_{\mu\nu}}{\partial x^\kappa}= \frac{1}{2}g^{\lambda\rho}\left(\frac{\partial^2 g_{\rho\mu}}{\partial x^\kappa\partial x^{\nu}} +\frac{\partial^2 g_{\rho\nu}}{\partial x^\kappa\partial x^{\mu}}-\frac{\partial^2 g_{\mu\nu}}{\partial x^\kappa\partial x^{\rho}}\right )\quad,
\label{eq:4,3}
\end{equation}

\noindent where we used the fact that $\frac{\partial g^{\lambda\rho}}{\partial x^\kappa}=0$. Thus

 \begin{align}
R_{\mu\nu\kappa}^{\lambda}=& \frac{1}{2}g^{\lambda\rho}\left(\frac{\partial^2 g_{\rho\mu}}{\partial x^\kappa\partial x^{\nu}} +\frac{\partial^2 g_{\rho\nu}}{\partial x^\kappa\partial x^{\mu}}-\frac{\partial^2 g_{\mu\nu}}{\partial x^\kappa\partial x^{\rho}}-  \frac{\partial^2 g_{\rho\mu}}{\partial x^{\nu}\partial x^\kappa} -\frac{\partial^2 g_{\rho\kappa}}{\partial x^\nu\partial x^{\mu}}+\frac{\partial^2 g_{\mu\kappa}}{\partial x^\nu\partial x^{\rho}}\right )\quad,\notag\\
=& \frac{1}{2}g^{\lambda\rho}\left(\frac{\partial^2 g_{\rho\nu}}{\partial x^\kappa\partial x^{\mu}}-\frac{\partial^2 g_{\mu\nu}}{\partial x^\kappa\partial x^{\rho}} -\frac{\partial^2 g_{\rho\kappa}}{\partial x^\nu\partial x^{\mu}}+\frac{\partial^2 g_{\mu\kappa}}{\partial x^\nu\partial x^{\rho}}\right )\quad.
\label{eq:4,5}
\end{align}

\noindent But
\begin{equation}
g_{\alpha\lambda}R_{\mu\nu\kappa}^{\lambda}= R_{\alpha\mu\nu\kappa}=  \frac{1}{2}\delta_\alpha^\rho\left(\frac{\partial^2 g_{\rho\nu}}{\partial x^\kappa\partial x^{\mu}}-\frac{\partial^2 g_{\mu\nu}}{\partial x^\kappa\partial x^{\rho}} -\frac{\partial^2 g_{\rho\kappa}}{\partial x^\nu\partial x^{\mu}}+\frac{\partial^2 g_{\mu\kappa}}{\partial x^\nu\partial x^{\rho}}\right )\quad,
\label{eq:4,6}
\end{equation}

\noindent from where we get the result
\begin{equation}
R_{\alpha\mu\nu\kappa;\sigma}=  \frac{1}{2}\frac{\partial}{\partial x^\sigma}\left(\frac{\partial^2 g_{\alpha\nu}}{\partial x^\kappa\partial x^{\mu}}-\frac{\partial^2 g_{\mu\nu}}{\partial x^\kappa\partial x^{\alpha}} -\frac{\partial^2 g_{\alpha\kappa}}{\partial x^\nu\partial x^{\mu}}+\frac{\partial^2 g_{\mu\kappa}}{\partial x^\nu\partial x^{\alpha}}\right )\quad.
\label{eq:4,7}
\end{equation}

Writing the Eq. (\ref{eq:4,7}) with  $\nu$, $\kappa$ and $\sigma$ permuted cyclically we have
\begin{equation}
\begin{split}
 R_{\alpha\mu\nu\kappa;\sigma}+ R_{\alpha\mu\sigma\nu;\kappa} +R_{\alpha\mu\kappa\sigma;\nu} =
 \frac{1}{2}\frac{\partial}{\partial x^\sigma}(g_{\alpha\nu; \kappa\mu}-g_{\mu\nu; \kappa\alpha}-g_{\alpha\kappa; \nu\mu}+g_{\mu\kappa;\nu\alpha})+ \\
\frac{1}{2} \frac{\partial}{ \partial x^\kappa}(g_{\alpha\sigma; \nu\mu}-g_{\mu\sigma; \nu\alpha}-g_{\alpha\nu; \sigma\mu}+g_{\mu\nu; \sigma\alpha})+\\
 \frac{1}{2}\frac{\partial}{\partial x^\nu}(g_{\alpha\kappa; \sigma\mu}-g_{\mu\kappa; \sigma\alpha}-g_{\alpha\sigma; \kappa\mu}+g_{\mu\sigma, \kappa\alpha})\quad.
 \end{split}
\label{eq:4,8}
\end{equation}

\noindent After applying the derivatives we get
\begin{equation}
\begin{split}
 R_{\alpha\mu\nu\kappa;\sigma}+ R_{\alpha\mu\sigma\nu;\kappa} +R_{\alpha\mu\kappa\sigma;\nu} =\frac{1}{2}(g_{\alpha\nu; \sigma\kappa\mu}-g_{\mu\nu; \sigma\kappa\alpha} -g_{\alpha\kappa; \sigma\nu\mu}  +g_{\mu\kappa;\sigma\nu\alpha} + \\ g_{\alpha\sigma; \kappa\nu\mu} -g_{\mu\sigma; \kappa\nu\alpha}-g_{\alpha\nu; \kappa\sigma\mu} +g_{\mu\nu; \kappa\sigma\alpha}+\\
 g_{\alpha\kappa; \nu\sigma\mu}-g_{\mu\kappa; \nu\sigma\alpha}-g_{\alpha\sigma; \nu\kappa\mu} +g_{\mu\sigma; \nu\kappa\alpha})\quad.
 \end{split}
\label{eq:4,9}
\end{equation}

 The terms cancel in pairs, therefore
\begin{equation}
 R_{\alpha\mu\nu\kappa;\sigma}+ R_{\alpha\mu\sigma\nu;\kappa} +R_{\alpha\mu\kappa\sigma;\nu} =0\quad,
\label{eq:4,10}
\end{equation}

\noindent which are known as the Bianchi identities.

Multiplying Eq. (\ref{eq:4,10}) by $g^{\alpha\nu}$ and using the fact that the covariant derivative of the metric is zero, we have
\begin{align}
(g^{\alpha\nu}R_{\alpha\mu\nu\kappa})_{;\sigma}+ (g^{\alpha\nu}R_{\alpha\mu\sigma\nu})_{;\kappa} +(g^{\alpha\nu}R_{\alpha\mu\kappa\sigma})_{;\nu} &=0\quad,\\
R^\nu_{\mu\nu\kappa;\sigma}+ R^\nu_{\mu\sigma\nu;\kappa}+ R^\nu_{\mu\kappa\sigma;\nu} &=0\quad,\\
R^\nu_{\mu\nu\kappa;\sigma}- R^\nu_{\mu\nu\sigma;\kappa} +R^\nu_{\mu\kappa\sigma;\nu} &=0\quad.
\label{eq:4,11.2}
\end{align}

\noindent Thus 
\begin{equation}
 R_{\mu\kappa;\sigma}- R_{\mu\sigma;\kappa} +R_{\mu\kappa\sigma;\nu}^\nu =0\quad.
\label{eq:4,12}
\end{equation}

Contracting Eq. (\ref{eq:4,12}) once more  yields
\begin{equation}
\begin{split}
g^{\mu\kappa}(R_{\mu\kappa;\sigma}- R_{\mu\sigma;\kappa} +R_{\mu\kappa\sigma;\nu}^\nu) =0\quad,\\
R^\kappa_{\kappa;\sigma}- R^\kappa_{\sigma;\kappa} +R_{\kappa\sigma;\nu}^{\kappa\nu} =0\quad,\\
R^\kappa_{\kappa;\sigma}- R^\kappa_{\sigma;\kappa} -R_{\kappa\sigma;\nu}^{\nu\kappa} =0\quad,
\end{split}
\label{eq:4,13}
\end{equation}

\noindent thus,
\begin{equation}
\begin{split}
R_{;\sigma}- 2R^\kappa_{\sigma;\kappa} =0\quad,\\
R^\mu_{\sigma;\mu}-\frac{1}{2}\delta^\mu_\sigma R_{;\mu} =0\quad,\\
\left[g^{\nu\sigma}R^\mu_{\sigma}-\frac{1}{2}g^{\nu\sigma}g^\mu_\sigma R\right]_{;\mu} =0\quad.
\end{split}
\label{eq:4,14}
\end{equation}

\noindent We finally get
\begin{equation}
\left[R^{\mu\nu}-\frac{1}{2}g^{\mu\nu} R\right]_{;\mu} =0\quad.
\label{eq:4.15}
\end{equation}

This result shows that the covariant derivative of the left-hand side of Eq. (\ref{eq:4.19}) is null, so it must be so for the right-hand side. Therefore

\begin{equation}
(T^{\mu\nu})_{;\mu} =0\quad.
\label{eq:4,15}
\end{equation}

\noindent The Bianchi identities in GR imply the conservation of the energy-momentum tensor.

\section{Energy-momentum tensor of a perfect fluid}

The energy-momentum tensor for a continuous distribution (dense collection) of particles is given by

\begin{equation}\label{tmunu1}
T^{\mu\nu}=\frac{1}{\Delta V_4}\int T^{\mu\nu}_{p} \,dV = \frac{1}{\sqrt{-g}d^3x^idx^0}\int_{\Delta V_4} T^{\mu\nu}_{p}\sqrt{-g}\,d^3x^i\,dx^0\quad,
\end{equation}

\noindent where 
\begin{equation}\label{Tmunup2}
T^{\mu\nu}_{p}= \frac{1}{d^3x^idx^0}\sum_a{\frac{p_a^{\mu}p_a^{\nu}}{p^0_a}} \delta^{(3)}(\mathbf{x}-\mathbf{x^{(a)}})
\end{equation}

\noindent is the energy-momentum of a system of discrete particles. Using Eq. (\ref{Tmunup2}) into Eq. (\ref{tmunu1}) we have
\begin{equation}
T^{\mu\nu}= \frac{1}{d^3x^idx^0}\sum_a{\frac{p_a^{\mu}p_a^{\nu}}{p^0_a}}\int_{\Delta V_4} \delta^{(3)}(\mathbf{x}-\mathbf{x^{(a)}})\,d^3x^i\,dx^0 = \frac{1}{d^3x^i}\sum_a{\frac{p_a^{\mu}p_a^{\nu}}{p^0_a}}\quad.
\end{equation}

\noindent Considering the element $\mu=\nu=0$ we have
\begin{equation}
T^{00}=  \frac{1}{d^3x^i}\sum_a{p_a^{0}}\equiv \rho\quad,
\label{eq:2.100}
\end{equation}

\noindent since $p_a^{0}=E$. For
\begin{equation}
T^{0i}=  \frac{1}{d^3x^i}\sum_a{p_a^{i}}\quad,
\end{equation}

\noindent it should be zero due to isotropy, because otherwise it would have a preferable direction. Considering now the spatial terms
\begin{equation}
T^{ij}=  \frac{1}{d^3x^i}\sum_a{\frac{p_a^{i}p_a^{j}}{p^0_a}}\quad,
\end{equation}

\noindent which once again is different from zero only for $i=j$, since the particles have isotropically distributed velocities. The second term in the equation above is dimensionally equal to pressure (where we considered $c=1$), thus 
\begin{equation}
T^{ij}=  \delta^{ij} p\quad.
\label{eq:2.2}
\end{equation}

\noindent We get the perfect fluid energy-momentum tensor for the rest frame
\begin{equation}
T^{\mu\nu} = 
\begin{pmatrix}
	\rho && 0 && 0 && 0
	\\ 0 && p && 0 && 0
	\\ 0 && 0 && p && 0
	\\0 && 0&& 0&& p
\end{pmatrix}\quad. 
\label{eq:2.3}
\end{equation}

\noindent Improving this tensor for any frame we have
\begin{equation}
T^{\mu\nu}=  (\rho + p)U^{\mu}U^{\nu} + pg^{\mu\nu}\quad,
\end{equation}

\noindent which is properly reduced to Eq. (\ref{eq:2.3}) in the rest frame.

\section{Continuity equation }

The energy-momentum conservation (\ref{eq:4,15})  for a perfect fluid described by Eq. (\ref{eq:2.3}) leads to the  continuity equation 
\begin{equation}\label{eq:continuity}
  \dot{\rho}+3H(\rho+p)=0\quad,
\end{equation}

\noindent where $H=\dot{a}/a$. The equation above can be written as 
\begin{equation}\label{eq:continuity2}
 a^{-3} \frac{d(\rho a^{3})}{dt}=-3Hp\quad.
\end{equation}

Cosmological fluids are assumed to be barotropic, thus the energy density depends only on pressure and  is written as $p=w\rho$, where $w$ is called equation of state. For radiation, such as photon and massless neutrinos, the equation of state is $w=1/3$. This can be easily seen because the energy-momentum tensor is traceless for theories that present the invariance of the action under a Weyl transformation $\delta g_{\mu\nu}=\Omega(x) g_{\mu\nu}$, then $\rho=3p$. Ordinary matter and the majority of dark matter (called `cold dark mater' -- CDM) are described by fluids which are non-relativistic and therefore their pressure is very close to zero, hence $w\approx 0$. For the cases described in this paragraph, the continuity equation (\ref{eq:continuity}) has the following solutions

\begin{align}\label{eq:continuityRad}
 \rho &\propto a^{-4} \quad \text{for hot matter (or radiation)}\quad,\\
 \rho &\propto a^{-3} \quad \text{for cold matter (or dust)}\quad.
\end{align}

Another case of interest is for a fluid that has negative pressure and equation of state $w=-1$. It is known as `cosmological constant' ($\Lambda$) or `vacuum energy' and it has constant energy density $\rho$. Alternatives to this fluid will be presented as candidates to explain the current acceleration of the universe.

\subsection{Interacting dark energy}\label{deint}

We assume that dark energy is coupled with dark matter, in such a way that total energy-momentum is still conserved. In the flat FLRW background with a scale factor $a$, the continuity equations for both components and for radiation are \cite{Amendola:1999er}
\begin{align}\label{contide1}
\dot{\rho_{de}}+3H(\rho_{de}+p_{de})&=-\mathcal{Q}\quad,\\
\dot{\rho_m}+3H\rho_m &=\mathcal{Q}\quad,\\
\dot{\rho_r}+4H\rho_r &=0\quad,
\end{align}

\noindent respectively, where $H=\dot{a}/a$ is the Hubble rate,  $\mathcal{Q}$ is the coupling and the dot is a derivative with respect to the cosmic time $t$. The indices $de$, $m$ and $r$ stand for dark energy, dark matter and radiation, respectively. The case of $\mathcal{Q}>0$ corresponds to a dark-energy transformation into dark matter, while $\mathcal{Q}<0$ is the transformation in the opposite direction. In principle, the coupling  can depend on several variables $\mathcal{Q}=\mathcal{Q}(\rho_m,\rho_\phi, \dot{\phi},H,t,\dots)$, but assuming it is a small quantity, the interaction term can be written as a Taylor expansion, giving rise to three kernels, for instance: (i) $Q\propto H\rho_M$, (ii)   $Q\propto H\rho_D$ and (iii)  $Q\propto H(\rho_M+\rho_D)$. In the next chapters we will explore the interacting dark energy.

%%%%%%%%%%%%%%%%%%%%%%%%%%%%%%%%%%%%%%%%%%%%%%%%%%%%%%%%%%%%%%%%%%%%%%%%%%%%%%%%%%%%%%%%%%%%%%%%%%%%%%%%%%%%%%%%%%
\section {Friedmann equations }

The FLRW metric (\ref{FLRWmetric}) is written in a matrix form as

\begin{equation}
g_{\mu\nu} = 
\begin{pmatrix}
	-1 & 0 & 0 & 0
	\\ 0 & \frac{a^2(t)}{1+\Omega_kH^2_0r^2} & 0 & 0
	\\ 0 & 0 &a^2(t)r^2 & 0
	\\0 & 0& 0& a^2(t)r^ 2\sin^2\theta
\end{pmatrix}\quad. 
\label{eq:2.1}
\end{equation}
\noindent The Christoffel symbols (\ref{eq:1.2}) for $\lambda=i$, $\mu=0$ and $\nu=j$ are
\begin{equation}
\Gamma^{i}_{0j}= \frac{1}{2}g^{ii}(g_{i0,j}+g_{ij,0}-g_{0j,i})\quad,
\label{eq:2.31}
\end{equation}

\noindent where the first and last terms are zero. Thus
\begin{equation}
\Gamma^{i}_{0j} = \delta_{ij}\frac{\dot{a}(t)}{a(t)}=\delta_j^iH\quad.
\label{eq:2.5}
\end{equation}

\noindent It is easy to see that if $j=0$ in Eq. (\ref{eq:2.31}) $\Gamma$ would be zero. For $\lambda=0$, $\mu=i$ and $\nu=j$ we have
\begin{equation}
\Gamma^{0}_{ij}= \frac{1}{2}g^{00}(g_{0i,j}+g_{0j,i}-g_{ij,0})\quad,
\label{eq:2.6}
\end{equation}

\noindent where the first and second terms are zero. Thus
\begin{equation}
\Gamma^{0}_{ij}= \frac{\dot{a}(t)}{a(t)}g_{ij}=Hg_{ij}\quad.   
\label{eq:2.7}
\end{equation}

\noindent From Eq. (\ref{eq:2.6}) we see that if $i=0$ and/or $j=0$, $\Gamma$ would be zero. For $\lambda=i$, $\mu=j$ and $\nu=k$ we have
\begin{equation}
\Gamma^{i}_{jk}= \frac{1}{2}g^{il}(g_{lj,k}+g_{lk,j}-g_{jk,l})\quad.
\label{eq:2.8}
\end{equation}

\noindent We first calculate  $R_{00}$
\begin{equation}
R_{00}=  \Gamma_{00,\lambda}^{\lambda}- \Gamma_{0\lambda,0}^{\lambda}  +  \Gamma_{00}^{\eta} \Gamma_{\lambda\eta}^{\lambda}-\Gamma_{0\lambda}^{\eta} \Gamma_{0\eta}^{\lambda}\quad.
\label{eq:2.9}
\end{equation}

\noindent The first and the third terms are zero, since there is no $\Gamma_{00}$ and $\eta=\lambda$. Thus
\begin{equation}
\begin{split}
R_{00}=  -\delta^i_j\frac{\ddot{a}}{a}+\delta^i_j\frac{\dot{a}^2}{a^2}  -\delta^ i_j\left(\frac{\dot{a}}{a}\right)^2=-3\frac{\ddot{a}}{a}\quad,
\end{split}
\label{eq:2.10}
\end{equation}

\noindent For $R_{ij}$ we have
\begin{equation}
R_{ij}=  \Gamma_{ij,\lambda}^{\lambda}- \Gamma_{i\lambda,j}^{\lambda}  +  \Gamma_{ij}^{\eta} \Gamma_{\lambda\eta}^{\lambda}-\Gamma_{j\lambda}^{\eta} \Gamma_{i\eta}^{\lambda}\quad.
\label{eq:2.12}
\end{equation}

\noindent Evaluating the Christoffel symbols we have
\begin{equation}
\Gamma^{i}_{jk}= \frac{1}{2}g^{il}(g_{lj,k}+g_{lk,j}-g_{jk,l})\quad,
\label{eq:2.121}
\end{equation}

\noindent where we can see that $\Gamma$ is zero if $i\neq j\neq k$. We get 
\begin{align}
R_{ij}=g_{ij}\left(\frac{\ddot{a}}{a}+2H^2-\frac{2\Omega_kH_0^2}{a^2}\right)\quad.
\end{align}

The Ricci scalar is
\begin{equation}
R=  g^{\mu\nu}R_{\mu\nu} = g^{00}R_{00} + g^{ij}R_{ij} = 3\frac{\ddot{a}}{a} + 3\left[\frac{\ddot{a}}{a} + 2H^2 - \frac{2\Omega_kH_0^2}{a^2}\right]\quad,
\label{eq:2.c1}
\end{equation}

\noindent therefore
\begin{equation}
R=  6\left[\frac{\ddot{a}}{a} + H^2 - \frac{\Omega_kH_0^2}{a^2}\right]\quad.
\label{eq:2.c2}
\end{equation}

\noindent The time-time component of Einstein equations are
\begin{equation}
R_{00} -\frac{1}{2}g_{00}R=8\pi GT_{00}\quad,
\label{eq:2.c3}
\end{equation}

\noindent and using Eqs. (\ref{eq:2.10}) and (\ref{eq:2.c2}) we get
\begin{align}
H^2 =& \frac{8\pi G\rho}{3}+\frac{\Omega_kH_0^2}{a^2}\quad, \notag \\
 =& \frac{8\pi G}{3}\left[\rho+\frac{\Omega_k\rho_{cr}}{a^2}\right]\quad,
\label{eq:2.c4}
\end{align}

\noindent where we used $H_0^2=8\pi G\rho_{cr}/3$ and $\rho_{cr}$ is the critical density. This is known as the first Friedmann equation. The spatial components of the Einstein  equations is 
\begin{align}
R_{ij} -\frac{1}{2}g_{ij}R=8\pi GT_{ij}
\label{eq:2.c5}
\end{align}

\noindent and using Eq. (\ref{eq:2.c4}) they lead to the second Friedmann equation
\begin{align}
\frac{\ddot{a}}{a} =- \frac{4\pi G}{3}(\rho+3p)\quad.
\label{eq:2.c6}
\end{align}

\noindent  The accelerated expansion occurs for $\rho+3p<0$. For a barotropic fluid ($p=w\rho$)  the accelerated expansion occurs for $w< -1/3$.

Observations show that the universe is close to spatially flat geometry ($|\Omega_k|<0.005$) \cite{Ade:2015xua}, so the curvature will be ignored in the rest of the thesis.  

For non-relativistic matter $\rho_m=\rho_{m0} a^{-3}$ (with $a_0=1$) and the solution of Eq. (\ref{eq:2.c4}) for zero curvature is
\begin{align}
a \propto t^{2/3}\quad,
\label{eq:2.c7}
\end{align}
\noindent thus the Hubble parameter becomes
\begin{align}
\label{eq:2.c8}
H=\frac{2}{3t}\quad.\end{align}
For relativistic matter $\rho_r=\rho_{r0} a^{-4}$ (with $a_0=1$) and the solution of Eq. (\ref{eq:2.c4}) for zero curvature is
\begin{align}
a \propto t^{1/2}\quad,
\label{eq:2.c9}
\end{align}

\noindent and the Hubble parameter is given by
\begin{align}
\label{eq:2.c10}
H=\frac{1}{2t}\quad.\end{align}

Since for the vacuum energy $\rho_\Lambda$ is constant, as so $H$, the solution of Eq. (\ref{eq:2.c4}) for this case is
\begin{align}
a\propto \exp(Ht)\quad.
\label{eq:2.c11}
\end{align}

More generally, for  a mixture of vacuum energy and relativistic and non-relativistic matter, as fractions of the critical density $\rho_{cr0}$, we have
\begin{align}
\rho=\frac{3H_0^2}{8\pi G}\left[\Omega_\Lambda+\Omega_m\left(\frac{a_0}{a}\right)^3+\Omega_r\left(\frac{a_0}{a}\right)^4\right]\quad,
\label{eq:2.c12}
\end{align}

\noindent where the quantity $\Omega_i$ is called density parameter and it is defined as
\begin{align}
\Omega_i=\frac{\rho_{i0}}{\rho_{cr0}}\quad.
\label{eq:2.c13}
\end{align}

The first Friedmann equation is written in terms of the density parameters as
\begin{align}
\Omega_\Lambda+\Omega_r+\Omega_m=1\quad.
\label{eq:2.c14}
\end{align}

%%%%%%%%%%%%%%%%%%%%%%%%%%%%%%%%%%%%%%%%%%%%%%%%%%%%%%%%%%%%%%%%%%%%%%%%%%%%%%%%%%%%%%%%%%%%%%%%%%%%%%%%%%%%%%%%%%%%%%%%%
\section{Beyond thermal equilibrium}

The temperature of the universe is usually taken as the cosmic microwave background (CMB) temperature. The CMB are microwave photons that have today a temperature of $2.75$ K. Besides photons, radiation is composed also by the three neutrinos species with a temperature of $1.96$ K. Although massive, their masses are supposed to be of the order of a electron-volt and can be regarded as relativistic. The early universe was to a good approximation in thermal equilibrium, thus there should have been other relativistic particles present, with comparable abundances. 

The general relativistic expression for the energy-momentum tensor in terms of the distribution functions $f_i(\vec{p},t)$ is given by
\begin{equation}
T_\nu^\mu= g_i\int \,\frac{dP_1dP_2dP_3}{(2\pi)^3\sqrt{-g}}\frac{P^\mu P_\nu}{P^0}f_i(\vec{p},t)\quad,
\label{eq:3.100}
\end{equation}

\noindent where $g_i$ is the number of spin states for species $i$  or species in kinetic equilibrium, $P^\mu$ is the four-momentum and $f_i(\vec{p},t)$ is given by
\begin{equation}
f_i(\vec{p},t)=\frac{1}{e^{(E-\mu)/T}\pm 1}\quad,
\label{eq:3.101}
\end{equation}

\noindent where $+1$ is for the  Fermi-Dirac distribution and $-1$ for the Bose-Einstein distribution. 

The time-time component of Eq. (\ref{eq:3.100}) is
\begin{equation}
T_0^0\equiv \rho_i= g_i\int \,\frac{dP_1dP_2dP_3}{(2\pi)^3\sqrt{-g}}{P^0f_i(\vec{p},t)}\quad.
\label{eq:3.1}
\end{equation}

\noindent Since for the flat FLRW the determinant of the metric is $-a(t)^6$ and $p^2=g^{ij}P_i P_j=a^{-2}(t)\delta^{ij}P_i P_j$, we have $p_i=P_i/a(t)$, therefore $dP_1dP_2dP_3=a^3(t)d^3p$. In this case we get
\begin{equation}
T_0^0= g_i\int \,\frac{a^3(t)d^3p}{(2\pi)^3a^3(t)}{P^0f_i(\vec{p},t)}\quad.
\label{eq:3.2}
\end{equation}

\noindent Moreover, the on-shell condition gives
\begin{align}
P^2&=-m^2=g_{\mu\nu}P^\mu P^\nu=g_{00}P^0P^0+g_{ij}P^iP^j=-(P^0)^2 + p^2\quad,\notag\\
P^0&=\sqrt{m^2 + p^2}= E\quad.
\label{eq:3.4}
\end{align}

\noindent Hence
\begin{equation}
\rho= g_i\int \,\frac{d^3p}{(2\pi)^3}{E(p)f_i(\vec{p},t)}\quad.
\label{eq:3.5}
\end{equation}

Similar reasoning can be done to express the pressure $P$  as an integral over the distribution
function
\begin{equation}
P_i= g_i\int \,\frac{d^3p}{(2\pi)^3}\frac{p^2}{3E(p)}f_i(\vec{p},t)\quad.
\label{eq:3.56}
\end{equation}

The density number $n_i$ is defined in terms of the distribution function as 
\begin{equation}
\begin{split}
n_i=g_i\int{\frac{d^3p}{(2\pi)^3}\frac{1}{e^{\sqrt{p^2+m^2}/T}\pm 1}}\quad.
\label{eq:4.111}
\end{split}
\end{equation}

\noindent The Maxwell-Boltzmann distribution does not have the 1 in the denominator. We will compute the equilibrium number density (i.e. with zero chemical potential) of a relativistic and non-relativistic species. Considering first the non-relativistic limit  ($m\gg T$), we  expand in Taylor series $\sqrt{p^2+m^2}/T$ which yields
\begin{equation}
\begin{split}
\frac{\sqrt{p^2+m^2}}{T} \approx \frac{m}{T}+\frac{p^2}{2mT}\quad,
\label{eq:4.21}
\end{split}
\end{equation}

\noindent where we kept only terms in first order of $m$. Then Eq. (\ref{eq:4.111}) yields
\begin{equation}
\begin{split}
n_{i\text{ dust}}=g_i\int{\frac{d^3p}{(2\pi)^3}\frac{1}{e^{\left(m+\frac{p^2}{2m}\right)/T}}}=\frac{e^{-m/T}}{\pi^2}\int_0^\infty{p^2e^{-\frac{p^2}{2mT}}}\,dp\quad,
\label{eq:4.31}
\end{split}
\end{equation}

\noindent where we  neglected $\pm 1$ because of the high $m/T$ limit. Then, the three distributions are equal for this limit,
\begin{align}
n_{i\text{ dust}}=&g_i\frac{e^{-m/T}}{2\pi^2}\int_0^\infty{p^2e^{-\frac{p^2}{2mT}}}\,dp=g_i\frac{e^{-m/T}}{2\pi^2}\sqrt{\frac{(2mT)^3\pi}{16}}\notag\\
=&g_i e^{-m/T}\left(\frac{mT}{2\pi}\right)^{3/2}\quad.
\label{eq:4.41}
\end{align}

\noindent Now, considering the relativistic limit ($m\ll T$), Eq. (\ref{eq:4.111}) becomes
\begin{equation}
\begin{split}
n_{i\text{ rad}}=\frac{g_i}{2\pi^2}\int_0^\infty{\frac{p^2}{e^{p/T}\pm 1}}\,dp\quad.
\label{eq:4.51}
\end{split}
\end{equation}

\noindent The Maxwell-Boltzmann distribution for this case is
\begin{align}
n_{i\text{ rad MB}}=&\frac{g_i}{2\pi^2}\int_0^\infty{p^2e^{-p/T}}\,dp \notag\\
= &\frac{g_iT^3}{\pi^2}\quad,
\label{eq:4.61}
\end{align}

\noindent where an integration  by parts was performed. For Fermi-Dirac distribution, using the Riemann zeta function $\zeta(s).$\footnote{$\zeta(s)=[(1 - 2^{1-s})\Gamma(s)]^{-1}\int_0^\infty{\frac{x^{s-1}}{e^{x}+ 1}}\,dx=\Gamma(s)^{-1}\int_0^\infty{\frac{x^{s-1}}{e^{x}- 1}}\,dx$.}
\begin{align}
n_{i \text{ rad FD}}=&g_i\frac{1}{2\pi^2}\int_0^\infty{\frac{p^2}{e^{p/T}+ 1}}\,dp=g_i\frac{T^3}{2\pi^2}(1-2^{-2})\zeta(3)\Gamma(3)\notag\\
=&g_i\frac{3\zeta(3)}{4\pi^2}T^3\quad.
\label{eq:4.71}
\end{align}

\noindent For the Bose-Einstein distribution
\begin{align}
n_{i \text{ rad BE}}=&\frac{g_i}{2\pi^2}\int_0^\infty{\frac{p^2}{e^{p/T}- 1}}\,dp\notag\\
=&\frac{g_i\zeta(3)}{\pi^2}T^3\quad.
\label{eq:4.81}
\end{align}

Similar calculations can be done for $\rho$ and $P$ in both limits. The results are 
\begin {itemize}
\item non-relativistic limit ($m\gg T$)
\end{itemize}

\begin{align}\label{neq2}
\rho =& mn\quad,\\
 n=& g e^{-m/T}\left(\frac{mT}{2\pi}\right)^{3/2}\quad,\\
 P=&nT\ll\rho\quad.
\end{align}

\begin {itemize}
\item relativistic limit ($m\ll T$)
\end{itemize}
\begin{align}\label{relativrho}
\rho = &    \frac{\pi^2}{30}gT^4 \qquad \text{Bose-Einstein}\\
\rho = & 	\frac{7}{8}\frac{\pi^2}{30}gT^4  \qquad\text{Fermi-Dirac}\\
  n=&\frac{\zeta(3)}{\pi^2}gT^3  \qquad \text{Bose-Einstein}\label{nrelat}\\
n=&\frac{3}{4}\frac{\zeta(3)}{\pi^2}gT^3  \qquad \text{Fermi-Dirac}\\
P=&\frac{\rho}{3}\quad.
\end{align}

\subsection{The Boltzmann equation}
The constituents of the universe remained in thermal equilibrium until the expansion rate ($H$) becomes roughly of the same order of the interaction rate ($\Gamma$) of the particle species. Then, when $\Gamma\lesssim H$, the particle species decouples and it evolves independently of the thermal bath. The out-of-equilibrium phenomena played an essential role in three important production processes: i) the formation of the light elements during Big Bang Nucleosynthesis (BBN); (ii) recombination of electrons and protons into neutral hydrogen when the temperature was of order $1/4$ eV; and possibly in (iii) production of dark matter in the early universe.

To analyze the abundances of such elements we should use the Boltzmann equation, which can be  written as 
\begin{align}\label{liouville}
\hat{L}[f]=C[f]\quad,
\end{align}

\noindent where $C$ is the collision operator and $\hat{L}$ is the Liouville operator, which in turn, has 	the covariant, relativistic form given by
\begin{align}\label{liouvilleop}
\hat{L}=P^{\alpha}\frac{\partial }{\partial x^\alpha}-\Gamma^\alpha_{\beta\gamma}P^\beta P^\gamma\frac{\partial }{\partial P^\alpha}\quad.
\end{align}

For the FLRW metric and with the definition of number density (\ref{eq:4.111}) we can integrate the Boltzmann equation in the phase space. Since the distribution function depends only on the energy (or on the magnitude of the three momentum), because the FLRW metric describes a homogeneous and isotropic universe, its spatial derivatives and three momentum derivatives vanish. Using  Eqs. (\ref{eq:2.5}), (\ref{eq:2.7}) and (\ref{liouvilleop}), Eq. (\ref{liouville}) gives
\begin{equation}\label{BOlteq1}
E\frac{\partial f }{\partial t}-H p^2\frac{\partial f }{\partial E}=C[f]\quad.
\end{equation}

We multiply both sides of Eq. (\ref{BOlteq1}) by the phase space volume $d^3p/(2\pi)^3$ and integrate, which leads to
\begin{equation}\label{BOlteq2}
\frac{\partial  }{\partial t}\int{\frac{d^3p}{(2\pi)^3	} f}\,-H\int \frac{d^3p}{(2\pi)^3	} \frac{p^2}{E}\frac{\partial f }{\partial E}=\int{\frac{d^3p}{(2\pi)^3}\frac{C[f]}{E}}\,\quad.
\end{equation}

The second term can be simplified if we use $EdE=pdp$, which comes from $E^2=p^2+m^2$. Thus  $\frac{p^2}{E}\frac{\partial f }{\partial E}=p\frac{\partial f }{\partial p}$ and the integral becomes
\begin{align}\label{BOlteq3}
\int \frac{d^3p}{(2\pi)^3	} \frac{p^2}{E}\frac{\partial f }{\partial E}&=\int \frac{d^3p}{(2\pi)^3	} p\frac{\partial f }{\partial p}\notag\\
&=\int_0^\infty \frac{4\pi}{(2\pi)^3	} p^3dp\frac{\partial f }{\partial p}\notag\\\notag
&= -3\frac{4\pi}{(2\pi)^3	}\int_0^\infty  p^2dp f \\
&=-3n\quad,
\end{align}

\noindent where in the third line we performed an integration by parts and used the definition of number density (\ref{eq:4.111}). Therefore, the Eq. (\ref{BOlteq2}) becomes
\begin{equation}\label{BOlteq4}
\frac{\partial n}{\partial t} +3Hn=\int{\frac{d^3p}{(2\pi)^3}\frac{C[f]}{E}}\,\quad,
\end{equation}

\noindent which in turn can be written as
\begin{equation}\label{BOlteq}
a^{-3}\frac{d(na^{3}) }{dt}=\int{\frac{d^3p}{(2\pi)^3}\frac{C[f]}{E}}\,\quad. 
\end{equation}

The collision term for the process $1+2 \leftrightarrow 3+4$ is
\begin{align}
\frac{1}{(2\pi)^3}\int{\frac{C[f]}{E_1}}\,d^3p_1=&\int{\frac{d^3p_1}{(2\pi)^32E_1}}\int{\frac{d^3p_2}{(2\pi)^32E_2}}\int{\frac{d^3p_3}{(2\pi)^32E_3}}\int{\frac{d^3p_4}{(2\pi)^32E_4}}\notag\\
&\times(2\pi)^4\delta^3(p_1+p_2-p_3-p_4)\delta(E_1+E_2-E_3-E_4)\left|\mathcal{M}\right|^2\notag\\
&\times [f_3f_4(1\pm f_1)(1\pm f_2)-f_1f_2(1\pm f_3)(1\pm f_4)]\quad,
\label{eq:1.1111}
\end{align}

\noindent where the plus sign is for bosons and the minus sign for fermions. In the absence of Bose condensation and Fermi degeneracy, the blocking and stimulated factors can be ignored $1\pm f\approx f$ and $f=\exp[-(E-\mu)/T$ for all species in kinetic equilibrium. Using the Eqs. (\ref{eq:4.41}) and (\ref{eq:4.61}) as the definition of the equilibrium  number density  of the $i$ species, we can rewrite $e^{\mu_i/T}$ as $n_i/n^{(0)}_i$, thus the  Eq. (\ref{eq:1.1111}) becomes
\begin{align}
\left\langle \sigma v\right\rangle\equiv &\int{\frac{C[f]}{(2\pi)^3E_1}}\,d^3p_1=\frac{1}{n^{(0)}_{1}n^{(0)}_{2}}\int{\frac{d^3p_1}{(2\pi)^32E_1}}\int{\frac{d^3p_2}{(2\pi)^32E_2}}\int{\frac{d^3p_3}{(2\pi)^32E_3}}\int{\frac{d^3p_4}{(2\pi)^32E_4}}\notag\\
&\times e^{-\frac{(E_1+E_2)}{T}}(2\pi)^4\delta^3(p_1+p_2-p_3-p_4)\delta(E_1+E_2-E_3-E_4)\left|\mathcal{M}\right|^2\quad.
\label{eq:1.1112}
\end{align}

Then, the Boltzmann equation becomes
\begin{align}
a^{-3}\frac{d(n_1 a^{3}) }{dt}=n^{(0)}_{1}n^{(0)}_{2}\left\langle \sigma v\right\rangle\left(\frac{n_3n_4}{n^{(0)}_{3}n^{(0)}_{4}}-\frac{n_1n_2}{n^{(0)}_{1}n^{(0)}_{2}}\right)\quad.
\label{eq:1.1113}
\end{align}

Unless $\langle \sigma v\rangle=0$, the equilibrium is maintained when 
\begin{align}
\frac{n_3n_4}{n^{(0)}_{3}n^{(0)}_{4}}=\frac{n_1n_2}{n^{(0)}_{1}n^{(0)}_{2}}\quad.
\label{eq:1.1114}
\end{align}

\subsection{Big Bang Nucleosynthesis}

Nuleosynthesis occurred during the first minutes of the universe, when the temperature was around $1$ MeV.
The first nucleus that was produced was the hydrogen, followed by helium and a small amount of lithium, and their isotopes: deuterium, tritium, and helium-3.   A nucleus with $Z$ protons and $A-Z$ neutrons has a slightly different mass compared with the sum of the masses of the individual particles. This difference is the binding energy, defined as 
\begin{equation}
B\equiv Zm_p+(A-Z)m_n-m\quad,
\label{eq:}
\end{equation}

\noindent where $m$ is the nucleus mass. Due to a lack of tightly bound isotopes at an atomic number between five and eight and the low baryon number, there are no heavier nuclei produced in the early universe. Deuterium is produced through the process $p+n \rightarrow D+\gamma$ and two deuterium form helium-3 plus one neutron. Finally, helium-3 and one deuterium produce helium and one proton.

First, let's  consider the deuterium production, where the equilibrium condition (\ref{eq:1.1114}) gives
\begin{align}
\frac{n_D}{n_{p}n_{n}}=\frac{n_D^{(0)}}{n^{(0)}_{p}n^{(0)}_{n}}\quad,
\label{eq:1.1114d}
\end{align}

\noindent since $n_\gamma=n_\gamma^{(0)}$. From Eq. (\ref{eq:4.41}) we have 
\begin{align}
\frac{n_D}{n_{p}n_{n}}=\frac{3}{4}\left(\frac{2\pi m_D}{m_n m_p T}\right)^{3/2}e^{(m_n+m_p-m_D)/T}\quad,
\label{eq:1.1114dd}
\end{align}

\noindent where the spin states are 3 for $D$ and 2 each for $p$ and $n$. We can set $m_D=2m_p=2m_n$ in the  prefactor but not in the exponential because it contains the binding energy of deuterium ($B_D=2.2 $ MeV). Thus
\begin{align}
\frac{n_D}{n_{p}n_{n}}=\frac{3}{4}\left(\frac{4\pi}{ m_p T}\right)^{3/2}e^{B_D/T}\quad.
\label{eq:1.1114d2}
\end{align}

The equation above can be written in terms of the baryon-to-photon ratio ($\eta_b\equiv n_b/n_\gamma$). Once $n_p\sim n_n\sim n_b$ and using Eq. (\ref{nrelat}) we have
\begin{equation}
\frac{n_D}{n_{b}}\sim \eta_b \left(\frac{T}{ m_p}\right)^{3/2}e^{B_D/T}\quad.
\label{eq:1.1114d3}
\end{equation}

The exponential is not too large because of $B_D$ and $\eta_b \sim 10^{-10}$ when the universe had a temperature of about MeV. Therefore, the small baryon-to-photon ratio inhibits nuclei production until the temperature drops below the nuclear binding energy.

At the temperature of $\sim 1$ MeV protons, electrons and positrons are in equilibrium.  Protons can be converted into neutrons via weak interactions, such as $p+e^-\rightarrow n+\nu_e$, for instance. Since protons and neutrons have approximately the same mass, from Eq. (\ref{eq:4.41}) we have $(m_p/m_n)^{3/2}\approx 1$ and the ratio $n_p^{(0)}/n_n^{(0)}$ is
\begin{align}
\frac{n_p^{(0)}}{n_n^{(0)}}=e^{\mathcal{Q}/T}\quad,
\label{eq:1.1116}
\end{align}

\noindent with $\mathcal{Q}\equiv m_n-m_p= 1.293$ MeV. At high temperature there are as many neutrons as protons and as the temperature decreases to 1 MeV the neutron fraction  goes down. It is convenient to define the ratio of neutrons to total nuclei
\begin{align}
X_n\equiv \frac{n_n}{n_n+n_p}\quad,
\label{eq:1.11171}
\end{align}

\noindent where in equilibrium
\begin{equation}
X_n^{EQ}=\frac{1}{1+n^{(0)}_p/n^{(0)}_n}\quad.
\label{eq:XnEQ}
\end{equation}

 Using Eqs. (\ref{eq:1.11171}) and (\ref{eq:1.1116}) the Boltzmann equation for the process $p+e^-\rightarrow n+\nu_e$ (where the leptons are in complete equilibrium $n_l=n_l^{(0)}$) becomes 
\begin{align}
\frac{dX_n}{dt}= \lambda_{np}\left[(1-X_n)e^{-\mathcal{Q}/T}-X_n\right]\quad,
\label{eq:1.1117}
\end{align}

\noindent where 
\begin{align}
\lambda_{np}= n^{(0)}_{\nu_e}\left\langle \sigma v\right\rangle&= n^{(0)}_{\nu_e}\frac{1}{n^{(0)}_{n}n^{(0)}_{\nu_e}}\int{\frac{d^3p_n}{(2\pi)^32E_n}}\int{\frac{d^3p_\nu}{(2\pi)^32E_\nu}}\int{\frac{d^3p_p}{(2\pi)^32E_p}}\int{\frac{d^3p_e}{(2\pi)^32E_e}}\notag\\
&\times e^{-{(E_n+E_\nu)}{T}}(2\pi)^4\delta^3(p_n+p_\nu-p_p-p_e)\delta(E_n+E_\nu-E_p-E_e)\left|\mathcal{M}\right|^2\quad.
\label{eq:1.1118}
\end{align}

In Eq. (\ref{eq:1.1118}) we consider the non-relativistic limit for both protons and neutrons, so we have  $E_\nu=p_\nu$, $E_e=p_e$, $E_p=m_p+p^2/2m_p$ and $E_n=m_n+p^2/2m_n$. Performing the integral first over $p_p$, we have
\begin{align}
\lambda_{np}&= \frac{1}{n^{(0)}_{n}}\int{\frac{d^3p_n}{(2\pi)^64E_nE_p}}\int{\frac{d^3p_\nu}{(2\pi)^32p_\nu}}\int{\frac{d^3p_e}{(2\pi)^32p_e}}
e^{-\frac{(E_n+p_\nu)}{T}}\notag\\
&\ \ \times(2\pi)^4\delta(m_n+p_\nu-m_p-p_e)\left|\mathcal{M}\right|^2\quad,
\label{eq:1.1119}
\end{align}

\noindent since $p_p\approx p_n$. Considering $m_p \approx m_n$, the product of the proton energy by the neutron energy yields $E_nE_p = (m_p+p^2/2m_p)(m_n+p^2/2m_n)\approx m^2$, since $m^2 \gg 2p^2+p^4/m^2$. Thus Eq. (\ref{eq:1.1119}) becomes 
\begin{align}
\lambda_{np}&=  \frac{1}{n^{(0)}_{n}}4\pi e^{-m/T}\int_0^\infty{\frac{p^2 e^{-\frac{p^2}{2mT}}dp}{(2\pi)^24m^2}}\int{\frac{d^3p_\nu e^{-p_\nu/T}}{(2\pi)^32p_\nu}}\int{\frac{d^3p_e}{(2\pi)^32p_e}}\notag\\
&\ \ \times \delta(\mathcal{Q}+p_\nu-p_e)\left|\mathcal{M}\right|^2 \notag\\
&= \left(\frac{2\pi}{mT}\right)^{3/2}\frac{1}{4m^2\pi g_n}(\pi)^{1/2}(2mT)^{3/2}\int{\frac{d^3p_\nu e^{-p_\nu/T}}{(2\pi)^32p_\nu}}\int{\frac{d^3p_e}{(2\pi)^32p_e}}\notag\\
&\ \ \times \delta(\mathcal{Q}+p_\nu-p_e)\left|\mathcal{M}\right|^2\quad.
\label{eq:1.11110}
\end{align}

\noindent Therefore, for $g_n=2$,

\begin{align}
\lambda_{np}=\frac{\pi}{4m^2}\int{\frac{d^3p_\nu}{(2\pi)^32p_\nu}e^{-p_\nu/T}}\int{\frac{d^3p_e}{(2\pi)^32p_e}}\delta(\mathcal{Q}+p_\nu-p_e)\left|\mathcal{M}\right|^2\quad.
\label{eq:1.11111}
\end{align}

 Evaluating Eq. (\ref{eq:1.11111}) with $|\mathcal{M}|^2=32G_F^2(1+3g_A^2)m_p^2p_\nu p_e$ where $g_A$ is the axial-vector coupling of the nucleon and $G_F$ is the Fermi constant, we have
\begin{align}
\lambda_{np}&=2\frac{\pi}{4m^2}4\pi\int_0^\infty{\frac{p_\nu^2dp_\nu}{(2\pi)^32p_\nu}e^{-p_\nu/T}}4\pi\int_0^\infty{\frac{p_e^2dp_e}{(2\pi)^32p_e}}\delta(\mathcal{Q}+p_\nu-p_e)\notag\\
&\ \ \times 32G_F^2(1+3g_A^2)m_p^2p_\nu p_e\notag\\
&= 2\frac{(2\pi)^3}{4m^2(2\pi)^6}\int_0^\infty {p_\nu^2dp_\nu e^{-p_\nu/T}}\int_0^\infty{p_e^2dp_e}\delta(\mathcal{Q}+p_\nu-p_e)\frac{32m_p^22\pi^3}{\tau_n \lambda_0 m_e^5}\notag\\
&= \frac{4}{\tau_n \lambda_0 m_e^5}\int_0^\infty{\frac{1}{2}(\mathcal{Q}+p_\nu)^2p_\nu^2dp_\nu e^{-p_\nu/T}}\notag\\
&=\frac{2}{\tau_n \lambda_0 m_e^5}\int_0^\infty{(p_\nu^4+2\mathcal{Q}p_\nu^3+\mathcal{Q}^2p_\nu^2)dp_\nu e^{-p_\nu/T}}\notag\\
&=\frac{2}{\tau_n \lambda_0 m_e^5}(24T^5+2\mathcal{Q}6T^4+\mathcal{Q}^22T^2)\quad,
\label{eq:1.666666}
\end{align}

\noindent where we used the neutron lifetime $\tau_n=\lambda_0 G_F^2(1+3g_A^2)m_e^5/(2\pi^3)=886.7$s, with $\lambda_0=1.636$.
Finally, we have
\begin{align}
\lambda_{np}=\frac{4T^5}{\tau_n \lambda_0 m_e^5}\left(12+6\frac{\mathcal{Q}}{T}+\frac{\mathcal{Q}^2}{T^2}\right)\quad.
\label{eq:1.71111111}
\end{align}

  Eq. (\ref{eq:1.71111111}) together with Eq. (\ref{eq:1.1117}) can be solved numerically to give the neutron abundance. To do so, we first change the variables

\begin{align}
x\equiv \frac{\mathcal {Q}}{T}\quad.
\label{eq:1.711111112}
\end{align}

\noindent The left-hand side of Eq. (\ref{eq:1.1117}) becomes $\dot{x}dX_n/dx$. The equation for $\dot{x}$ comes from Eq.  	(\ref{eq:1.711111112})

\begin{align}
\frac{dx}{dt} =-\frac{\mathcal {Q}}{T^2}\frac{dT}{dt}=-x\frac{\dot{T}}{T}\quad
\label{eq:1.711111112311}
\end{align}

\noindent and the expression for $\dot{T}$ comes, in turn, from the fact that $T\propto a^{-1}$, thus	

\begin{align}
T^{-1}\frac{dT}{dt}=a\left(-\frac{1}{a^2}\right)\frac{da}{dt}=-H=\sqrt{\frac{8\pi G \rho}{3}}\quad,
\label{eq:1.71111111231}
\end{align}

 \noindent where the last equality follows from the first Friedmann equation. Since BBN occurred in the radiation era, the main contribution to the energy density comes from relativistic particles (\ref{relativrho}). Combining the contributions of fermions and bosons, we have

\begin{align}
\rho&=\frac{	\pi^2}{30}T^4\left[\sum_{i=\text{bosons}}{g_i}+\frac{7}{8}\sum_{i=\text{fermions}}{g_i}\right] \quad (i \quad \text{relativistic})\notag\\
&\equiv g_*\frac{	\pi^2}{30}T^4\quad,
\label{eq:1.7111111123}
\end{align}

\noindent where $g_*$ is the effective degrees of freedom, in the period of interest and it depends on the temperature. In the BBN there were photons ($g_\gamma=2$), neutrinos ($g_\nu=6$), electrons and positrons ($g_{e^+}=g_{e_-}=2$). These particles gives the value (roughly independent of the temperature) $g_*\simeq 10.75$. Finally, Eq. (\ref{eq:1.1117}) becomes 

\begin{align}
\frac{dX_n}{dx}=\frac{x\lambda_{np}}{H(x=1)}\left[e^{-x}-X_n(1+e^{-x})\right]\quad,
\label{eq:1.7111111123dx}
\end{align}

\noindent with $H(x=1)=\sqrt{\frac{4\pi^3 G \mathcal {Q}^4}{45}10.75}\approx 1.13$ s$^{-1}$. From Eq. (\ref{eq:1.71111111}) the neutron-proton conversion rate is 

\begin{align}
\lambda_{np}=\frac{255}{\tau_n x^5}(12+6x+x^2)\quad.
\label{eq:1.6666662}
\end{align}

Notice that we have a first-order non-homogeneous differential equation (\ref{eq:1.7111111123dx}), which can be written in the form

\begin{align}
X'_n(x)+\frac{\lambda_{np}(x)x}{H(x=1)}(1+e^{-x})X_n(x)=\frac{\lambda_n(x)x}{H(x=1)}e^{-x}\quad.
\label{eq:2.1111111}
\end{align}

\noindent We define the integrating factor as

\begin{align}
\mu(x)\equiv\int_{x_i}^x{\frac{dx'\lambda_{np}(x')x'}{H(x')}(1+e^{-x'})}\quad
\label{eq:2.22222}
\end{align}

\noindent and multiply both sides of Eq. (\ref{eq:2.1111111}) by $e^{\mu(x)}$. Then,
\begin{align}
e^{\mu(x)}X'_n(x)+\frac{\lambda_{np}(x)x}{H(x=1)}(1+e^{-x})X_n(x)e^{\mu(x)}=\frac{\lambda_n(x)x}{H(x=1)}e^{-x}e^{\mu(x)}\quad,
\label{eq:2.33232}
\end{align}

\noindent which can be rewritten as
\begin{align}
[e^{\mu(x)}X_n(x)]'=\frac{\lambda_{np}(x)x}{H(x=1)}e^{-x}e^{\mu(x)}\quad,
\label{eq:2.423232}
\end{align}

\begin{align}
e^{\mu(x)}X_n(x)=\int_{x_i}^x{dx'\frac{\lambda_{np}(x')e^{-x'}}{x'H(x')}e^{\mu(x')}}\quad,
\label{eq:2.523232}
\end{align}

\noindent since $H(x)=H(x=1)/x^2$. Hence, with $\mu(x)$ defined by Eq. (\ref{eq:2.22222}),
\begin{align}
X_n(x)=\int_{x_i}^x{dx'\frac{\lambda_{np}(x')x'e^{-x'}}{H(x')}e^{\mu(x')-\mu(x)}}\quad.
\label{eq:2.6322322}
\end{align}

\noindent  Since
\begin{align}
\mu(x)=\int_{x_i}^x{\frac{dx'x'255}{\tau_n x'^5H(x'=1)}(12+6x'+x'^2)(1+e^{-x'})}\quad,
\label{eq:2.7323232}
\end{align}

\begin{align}
\mu(x)&=\int_{x_i}^x{\frac{dx'255}{\tau_n H(x'=1)}\left(\frac{12}{x'^4}+\frac{6}{x'^3}+\frac{1}{x'^2}\right)(1+e^{-x'})}\notag\\
&=\frac{255}{\tau_n H(x'=1)}\left[ \left(-\frac{4}{x^3}-\frac{3}{x^2}-\frac{1}{x}\right)\bigg|_{x_i}^x+\int_{x_i}^x{dx'\left(\frac{12}{x'^4}+\frac{6}{x'^3}+\frac{1}{x'^2}\right)e^{-x'}}\right]\notag\\
&=\frac{255}{\tau_n H(x'=1)}\left[ \left(-\frac{4}{x^3}-\frac{3}{x^2}-\frac{1}{x}\right)+\left(-\frac{4}{x^3}-\frac{1}{x^2}\right)e^{-x'}\right]\bigg|_{x_i}^x\notag\\
&=-\frac{255}{\tau_n H(x'=1)}\left[ \left(\frac{4}{x^3}+\frac{3}{x^2}+\frac{1}{x}\right)+\left(\frac{4}{x^3}+\frac{1}{x^2}\right)e^{-x'}\right]\bigg|_{x_i}^x\quad.
\label{eq:2.82323232}
\end{align}

\noindent Therefore,
\begin{align}
\mu(x)=-\frac{255}{\tau_n}\left[\frac{4\pi^3G\mathcal{Q}^4g_*}{45}\right]^{-1/2}\left[ \left(\frac{4}{x^3}+\frac{3}{x^2}+\frac{1}{x}\right)+\left(\frac{4}{x^3}+\frac{1}{x^2}\right)e^{-x'}\right]\bigg|_{x_i}^x\quad.
\label{eq:2.90000}
\end{align}

\begin{figure}\centering
\includegraphics[scale=0.4]{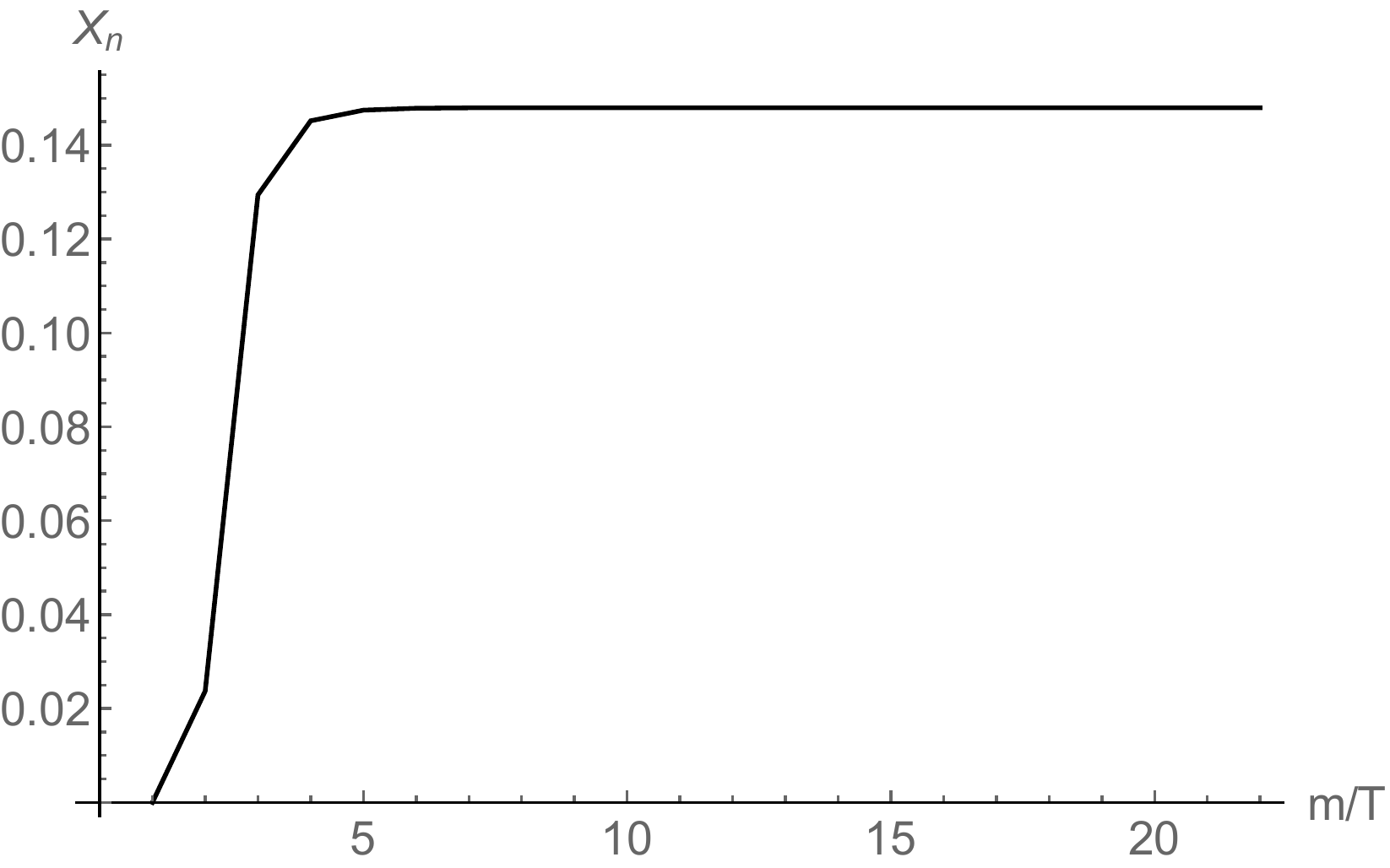}%
\caption{The figure shows  $X_n$ as a function of $x$, that is, with the inverse of 
temperature ($m/T$). }%
\label{neutronabundance}%
\end{figure}

The freeze-out occurred for the asymptotic value of $X_n$ at $x\rightarrow \infty$. The value is found numerically, solving the Eq.(\ref{eq:2.6322322}). We find $X_n=0.148$ at $x\rightarrow \infty$ and the evolution of $X_n$ is showed in Fig. \ref{neutronabundance}.

The production of light elements is well constrained from observed data, in such a way that a hypothetical unstable particle, for instance,  has its mass and decay rate also constrained. As an example, the gravitino should decay before the BBN in order not to destroy the light elements, for instance, due to the high energy photons emitted from its decay.

\subsection{Recombination}

As the universe expands its temperature decreases. When the temperature drops to around one electron-volt,  electrons, protons and photons are coupled via Coulomb and Compton scattering. Although the production of neutral hydrogen demands a binding energy of $\epsilon_0=13.6$ eV, the high photon per baryon ratio ensures that any hydrogen atom produced will be instantaneously ionized. In order to study the out-of-equilibrium process for the reaction $e^-+p\leftrightarrow H+\gamma$, we use the fact that the neutrality of the universe ensures $n_p=n_e$ and we define the free electron fraction
\begin{equation}
X_e\equiv \frac{n_e}{n_e+n_H}=\frac{n_p}{n_p+n_H}\quad.
\label{eq:Xe}
\end{equation}

 The Boltzmann equation (\ref{eq:1.1113}) becomes
\begin{align}
a^{-3}\frac{d(n_ea^{3})}{dt}&=n^{(0)}_en^{(0)}_p \langle \sigma v \rangle \left(\frac{n_H}{n^{(0)}_H}-\frac{n_e^2}{n^{(0)}_en^{(0)}_p} \right) \nonumber\\
&= (1-X_e)\left\langle \sigma v \right\rangle n_b\left[\left(\frac{m_e T}{2\pi}\right)^{3/2}e^{-\epsilon_0/T}-X^2_en_b\right]\quad.
\label{eq:4.11111}
\end{align}

\noindent Expressing $n_e$ as $X_en_b$, where $n_e+n_H=n_b$, $n_b a^3$ is constant and can be passed through the time derivative, and using $x\equiv \frac{\epsilon_0}{T}$ we have
\begin{align}
\frac{dX_e}{dt}=\dot{x}\frac{dX_e}{dx}=-x\frac{\dot{T}}{T}\frac{dX_e}{dx}\quad,
\label{eq:4.233333}
\end{align}

\noindent since $\dot{x}=-\epsilon_0\dot{T}/T^2=-x \dot{T}/T.$ As $T \propto 1/\alpha$
\begin{align}
\frac{\dot{T}}{T}=-H=-\frac{1}{x^2}\sqrt{\frac{4\pi^3G\epsilon_0^4g_*}{45}}=-\frac{1}{x^2}H(x=1)\quad,
\label{eq:4.33333}
\end{align}

\noindent thus Eq. (\ref{eq:4.11111}) yields
\begin{align}
\frac{dX_e}{dx}=\frac{x}{H(x=1)}\left[(1-X_e)\beta-X^2_en_b\alpha\right]\quad,
\label{eq:4.5333}
\end{align}

\noindent with $\beta\equiv\left\langle \sigma v \right\rangle\left(\frac{m_e T}{2\pi }\right)^{3/2}e^{-x}$ and $\alpha\equiv \left\langle \sigma v \right\rangle$.

Writing down Eq. (\ref{eq:4.5333}) we have
\begin{align}
\frac{dX_e}{dx}&=\frac{x}{H(x=1)}\left[(1-X_e)\beta-X^2_en_b\alpha\right]\notag\\
&=\frac{x}{H(x=1)}\left[(n_e+n_H)(X^{(0)}_e)^2\left\langle \sigma v \right\rangle-X^2_en_b\left\langle \sigma v \right\rangle\right]\notag\\
&=\frac{x}{H(x=1)}\left\langle \sigma v \right\rangle n_b[(X^{(0)}_e)^2-X^2_e]\quad,
\label{eq:4.62}
\end{align}

\noindent where we used 
\begin{equation}
\frac{X_e^2}{1-X_e}=\frac{1}{n_e+n_H}\left[ \left(\frac{m_e T}{2\pi }\right)^{3/2} e^{-(m_e+m_p-m_H)/T}\right]\quad,
\label{}
\end{equation}

\noindent which in turn is obtained from the condition of equilibrium (\ref{eq:1.1114}) with $1=e^-$, $2=p$ and $3=H$.

 Denoting $\lambda\equiv n_b x^3\left\langle \sigma v \right\rangle/H(x=1)$ we have
\begin{align}
\frac{dX_e}{dx}=-\frac{\lambda}{x^2}[X^2_e-(X^{(0)}_e)^2]\quad.
\label{eq:4.72222}
\end{align}

The equation above does not have analytic solutions, but we can find a solution for the extreme cases. During the thermal equilibrium $X_e=X^{(0)}_e$, while after the freeze-out, $X^{(0)}_e$ decreases and $X_e\gg X^{(0)}_e$. So Eq. (\ref{eq:4.72222}) yields
\begin{align}
\frac{dX_e}{dx}\approx -\frac{\lambda}{x^2}X^2_e \ \ \quad (x \gg 1)\quad.
\label{eq:4.82222}
\end{align}

\noindent Now, integrating this from the epoch of freeze-out $x_f$ until very late times $x=\infty$
\begin{align}
\frac{1}{X_{\infty}}-\frac{1}{X_f}=\frac{\lambda}{x_f}\quad.
\label{eq:4.92222}
\end{align}

\noindent Since $X_e$ at freeze-out $X_f$ is significantly larger than $X_\infty$ we neglect the second term in Eq. (\ref{eq:4.92222}) and we get
\begin{align}
X_{\infty}\approx \frac{x_f}{\lambda}\quad.
\label{eq:4.102222}
\end{align}

\subsection{Dark Matter}\label{DMboltz}

There is a strong evidence of non-baryonic and non-relativistic dark matter with a density parameter today of  $\Omega_{dm}=0.27$ \cite{Ade:2015xua}. As we have seen in the previous sections the energy density of a non-relativistic matter is proportional to $e^{-m/T}$, so if dark matter remained in thermal equilibrium its abundance would be totally suppressed.  

Here we reproduce the main mechanism of thermal production of dark matter in the young universe. To do so, the dark matter particle is assumed to be in close contact with the cosmic plasma at high temperatures, but then it froze-out when the temperature dropped below its mass. The two heavy dark matter particles $X$ can annihilate	into two essentially massless particles $l$. These light particles are assumed to be tightly coupled to the cosmic plasma, so $n_l^{(0)}=n_l$. In this section there is not an interaction with dark energy.

With these assumptions the Boltzmann equation (\ref{eq:1.1113}) becomes 	
\begin{equation}
a^{-3}\frac{d(n_Xa^3)}{dt}=\langle \sigma v\rangle \left[(n^{(0)}_X)^2-	n_X^2\right]\quad.
\label{eq:DMBOltz}
\end{equation}

The freeze-out is believed to occur in the radiation era, so the temperature of the universe scales with $a^{-1}$. It is convenient to define here the new variable 
\begin{equation}
Y\equiv\frac{n_X}{T^3}\quad,
\label{}
\end{equation} 

\noindent thus the Boltzmann equation for cold dark matter becomes 
\begin{equation}
\frac{dY}{dt}=T^3\langle \sigma v\rangle \left[Y_{EQ}^2-Y^2\right]\quad,
\label{eq:boltz2}
\end{equation}  

\noindent were $Y_{EQ}\equiv n_X^{(0)}/T^3$.  

As previously done we define the quantity
\begin{equation}
x\equiv \frac{m}{T}\quad.
\label{}
\end{equation}

Similarly to what was done for recombination, the change of $t$ to $x$ is done using $dx/dt=Hx $. Since in the radiation era the energy density scales with $T^4$ (\ref{relativrho}) so $H=H(m)x^{-2}$ and Eq. (\ref{eq:boltz2}) can be written as
\begin{equation}
\frac{	x}{Y_{EQ}}\frac{dY}{dx}=-\frac{\Gamma}{	H(T)}\left[\left(\frac{Y}{Y_{EQ}}\right)^2-1 \right]\quad,
\label{boltzDM3}
\end{equation}

\noindent where $\Gamma\equiv n_{EQ}\langle \sigma v\rangle$ is the interaction rate.  When the effectiveness of annihilation $\Gamma/H$ is less than order unity the annihilation freezes-out and dark matter has a relic abundance. An example involving Yukawa interaction will be done soon, but before that we return to Eq. (\ref{boltzDM3})  and calculate the relic abundance, which is for $Y\gg Y_{EQ}$. In this limit we have
\begin{align}
\frac{dY}{dx}\simeq -\frac{\lambda}{x^2}Y^2 \ \ \quad (x \gg 1)\quad,
\label{eq:4.82223}
\end{align}

\noindent where $\lambda\equiv m^3 \langle \sigma v\rangle/H(m)$. Eq. (\ref{eq:4.82223})  has exactly the same form of Eq. (\ref{eq:4.82222}), so its solution is
\begin{align}
Y_{\infty}\simeq \frac{x_f}{\lambda}\quad,
\label{eq:4.102223}
\end{align}

\noindent where $Y_\infty$ is the abundance at very late times $x\rightarrow\infty$.  

After freeze-out, the dark matter density dilutes with $a^{-3}$ and from the Boltzmann equation $\rho_X a^3$ is constant, thus the energy density of dark matter today is $\rho_X(a_0)=mn_X(a_1)(a_1/a_0)^3$. The scale factor $a_1$ corresponds to a sufficiently late time, so that $Y(a_1)\approx Y_\infty$. Using $n_X(a_1)=Y_\infty T_1^3$ we have
\begin{align}
\rho_X(a_0)=m Y_\infty T_0^3\left(\frac{a_1T_1}{a_0T_0}\right)^3\quad.
\label{eq:4.1022233}
\end{align}

In principle one might think that the ratio $\frac{a_1T_1}{a_0T_0}$ is one, since $a\propto T^{-1}$. However, it does not because the photons  	are heated by the annihilation of several particles with mass between 1 MeV and 100 GeV. We find the ratio using  the fact that the entropy density scales as $a^{-3}$. Thus $[g_*(aT^3)]_1=[g_*(aT^3)]_0$. The effective degree of freedom $g_*$ for relativistic particles (defined in Eq. (\ref{eq:1.7111111123}) at high temperatures (10 GeV, for instance) has contributions of five quarks,\footnote{The top quark has a mass of 175 GeV, so it is too heavy to be in the universe at these temperatures.} six leptons, the photon and eight gluons, plus the correspondent anti-particles. All these degrees of freedom gives $g_*=91.5$. Today, the particles that contribute to the entropy are only photons and neutrinos, thus $g_*=3.36$. Therefore $\frac{a_1T_1}{a_0T_0}=\frac{3.36}{91.5}=\frac{1}{27}$. Using this result in Eq. (\ref{eq:4.1022233}) with  Eq. (\ref{eq:4.102223}) and $\lambda\equiv m^3 \langle \sigma v\rangle/H(m)$, we have 
\begin{align}
\rho_X(a_0)\simeq \frac{m Y_\infty T_0^3}{30}\sim \frac{x_fH_1T_0^3}{30m^2\langle \sigma v\rangle}\quad.
\label{eq:4.1022234}
\end{align}

The Hubble rate at the radiation era $H_1=H(m)$ is given by $\sqrt{8\pi G\rho/3}$ with the energy density given by Eq. (\ref{eq:1.7111111123}). The energy parameter for dark matter is 
\begin{align}
\Omega_X=\left[\frac{4\pi Gg_*(m)}{45}\right]^{1/2}\frac{x_fT_0^3}{30m^2\langle \sigma v\rangle\rho_{cr}}\sim 0.3\frac{x_f}{10}\left[\frac{g_*(m)}{100}\right]^{1/2}\frac{10^{-37}\text{cm}^2}{\langle \sigma v\rangle}\quad.
\label{eq:4.1022235}
\end{align}

\subsubsection{Sample calculation with Yukawa interaction}
We now calculate explicitly $\langle \sigma v\rangle$  from the scattering amplitude of quantum field theory. We consider the Yukawa interaction $g \varphi \psi \bar{\psi}$. One might think that the scalar field plays the role of the dark energy, as a possible explanation for the interacting term in coupled dark energy presented in Sect. \ref{deint}. However, we should recall that the equilibrium number density for dark matter is suppressed by $e^{-m/T}$ so the interaction would have stopped in the very early universe.

\begin{figure}\centering
\includegraphics[scale=0.28]{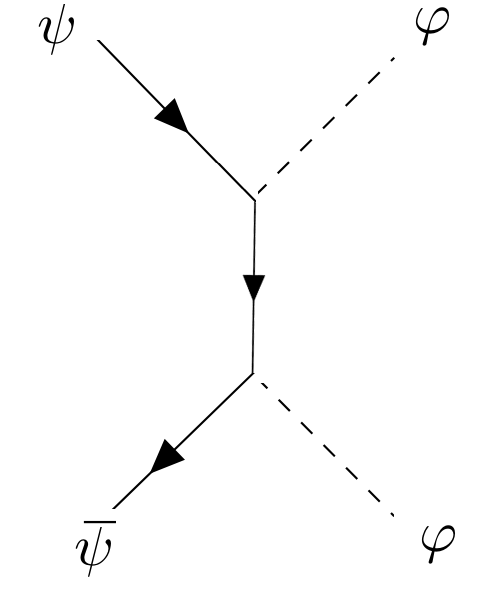}%
\caption{Feynman diagram for the process $\psi(p) + \bar{\psi}(p') \leftrightarrow \varphi(k) + \varphi(k')$. }%
\label{yukawaDM}%
\end{figure}

The complete squared-amplitude for the process $\psi(p) + \bar{\psi}(p') \leftrightarrow \varphi(k) + \varphi(k')$, whose Feynman diagram is shown in Fig. \ref{yukawaDM},  is given by

\begin{align}\label{ScattAmpl1}
g^{-4}|\mathcal{M}|^2=&2\frac{E^4-k^2p^2\cos^2\theta-E^2M^2+4m^2p^2}{((p-k)^2+m^2)^2} +2\frac{E^4-k^2p^2\cos^2\theta-E^2M^2+4m^2p^2}{((p+k)^2+m^2)^2}+\nonumber \\
&2\frac{-2E^4+2k^2p^2\cos^2\theta+2E^2M^2+8m^2p^2}{((p-k)^2+m^2)((p+k)^2+m^2)}\quad.
\end{align}

\noindent where $M$ is the scalar mass, $m$ is the fermion mass and $E^2=p^2+m^2=k^2+M^2$.

 Since we are taken CDM into account, we can simplify the expression above performing an expansion in powers of $p^2/m^2$. The result is
\begin{align}\label{ScattAmpl2}
g^{-4}|\mathcal{M}|^2= \frac{32(5  - 6  M^2/m^2 +   2 M^4/m^4)}{(2  - M^2/m^2)^{4}}\left(\frac{p^2}{m^2}\right)+\mathcal{O}\left(\frac{p^3}{m^3}\right)\quad.
\end{align}

We use the following equation for the total cross section 
\begin{align}\label{sigmav}
 \sigma v &= \frac{| \mathcal{M}|^2}{2E 2 E}\frac{4\pi k} {(2\pi)^24 E_{CM}}\quad,
\end{align}

\noindent where $E_{\text{CM}}$ is the energy in the center of mass frame. In terms of the energy $E$, we have $E_{CM}=2E$.

Using Eq. (\ref{ScattAmpl2}) into Eq. (\ref{sigmav}) we have
\begin{align}\label{ScattAmpl5}
 \sigma v\approx  \frac{g^{4}}{m^2\pi}\sqrt{1-\frac{M^2}{m^2}}\frac{(5  - 6  M^2/m^2 + 2  M^4/m^4)}{(2  - M^2/m^2)^{4}}\left(\frac{p^2}{m^2}\right)\quad.
\end{align}

%\begin{align}
%\langle \sigma v\rangle &= \int^\infty_0\int^\infty_0\frac{g^4}{\pi^5}\frac{k^3 p^2 }{E^3}e^{-2E/T}\left[ \frac{5/2  - 3  M^2/m^2 +   M^4/m^4}{(2  - M^2/m^2)^{4}}\left(\frac{p^2}{m^2}\right)\right]\,dp\,dk
%\end{align}

The cross section $\sigma v$ for CDM can be written as proportional to $v^p$, where  $p=0$ is known as s-wave, $p=2$ is the p-wave and so on. As a power of $v^2$ the cross section becomes $\sigma v =a + bv^2+ \dots$. In our case we have $a=0$ and 
\begin{align}\label{ScattAmpl3}
b=  \frac{g^{4}}{m^2\pi}\sqrt{1-\frac{M^2}{m^2}}\frac{(5  - 6  M^2/m^2 + 2  M^4/m^4)}{(2  - M^2/m^2)^{4}}\quad.
\end{align}

From the Maxwellian velocity distribution we have $\langle v^2\rangle = 3T/m$, hence we have $ \langle \sigma v\rangle= b\langle v^2\rangle =3bT/m$. For non-relativistic particles, the equilibrium number density is given by Eq. (\ref{neq2})
\begin{equation}\label{neq}
n^{EQ}=\left(\frac{mT}{2\pi}\right)^{3/2}e^{-m/T}\quad.
\end{equation}

Using Eq. (\ref{neq}) the interaction rate $\Gamma$ becomes
\begin{equation}
\Gamma= 3b\frac{T}{m}\left(\frac{mT}{2\pi}\right)^{3/2}e^{-m/T}\quad,
\label{eq:gamma}
\end{equation}

\noindent with $b$ given by Eq. (\ref{ScattAmpl3}). Since the temperature of non-relativistic matter scales with $a^{-2}$,\footnote{Using Eq. (\ref{BOlteq1}) we have 
\begin{align}\label{eq:3.199}
\frac{\partial f_{dm}}{\partial t}- \frac{\partial f_{dm}}{\partial E}H\frac{p^2}{E}=0\quad.
\end{align}

\noindent Since $f_{dm} \propto e^{-p^2/2mT}$ and $E=\sqrt{p^2 + m^2}$ we should expand $E$ and consider only zero-order term. Furthermore, $\partial f_{dm}/\partial t$ can be written as 
\begin{align}\label{eq:3.299}
\frac{\partial f_{dm}}{\partial t}=\frac{\partial f_{dm}}{\partial T}\dot{T}=-\frac{p}{2T}\frac{\partial f_{dm}}{\partial p}\dot{T}\quad.
\end{align}

\noindent Inserting Eq. (\ref{eq:3.299}) into Eq. (\ref{eq:3.199}) and taking the derivatives

\begin{align}\label{eq:3.499}
\frac{p^2}{2m}\frac{\dot{T}}{T^2} + \frac{1}{T}\frac{\dot{a}}{a}\frac{p^2}{m}=0\quad,
\end{align}

\begin{align}\label{eq:3.599}
&\frac{d{T}}{2T} + \frac{d{a}}{a}=0\notag\\ 
&\frac{d{T}}{T} =- 2\frac{d{a}}{a}\quad.
\end{align}

\noindent Hence
\begin{align}\label{eq:3.799}
T \propto \frac{1}{a^2}\quad.
\end{align} } we have $T=T_0a^{-2}$, where $T_0$ is the dark matter temperature today.\footnote{ An upper limit on the temperature to mass ratio of cold dark matter today (away from collapsed structures) was placed by \cite{Armendariz-Picon:2013jej}, giving the value $T_0\leq 10^{-14} m$ for non-interacting dark matter. }

\begin{figure}\centering
\includegraphics[scale=0.55]{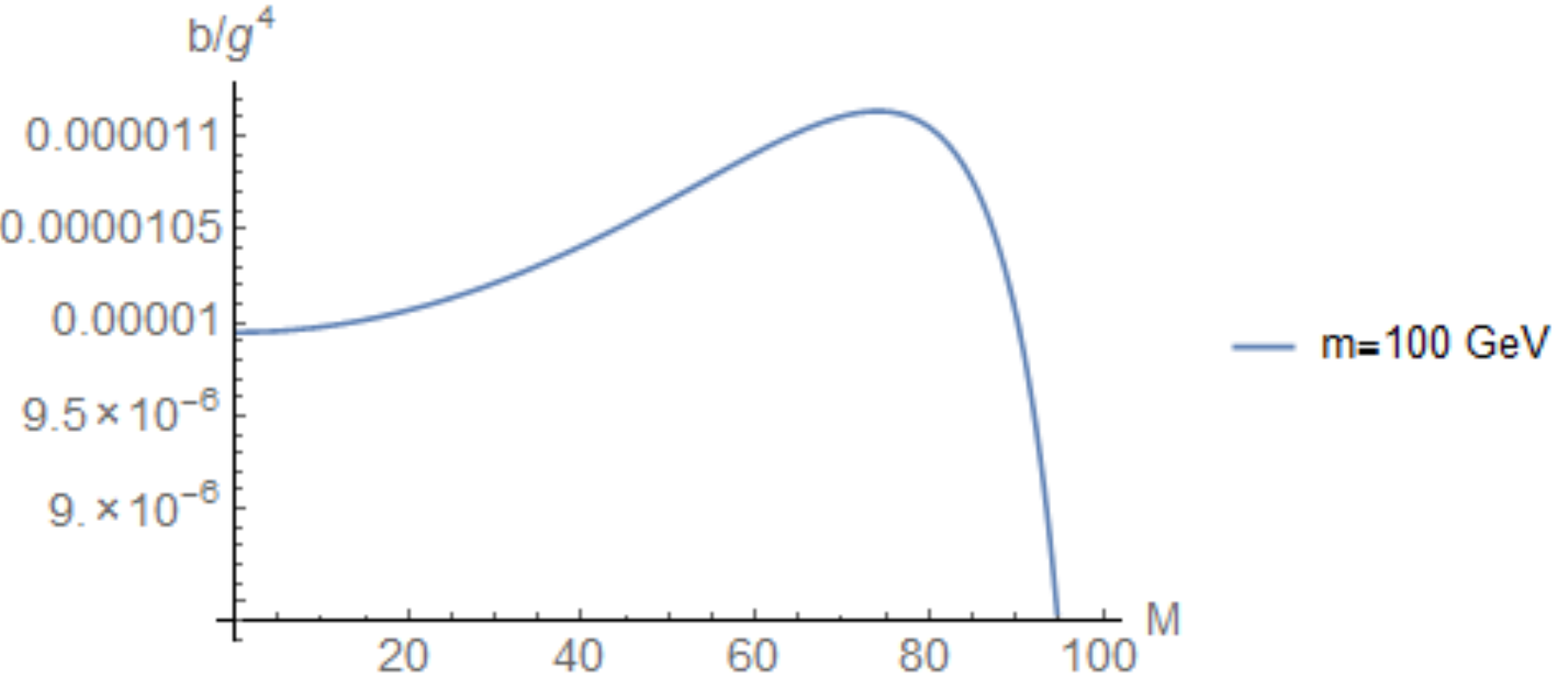}%
\caption{Plot of the parameter $b/g^4$ as a function of $M$ for $m=100$ GeV.  }%
\label{b}%
\end{figure}

We showed in Fig. \ref{b} the parameter $b$ (\ref{ScattAmpl3}) as a function of the scalar mass $M$, for the dark matter with $m=100$ GeV. Different values of masses only change the scale. 

If the interaction remained until today $H_0=10^{-42}$ GeV, the dark matter temperature $T_0$ would be $13$  MeV, $130$  MeV, $1.3$ GeV and $ 13$ GeV  for the range of dark matter mass  $1$ GeV, $10$ GeV, $100$ GeV and $1000$ GeV, respectively. This result leads to  $T_0/m\approx 0.01\ll 1$ for all the cases, which is consistent with the non-relativistic limit for the temperature to mass ratio. For comparison purposes, notice that the electroweak phase transition occurred around $T\sim 100$ GeV, while QCD phase transition happened at $ T\sim 150 $ MeV and neutrino decoupling at $T\sim 1$ MeV.

Due to the order of magnitude of the mass of the dark matter, it is expected that the particle decoupled during the radiation-dominated epoch, so $H(T)=1.66g_*^{1/2}T^2/M_{pl}$, where $g_*$ are the degrees of freedom of the particles that exist in the  universe for a temperature $T$. For $T\gtrsim  300$ GeV all particles of the standard model were relativistic, so $g_*=106.75$, while for $100$ MeV $\gtrsim T\gtrsim  1	$ MeV $g_*=10.75$. The temperature of decoupling $T$ at which $\Gamma = H$ is roughly independent of $g_*$ and it is $\sim $ $30$  MeV, $330$  MeV, $3.5$  GeV and $38$ GeV for the range of dark matter mass $1$ GeV, $10$ GeV, $100$ GeV and $1000$ GeV, respectively. These values of $m$ and $T$ lead to  $x_f=33$ for the lightest mass and  $x_f=26$ for the heaviest one.

%%%%%%%%%%%%%%%%%%%%%%%%%%%%%%%%%%%%%%%%%%%%%%%%%%%%%%%%%%%%%%%%%%%%%%%%%%%%%%%%%%%%%%%%%%%%%%%%%%%%%%%%%%%%%%%%%

\section{ Quintessence}\label{sec:quintessence}

Among a wide range of alternatives for the cosmological constant $\Lambda$, a scalar field is a viable candidate to be used  with a broad form of potentials. Its usage includes the canonical scalar field, called `quintessence' \cite{peebles1988,ratra1988,Frieman1992,Frieman1995,Caldwell:1997ii},  and the scalar field with the opposite-sign in the kinetic term, known as `phantom' \cite{Caldwell:1999ew,Caldwell:2003vq}. Both have similar equations, but different implications. The phantom case will be presented in the next chapter together with another possibility of non-canonical scalar field, the tachyon field. 

Quintessence is described by a canonical homogeneous field $\phi$ minimally coupled with gravity. In this section we do not consider any other coupling.

The action for quintessence is given by
\begin{equation}\label{eq:action}
  S = \int{d^4x\sqrt{-g}\left[-\frac{1}{2}g^{\mu\nu}\partial_\mu \phi\partial_\nu\phi-V(\phi)\right]}\quad,
\end{equation}

\noindent where $V(\phi)$ is the potential of the field. Considering the flat FLRW metric, its determinant is $\sqrt{-g}=a^3$, where $a\equiv a(t)$ is the scale factor. The variation of the action (\ref{eq:action}) with respect to $\phi$ gives the equation of motion
\begin{equation}\label{eq:fieldeq}
  \ddot{\phi} +3H\dot{\phi}+\frac{dV}{d\phi}=0\quad.
\end{equation}

The energy-momentum tensor  is given by Eq. (\ref{eq:3}). Since $\delta\sqrt{-g}=-(1/2)\sqrt{-g}g_{\mu\nu}\delta g^{\mu\nu}$, we find
\begin{equation}\label{eq:4}
  T_{\mu\nu}=\partial_\mu \phi\partial_\nu \phi - g_{\mu\nu}\left[\frac{1}{2}g^{\alpha\beta}\partial_\alpha\phi\partial_\beta\phi+V(\phi)\right]\quad.
\end{equation}

Once $\phi$ is homogeneous, spatial derivatives of the field yield zero, thus the energy density and pressure density obtained from Eq. (\ref{eq:4}) are
\begin{equation}\label{eq:rho}
  \rho_\phi=-T^0_0=\frac{1}{2}\dot{\phi}^2+V(\phi)\quad,
\end{equation}

\begin{equation}\label{eq:pressure}
  p_\phi=T^i_i=\frac{1}{2}\dot{\phi}^2-V(\phi)\quad.
\end{equation}

For this field the Friedmann equations become
\begin{equation}\label{eq:1stFE}
 H^2=\frac{8\pi G\rho_\phi}{3}=\frac{8\pi G}{3}\left[\frac{\dot{\phi}^2}{2}+V(\phi)\right]\quad,
\end{equation}

\begin{equation}\label{eq:2ndFE1}
\begin{split}
  \frac{\ddot{a}}{a}=-\frac{4\pi G}{3}(\rho_\phi+3p_\phi)=-\frac{8\pi G}{3}\left[\dot{\phi}^2-V(\phi)\right] \quad,
  \end{split}
\end{equation}

\begin{equation}\label{eq:2ndFE}
\begin{split}
 \dot{H}=-4\pi G(p_\phi+\rho_\phi)=-4\pi G\dot{\phi}^2\quad.
  \end{split}
\end{equation}

Inserting Eqs. (\ref{eq:rho}) and (\ref{eq:pressure}) in the continuity equation (\ref{eq:continuity})  we get Eq. (\ref{eq:fieldeq}) again. Using Eqs.  (\ref{eq:rho}) and (\ref{eq:pressure}) the equation of state for the field $\phi$ yields
\begin{equation}\label{eq:wphi}
  w_\phi= \frac{p_\phi}{\rho_\phi}=\frac{\dot{\phi}^2-2V(\phi)}{\dot{\phi}^2+2V(\phi)}\quad,
\end{equation}

\noindent whose  values are between $-1$ and $1$, with the accelerated expansion occurring for $w_\phi < -1/3$. For $\dot{\phi}^2 \ll V(\phi)$ we have $w_\phi=-1$, which is the equation of state for the cosmological constant.

Solving (\ref{eq:continuity}) we have
\begin{equation}\label{eq:continuitySolve}
  \rho_\phi=\rho_0\exp\left[-\int{3(1+w_\phi)\frac{da}{a}}\right]\quad.
\end{equation}

\noindent For $\dot{\phi}^2 \ll V(\phi)$, $w_\phi=-1$ and $\rho_\phi$ is constant. From the other limit $\dot{\phi}^2 \gg V(\phi)$, $w_\phi=1$ and $\rho_\phi \propto a^{-6}$. In the intermediary cases the energy density behaves as
 \begin{equation}\label{eq:continuitySolveInt}
  \rho_\phi \propto a^{-m}\quad, \ \ 0< m< 6\quad.
\end{equation}

\noindent Since $w_\phi=-1/3$ is the border of acceleration and deceleration, 
 \begin{equation}\label{eq:rhopropto}
  \rho_\phi \propto a^{-3(1-1/3)}=a^{-2}\quad,
\end{equation}

\noindent the universe exhibits an accelerated expansion for $0< m<2$. Using Eqs. (\ref{eq:1stFE}) and (\ref{eq:2ndFE}) we can write  $V(\phi)$ and $\dot{\phi}$ in terms of $H$ and $\dot{H}$
 \begin{equation}\label{eq:VintermofH}
  V=\frac{3H^2}{8\pi G}\left(1+\frac{\dot{H}}{3H^2}\right)\quad,
\end{equation}

 \begin{equation}\label{eq:phiintermofH}
  \phi=\int\,dt{\left[-\frac{\dot{H}}{4\pi G}\right]^{1/2}}\quad.
\end{equation}

It is of interest to derive a scalar-field potential that gives rise to a power-law expansion
 \begin{equation}\label{eq:aproptotp}
 a(t)\propto t^p\quad.
\end{equation}

\noindent where $p>1$ for an accelerated universe, since $\ddot{a}=p(p-1)t^{p-2}$. From Eq. (\ref{eq:aproptotp}) we have

 \begin{equation}\label{eq:HdotandH}
 H=\frac{p}{t}, \qquad \dot{H}=-\frac{p}{t^2}\quad,
\end{equation}

\noindent which is used in Eqs. (\ref{eq:VintermofH}) and (\ref{eq:phiintermofH}) to yield
 \begin{equation}\label{eq:phio}
 \phi=\sqrt{\frac{p}{4\pi G}}\ln t\quad,
\end{equation}

 \begin{equation}\label{eq:vo}
 V(\phi)=\frac{p}{8\pi G}(3p-1)e^{-2\sqrt{\frac{4\pi G}{p}}\phi}=V_0e^{-\sqrt{\frac{16\pi G}{p}}\phi}\quad.
\end{equation}

 In the next chapter it will be carried out a more detailed analysis of quintessence in a presence of a barotropic fluid.
 
 Before going ahead lets see how is the time evolution of the energy density (\ref{eq:rhopropto}), considering Eqs.  (\ref{eq:phio}) and (\ref{eq:vo}). Using these equations into Eq.  (\ref{eq:rho}) we have
   \begin{equation}\label{eq:rhot}
 \rho_\phi=\frac{3p^2}{8\pi Gt^2} \propto t^{-2} \propto a^{-m}\quad,\ \ 0<m<2\quad,
\end{equation}

\noindent where the last proportionality comes from Eq. (\ref{eq:continuitySolveInt}). Eq. (\ref{eq:rhot}) leads to
 \begin{equation}\label{eq:at}
a \propto t^{2/m}\quad, \qquad \frac{2}{m} >1\quad,
\end{equation}

 \noindent as required by Eq. (\ref{eq:aproptotp}). 

Besides the exponential potential (\ref{eq:vo}), one of the first proposals for the scalar potential was the inverse power-law $V(\phi)\sim \phi^{-\alpha}$, where $\alpha>0$. Although non-renormalizable, such potential has the remarkable property that it leads to the attractor-like behavior for the equation of state and density parameter of the  dark energy, which are nearly constants for a wide range of initial conditions \cite{Zlatev:1998tr,Steinhardt:1999nw}. This potential is an example of the so-called `tracker behavior', where the correspondent scalar field is a tracker field. 

Tracking occurs for any potential for which $w_\phi < w_m$ (where $w_m $ is the equation of state for radiation or matter) and $\Gamma\equiv VV''/(V')^2>1$ is nearly constant ($|d(\Gamma -1)/H dt| \ll |\Gamma -1|$) \cite{Steinhardt:1999nw}. In this case, the equation of state for the quintessence field is also nearly constant
\begin{equation}
w_\phi \approx \frac{w_m-2(\Gamma-1)}{1+2(\Gamma-1)}\quad .
\label{eq:trac}
\end{equation}

\noindent For the inverse power-law $\Gamma=(\alpha+1)/\alpha$. 

The independence of initial conditions alleviates another concern regarding the dark energy, namely, why dark energy and matter densities are of the same order today, known as `coincidence problem'. The two densities decrease at different rates, so it seems that the initial conditions in the early universe had to be set very carefully for those energy densities to be comparable today. 

As pointed in  \cite{bean2001}  the observed abundances of primordial elements put strong constraints on $\Omega_\phi^{BBN}$. We use this bound in chapter  \ref{dynanaly}.

%% file: dynanalysis.tex
\chapter{Dynamical analysis for coupled dark energy models}\label{dynanaly}

 In this chapter, equipped with dynamical system theory, we investigate the critical points that come from the evolution equations for the complex scalar field (quintessence, phantom and tachyon) and for the vector dark energy, considering the possibility of interaction between the two components of the dark sector. For the tachyonic scalar field, we first analyze the case of a real field for a new form of interaction between the two components of the dark sector, presented in our work \cite{Landim:2015poa}. Then, we proceed to the complex quintessence, phantom and tachyon field whose dynamical equations are derived and the critical point analyzed. This is a natural extension of the previous works \cite{Amendola:1999er,Gumjudpai:2005ry,ChenPhantom,Landim:2015poa} and the new results were presented in our papers \cite{Landim:2015poa, Landim:2015uda,Landim:2016dxh}.

In Sect. \ref{basicsDA} we present a quick review on dynamical system theory, necessary for our purposes. In Sect. \ref{CSF} we apply the formalism to quintessence (phantom) and tachyon fields. Sect. \ref{sec:vecDE} is reserved for the vector-like dark energy. 

\section{Basics}\label{basicsDA}

To deal with the dynamics of the system of evolution equations, we  define dimensionless variables. The new variables are going to characterize a system of differential equations in the form $\dot{x}=f(x_1,x_2,\dots,t)$, $\dot{y}=g(x_1,x_2\dots,t)$, $\dots$, so that $f$, $g$, $\dots$ do not depend explicitly on time. For this system (called \textit{autonomous}), a point ($x_{1c}$, $x_{2c}$, $\dots$) is called \textit{fixed} or \textit{critical point}  if $(f, g,\dots)|_{x_{1c},x_{2c},\dots}=0$ and it is an \textit{attractor} when $(x_{1},x_{2},\dots)\rightarrow (x_{1c},x_{2c},\dots)$ for $t\rightarrow \infty$. The equations of motion for the variables $x_1$, $x_2$, $x_3$, $y$, $z$  and $\lambda$ are obtained taking the derivatives  with respect to $N\equiv \log a$, where we set the present scale factor $a_0$ to be one. The fixed points of the system are obtained by setting $dx_1/dN=0$, $dx_2/dN=0$, $\dots$. The  system of differential equations can be written in the compact form

\begin{equation}
X'=f[X],
\end{equation}

\noindent where $X$ is a column vector of dimensionless variables and the prime is the derivative  with respect to $ \log a$. The critical points $X_c$ are those ones that satisfy $X'=0$. 

In order to study stability of the fixed points we consider linear perturbations $U$ around them, thus $X=X_c+U$. At the critical point the perturbations $U$ satisfy the following equation

\begin{equation}
U'=\mathcal{J}U,
\end{equation}

\noindent where $\mathcal{J}$ is the Jacobian matrix. The stability around the fixed points depends on the nature of the eigenvalues ($\mu$) of $\mathcal{J}$, in such a way that they are stable points if they if they are all negative, unstable points if they are all positive  and saddle points if at least one eigenvalue has positive (or negative) value, while some other has opposite sign.  In addition, if any eigenvalue is a complex number, the fixed point can be stable (Re $\mu<0$) or unstable (Re $\mu>0$) spiral, due to the oscillatory behavior of its imaginary part. If any fixed point has at least one eigenvalue equals zero, the  fixed point is not hyperbolic and a higher-order analysis is needed in order to have a complete description of their stability.

\section{A dynamical analysis with complex scalar field}\label{CSF}

\subsection{Quintessence and phantom dynamics}\label{quintphantom}

We will  present first the equations of a complex scalar field without interaction. The complex scalar field $\Phi$ can be written as $\Phi=\phi e^{i\theta}$, where $\phi$ is the absolute value of the scalar field and $\theta$ is a phase. Both canonical and phantom fields are described by the  Lagrangian
\begin{equation}\label{scalar}
 \mathcal{L}=-\sqrt{-g}\left(\frac{\epsilon}{2}\partial^\mu\Phi^*\partial_\mu\Phi+V(|\Phi|)\right)\quad,
\end{equation} 

\noindent where $V(|\Phi|)$ is the potential for the complex scalar and we consider it depends only on the absolute value  of the scalar field $\phi\equiv|\Phi|$. We have $\epsilon=+1$ for the canonical field (quintessence) and $\epsilon=-1$ for the phantom field. For a homogeneous field $\phi\equiv\phi(t)$ and $\theta\equiv\theta(t)$, in an expanding universe with FLRW metric with scale factor $a\equiv a(t)$, the equations of motion are
\begin{equation}\label{eqmotionscalar1}
 \epsilon \ddot{\phi}+3\epsilon H\dot{\phi}+V'(\phi)-\epsilon \phi \dot{\theta}^2=0\quad,
\end{equation}

\begin{equation}\label{eqmotionscalar2}
 \epsilon \ddot{\theta}+\left(3 H+\frac{2\dot{\phi}}{\phi}\right)\dot{\theta}=0\quad.
\end{equation} 

\noindent where the prime denotes derivative with respect to $\phi$. For the quintessence field, Eq. (\ref{eqmotionscalar2}) has the solution $\dot{\theta}=\frac{w}{a^3\phi^2}$, where $\omega$ is an integration constant. Using this result in Eq. (\ref{eqmotionscalar1}) leads to an effective potential, which is $\frac{d}{d\phi}\left(\frac{\omega^2}{2a^6}\frac{1}{\phi^2}+V(\phi)\right)$ \cite{Gu2001}. The first term in the brackets drives $\phi$ away  from zero  and the factor $a^{-6}$ may make the term decrease very fast, provided that $\phi$ does not decrease faster than $a^{-3/2}$.  

We assume the interaction between the scalar field with dark matter through the coupling  $Q \rho_m\dot{\phi}$ \cite{Wetterich:1994bg,Amendola:1999er} and it enters in the right-hand side of Eq. (\ref{eqmotionscalar1}), which in turn becomes

\begin{equation}\label{eqmotionscalar12}
 \epsilon \ddot{\phi}+3\epsilon H\dot{\phi}+V'(\phi)-\epsilon \phi \dot{\theta}^2=Q \rho_m\quad.
\end{equation} 

In the presence of matter and radiation, the Friedmann equations for the canonical (phantom) field are
\begin{align}\label{eq:1stFEmatterS}
  H^2&=\frac{1}{3}\left(\frac{\epsilon}{2}\dot{\phi}^2+\frac{\epsilon}{2}\phi^2\dot{\theta}^2+V(\phi)+ \rho_m+\rho_r\right)\quad,\\
  \dot{H}&=-\frac{1}{2}\left(\epsilon\dot{\phi}^2+\epsilon\phi^2\dot{\theta}^2+\rho_m+\frac{4}{3}\rho_r\right)\quad,
\end{align}

\noindent and the equation of state becomes
\begin{equation}\label{eqstateS}
 w_\phi=\frac{p_\phi}{\rho_\phi}=\frac{\dot{\phi}^2+\phi^2\dot{\theta}^2-2\epsilon V(\phi)}{\dot{\phi}^2+\phi^2\dot{\theta}^2+2\epsilon V(\phi)}\quad.
\end{equation}

We are now ready to proceed the dynamical analysis of the system.

\subsubsection{Autonomous system} 

The dimensionless variables are defined as
\begin{equation}\label{eq:dimensionlessXYS}
\begin{aligned}
 x_1\equiv  &\frac{\dot{\phi}}{\sqrt{6}H}, \quad x_2\equiv \frac{\phi\dot{\theta}}{\sqrt{6}H}, \quad x_3\equiv \frac{\sqrt{6}}{\phi}, \quad y\equiv \frac{\sqrt{V(\phi)}}{\sqrt{3}H}\quad, \\ 
 &z\equiv\frac{\sqrt{\rho_r}}{\sqrt{3}H}, \quad \lambda\equiv -\frac{V'}{V}, \quad \Gamma\equiv \frac{VV''}{V'^2}\quad.\\
\end{aligned}
\end{equation}

The dark energy density parameter is written in terms of these new variables as
\begin{equation}\label{eq:densityparameterXYS}
 \Omega_\phi \equiv \frac{\rho_\phi}{3H^2} =\epsilon x_1^2+\epsilon x_2^2+y^2\quad,
 \end{equation}

\noindent so that Eq. (\ref{eq:1stFEmatterS}) can be written as 
\begin{equation}\label{eq:SomaOmegasS}
\Omega_\phi+\Omega_m+\Omega_r=1\quad,
\end{equation}

\noindent where the matter and radiation density parameter are defined by $\Omega_i=\rho_i/(3H^2)$, with $i=m,r$. From Eqs. (\ref{eq:densityparameterXYS}) and (\ref{eq:SomaOmegasS}) we have that $x_1$, $x_2$ and $y$ are restricted in the phase plane by the relation
\begin{equation}\label{restrictionS}
0\leq \epsilon x_1^2+\epsilon x_2^2+y^2\leq 1\quad,
 \end{equation}
 
\noindent since $0\leq \Omega_\phi\leq 1$. Notice that if $y=0$ the restriction (\ref{restrictionS}) forbids the possibility of phantom field ($\epsilon=-1$) because for this case $\Omega_\phi<0$.

The equation of state $w_\phi$  becomes
\begin{equation}\label{eq:equationStateXYS}
 w_\phi =\frac{\epsilon x_1^2+\epsilon x_2^2-y^2}{\epsilon x_1^2+\epsilon x_2^2+y^2}\quad,
\end{equation}

\noindent which is a trivial extension of the real scalar field case. The total effective equation of state is
\begin{equation}\label{eq:weffS}
 w_{eff} = \frac{p_\phi+p_r}{\rho_\phi+\rho_m+\rho_r}=\epsilon x_1^2+\epsilon x_2^2-y^2+\frac{z^2}{3}\quad,
\end{equation}

\noindent with an accelerated expansion for  $w_{eff} < -1/3$.  The dynamical system for the variables  $x_1$, $x_2$,  $x_3$, $y$, $z$  and $\lambda$ are 
\begin{subequations}\label{dynsystemS}\begin{align}\label{eq:dx1/dnS}
\frac{dx_1}{dN}&=-3x_1+x_2^2x_3+\frac{\sqrt{6}}{2}\epsilon y^2\lambda-\frac{\sqrt{6}}{2}\epsilon Q(1-x_1^2-x_2^2-y^2-z^2)-x_1H^{-1}\frac{dH}{dN}\quad,\\
\frac{dx_2}{dN}&=-3x_2-x_1x_2x_3-x_2H^{-1}\frac{dH}{dN}\quad,\\
\frac{dx_3}{dN}&=-x_1x_3^2\quad,\\
\frac{dy}{dN}&=-\frac{\sqrt{6}}{2}x_1 y\lambda-yH^{-1}\frac{dH}{dN}\quad,\\
\frac{dz}{dN}&=-2z-zH^{-1}\frac{dH}{dN}\quad,\\
\frac{d\lambda}{dN}&=-\sqrt{6}\lambda^2 x_1\left(\Gamma-1\right)\quad,
\end{align}
\end{subequations}

\noindent where

\begin{equation}\label{}
H^{-1}\frac{dH}{dN}=-\frac{3}{2}(1+\epsilon x_1^2+\epsilon x_2^2-y^2)-\frac{z^2}{2}\quad.
\end{equation}

\subsubsection{Critical points}

 The fixed points of the system are obtained by setting $dx_1/dN=0$, $dx_2/dN=0$, $dx_3/dN=0$, $dy/dN=0$, $dz/dN$ and $d\lambda/dN=0$ in Eq. (\ref{dynsystemS}). When $\Gamma=1$, $\lambda$ is constant the potential is $V(\phi)=V_0e^{-\lambda \phi}$ \cite{copeland1998,ng2001}.\footnote{The equation for $\lambda$ is also equal zero when $x_1=0$ or $\lambda=0$, so that  $\lambda$ should not necessarily be constant, for the fixed points with this value of $x_1$. However, for the case of dynamical $\lambda$, the correspondent eigenvalue is equal zero, indicating that the  fixed points is not hyperbolic.} The fixed points are shown in Table \ref{criticalpointsS}. Notice that $x_3$ and $y$ cannot be negative and recall that $\Omega_r=z^2$. Some of the fixed points do not exist for the phantom field because for those cases $\Omega_\phi$ is negative.

 \begin{table}\footnotesize
\small\addtolength{\tabcolsep}{-3pt}
\begin{tabular}{cccccccccc}
\hline\noalign{\smallskip}
Point  & Existence & $x_1$ &$x_2$&$x_3$& $y$  &$z$& $w_\phi$ & $\Omega_\phi$& $w_{eff}$ \\
\noalign{\smallskip}\hline\noalign{\smallskip}
(a)   	&$ Q=0$	       & 0 & 0& any&0     & 0& --&0&0   \\
								(b) &$ \epsilon=+1$	 &$\frac{-\sqrt{6}Q}{3}$ & 0  & 0   & 0& 0& 1& $\frac{2Q^2}{3}$&  $\frac{2Q^2}{3}$ \\
		(c) & any & $0$&$0$ & any &0&1 &-- &0 & $\frac{1}{3}$ \\
		 (d) &$ \epsilon=+1$ & $\frac{-1}{\sqrt{6}Q}$ &0   &0& $0$ &$ \sqrt{1-\frac{1}{2Q^2}}$& $1$&$\frac{1}{6Q^2}$  &   $\frac{1}{3}$ \\ 
		 (e) &$ \epsilon=+1$ & $\frac{2\sqrt{6}}{3\lambda}$ & 0 &0 &$\frac{2\sqrt{3}}{3\lambda}$&$\sqrt{1-\frac{4}{\lambda^2}}$   &$\frac{1}{3}$  &$\frac{4}{\lambda^2}$ & $\frac{1}{3}$\\
		 
    (f) &$ \epsilon=+1$ & any&$\sqrt{1-x_1^2}$   &0& 0 & 0& 1&1  &   1\\ 
 
      (g) &any &$\frac{\sqrt{6}}{2(\lambda+Q)}$ & 0  & 0   & $\sqrt{\frac{2Q(Q+\lambda)+3\epsilon}{2(\lambda+Q)^2}}$ & $0$& $\frac{-Q(Q+\lambda)}{Q(Q+\lambda)+3\epsilon}$& $\frac{Q(Q+\lambda)+3\epsilon}{(\lambda+Q)^2}$& $\frac{-Q}{\lambda+Q}$\\
 (h) & any & $\frac{\epsilon \lambda}{\sqrt{6}}$ &0&0 & $\sqrt{1-\frac{\epsilon \lambda^2}{6}}$ & 0& $-1+\frac{\epsilon \lambda^2}{3}$ &1  & $-1+\frac{\epsilon \lambda^2}{3}$\\

 \noalign{\smallskip}\hline
\end{tabular}\caption{\label{criticalpointsS} Critical points ($x_1$, $x_2$, $x_3$, $y$ and $z$) of the Eq. (\ref{dynsystemS}) for the quintessence and phantom field. It is shown the condition of existence, if any,  of the fixed point (point (a) exists only for $Q=0$, for instance, while the point (b) does not exist for the phantom field.).  The table shows the correspondent equation of state for the dark energy (\ref{eq:equationStateXYS}), the effective equation of state (\ref{eq:weffS}) and the density parameter for dark energy (\ref{eq:densityparameterXYS}).}
\end{table}

The eigenvalues of the Jacobian matrix were found for each fixed point in Table \ref{criticalpointsS}.  The results are shown in Table \ref{stabilityS}.

\begin{table}[tbp]\footnotesize
\centering
\begin{tabular}{ccccccc}
\hline\noalign{\smallskip}
  Point & $\mu_1$ & $\mu_2$ & $\mu_3$&  $\mu_4$& $\mu_5$& Stability \\
\noalign{\smallskip}\hline\noalign{\smallskip}
  (a)        & $\frac{9}{2}$&$-\frac{3}{2}$& 0& $\frac{3}{2}$&$-\frac{1}{2}$&saddle\\
  (b)    & $Q^2-\frac{3}{2}$&$Q^2-\frac{3}{2}$&$0$& $Q(Q+\lambda)+\frac{3}{2}$& $Q^2+\frac{1}{2}$&unstable  or saddle\\
  (c)  & $-1$&$-1$  &$0$& 2& $1$&saddle\\ 
  (d)   & $-1+\frac{1}{2Q^2}$&$-1$   &   $0$    &$2+\frac{\lambda}{2Q}$ & $1-\frac{Q^2}{2}$&saddle    \\ 
   (e)   & & & see the main text    && & \\
   (f)   & $3x_1^2+\sqrt{6}Qx_1$& $3(x_1^2-1)$ &$\mp x_1\sqrt{1-x_1^2}$ &$3-\frac{\sqrt{6}x_1\lambda}{2}$& $1$ &unstable or saddle  \\
       (g)   & $-\frac{\lambda+4Q}{2(\lambda+Q)} $&$-\frac{3(\lambda+2Q)}{2(\lambda+Q)}$&0& $\mu_{4d}$& $\mu_{5d}$&saddle or stable\\
  (h) & $-3+\epsilon \lambda(\lambda+Q) $&   $-3+\frac{\epsilon\lambda^2}{2} $&0 &  $-3+\frac{\epsilon\lambda^2}{2}  $&$-2+\frac{\epsilon\lambda^2}{2}  $ &saddle or stable\\
 
     \noalign{\smallskip}\hline
\end{tabular}
\caption{\label{stabilityS} Eigenvalues and stability of the fixed points for the quintessence (phantom) field.}
\end{table}

The eigenvalues $\mu_{4e}$ and $\mu_{5e}$ are 
\begin{equation}\label{eigenh}
\mu_{4e,5e}=-\frac{3(\lambda+2Q)}{4(\lambda+Q)}\left(1\pm\sqrt{1+\frac{8[3-\epsilon\lambda(\lambda+Q)][3\epsilon+2Q(\lambda+Q)]}{3(\lambda+2Q)^2}}\right)\quad.
\end{equation}

All points shown here are similar to the ones found in the literature \cite{copeland1998,Amendola:1999er,Gumjudpai:2005ry}. At a first glance, one might think that  the linear analysis would not give a complete description of the stability, because all fixed points, but (f), have at least one eigenvalue equal zero. However, as pointed out in \cite{Boehmer:2011tp}, fixed points that have at least one positive and one negative eigenvalue are always unstable, and methods such as center manifold \cite{Boehmer:2011tp} should be used to analyze the stability of the critical points  that can be stable [((g) and (h)]. Even so, for almost all fixed points, but (a) and (c),  $x_3=0$, which means $\phi\rightarrow \infty$. However, this limit implies that $x_2\propto  \phi\dot{\theta}/H\rightarrow \infty$ as well,  provided that $H$ is finite.This issue occurred for the points (b) and (d) to (h), as can be seen in Table \ref{criticalpointsS} whose mathematical inconsistency indicates that the these critical points are  not physically acceptable.\footnote{This point could be thought as being an issue of the choice of variables definition. However, it does not seem viable to define $x_2$ and $x_3$ in a different way.}   We reproduce the main results of all the critical points below (for a real scalar field), for the sake of completeness. 

The fixed point (a) is a saddle point which describes a matter-dominated universe, however, it is valid only for $Q=0$. The other possibility of matter-dominated universe with $Q\neq 0$ arises from the fixed point (b). This point is called `$\phi$-matter-dominated epoch' ($\phi$MDE) \cite{Amendola:1999er} and it can be either unstable or a saddle point. However, due to $\Omega_\phi=2Q^2/3\ll 1$, the condition $Q^2\ll 1$ should hold in order to the point to be responsible for the matter era. Thus $\mu_1$ and $\mu_2$ are negative, while $\mu_5$ is always positive and $\mu_4$ is positive for $Q(\lambda+Q)>-3/2$. Therefore, (b) is a saddle point. 

The radiation-dominated universe is described by the critical points (c), (d) and (e), only for the quintessence field. The first two points are saddle, as it is easily seen in Table  \ref{stabilityS}, and the last one had its stability described numerically in \cite{Amendola:1999er}. However, both (d) and (e) are not suitable to describe the universe we live in, due to nucleosynthesis constraints \cite{Amendola,amendola2001}. The nucleosynthesis bound $\Omega_\phi^{BBN} < 0.045$ \cite{bean2001} implies $Q^2>3.7$ for the point (d) and $\lambda^2>88.9$ for the point (e). Thus the requirement for the point (d) is not consistent with the condition of point (b) and the constraint on $\lambda^2$ does not allow a scalar field attractor, as we will see soon. Therefore, the only viable cosmological critical point for the radiation era is (c).

The point (f) is an unstable or saddle point and it does not describe an accelerated universe.  The last possibility for the matter era is the point (g), with eigenvalues shown in Table \ref{stabilityS} and Eq. (\ref{eigenh}). Since $w_{eff}\simeq 0$ for $|\lambda|\gg |Q|$, the fixed point is either stable or stable spiral, hence the universe would not exit from the matter dominance. 

On the other hand, the point (g) can lead to an accelerated universe, for the quintessence field case ($\epsilon=+1$), provided that $3<\lambda(\lambda+Q)$, because $\Omega_\phi\leq 1$, and $Q>\lambda/2$, from $w_{eff}<-1/3$. Regarding $\lambda>0$, the two eigenvalues $\mu_1$ and $\mu_2$ are always negative and since $Q>3/\lambda-\lambda$, the behavior of $\mu_{4d,5d}$ depends on the second term in the square root of Eq. (\ref{eigenh})

\begin{equation}
A\equiv \frac{8[3-\lambda(\lambda+Q)][3\epsilon+2Q(\lambda+Q)]}{3(\lambda+2Q)^2}\quad.\end{equation}

 \noindent If $A<1$ the fixed point is stable and from the condition $3<\lambda(\lambda+Q)$ we have $A<0$. Otherwise, i.e. $A>1$, the critical point is a stable spiral. Thus the value of the coupling dictates which behavior the fixed point will have: stable for $3/\lambda-\lambda<Q<Q_*$ or stable spiral for $Q>Q_*$, where $Q_*$ is the solution of $A=1$. However, even in the case where one can get $\Omega_\phi\simeq 0.7$ \cite{Hebecker2000,amendola2001}, there are no allowed region in the $(Q,\lambda)$ plane corresponding to the transition from $\phi $MDE to scaling attractor \cite{Amendola:1999er}. Thus it is hard to gather the conditions for the point $\phi$MDE and the point (g). For the case of the phantom field ($\epsilon=-1$), the condition $y^2>0$ implies   $2Q(Q+\lambda)>3$. Hence, $\mu_4<0$ and $\mu_5>0$, and (g) is a saddle point.
 
 The last fixed point (h) leads to an accelerated universe provided that $\lambda^2<2$. With this condition, the eigenvalues $\mu_2$, $\mu_4$ and $\mu_5$ are always negative. The first eigenvalue $\mu_1$ is also always negative for the phantom field, and it is for the quintessence field with the condition $\lambda(\lambda+Q)<3$. Therefore, the point is stable if the previous conditions are satisfied. 

\begin{figure}\centering
\includegraphics[scale=0.52]{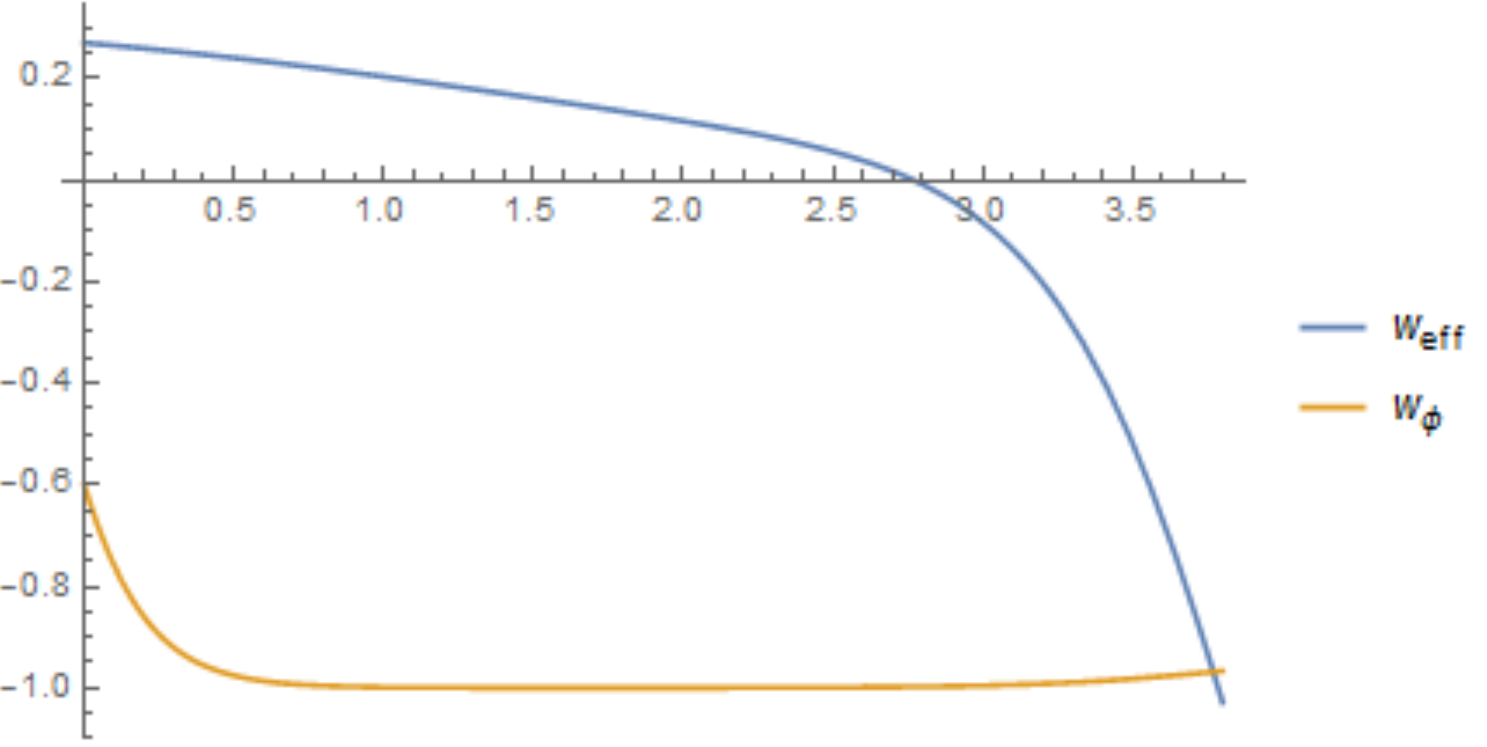}%
\caption{Behavior of the equation of state for the dark energy ($w_\phi$) and the effective equation of state ($w_{eff}$) for the quintessence field with $Q=0$. The time scale is arbitrary and the universe today is described by the value $t=4$.}%
\label{weffwphi}%
\end{figure}

\begin{figure}\centering
\includegraphics[scale=0.52]{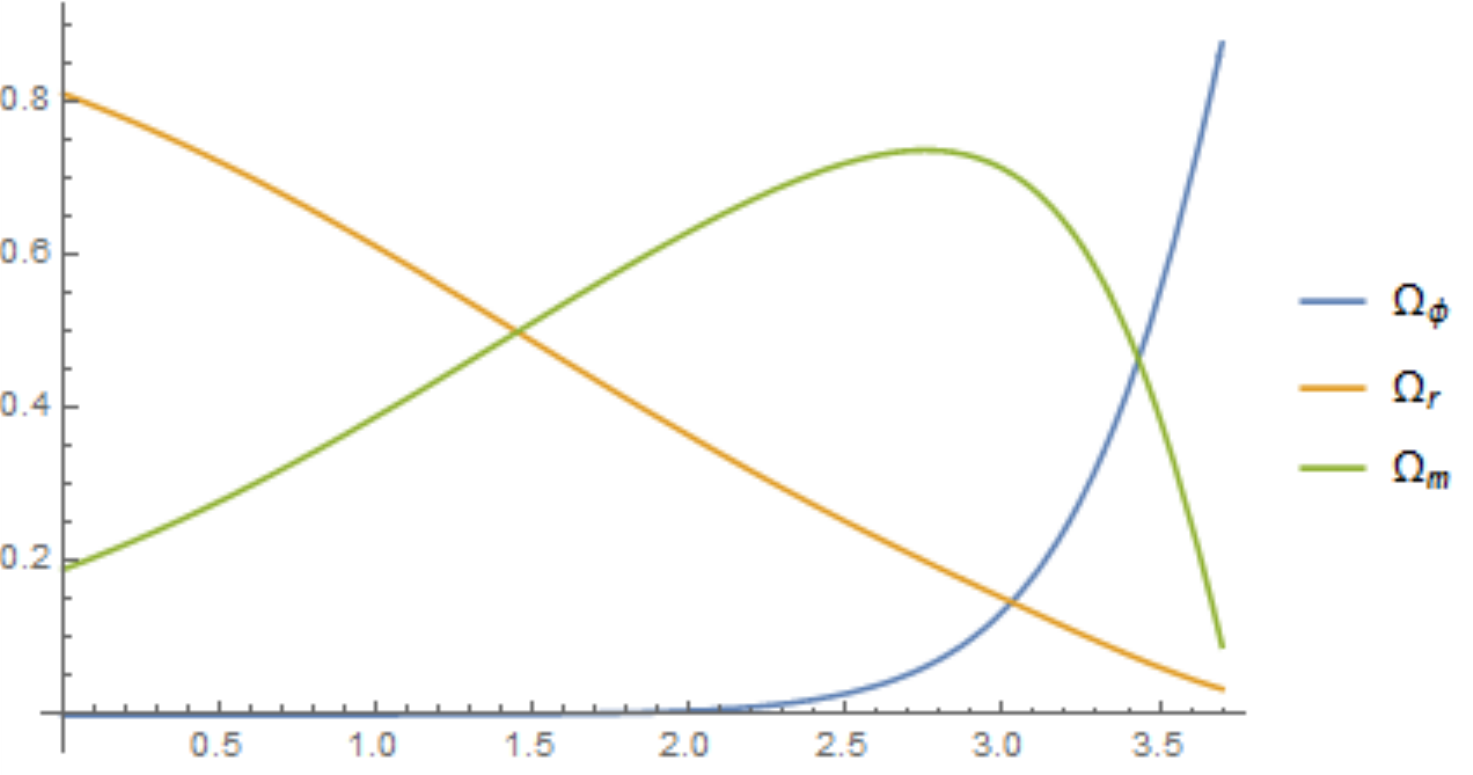}%
\caption{Behavior of the energy density parameter for the dark energy, radiation and matter, for the quintessence field with $Q=0$. The time scale is arbitrary and the universe today is described by the value $t=4$.}%
\label{omegas}%
\end{figure}

For illustrative purposes, the evolution of the relevant equations of state ($w_\phi$ and $w_{eff}$) and the energy density parameters, are shown respectively in Figs. \ref{weffwphi} and \ref{omegas}, for the real quintessence field with $Q=0$. 

\subsubsection{Summary}

From the eight fixed points presented in the quintessence (phantom) case, only (a) and (c) are physically viable and they describe  the sequence: radiation $\rightarrow$ matter. Both of them are unstable, however, there does not exist a point that describe the dark-energy-dominated universe. Thus the  extra degree of freedom due to the phase $\theta$ spoils the physically acceptable fixed points that exist for the case of real scalar field, indicating that the dynamical system theory is not a good tool when one tries to analyze the complex quintessence (phantom).

%%%%%%%%%%%%%%%%%%%%%%%%%%%%%%%%%%%%%%%%%%%%%%%%%%%%%%%%%%%%%%%%%%%%%%%%%%%%%%%%%%%%%%%%%%%%%%%%%%%%%%%%%%%%%%%%%%%%%%

\subsection{Tachyon dynamics}\label{tachdyn1}

 The role of the tachyon in theoretical physics has a long story. For the most part of it, tachyon has been considered an illness whose  treatment and cure have been hard to be achieved. In the field theory, tachyon is associated with a negative mass-squared particle, which means that the potential is expanded around its maximum point. The natural question that arises is  whether tachyon potential has a good minimum elsewhere. The Mexican hat potential in the spontaneous symmetry breaking is an example of this perspective. In the bosonic string, its ground state is the tachyon field, whereas in supersymmetric string theory a real tachyon is present in non-BPS Dp-branes,\footnote{Dp-branes are $d$-dimensional objects where there exists the matter multiplet. Stable branes are those ones with $d$ even for Type-IIA string theory, while $d$ is odd for Type-IIB.} while the complex tachyon appears in a brane-anti-brane system \cite{Sen:2004nf}.  However, the potential of such Dp-branes does have a minimum \cite{Sen:1998sm, Sen:1999xm} and at this minimum  the tachyon field behaves like a pressureless gas \cite{Sen:2002in}. As soon as tachyon condensation in string theory had been proposed, its role in cosmology was studied and plenty of works have thenceforth been done. It was also a dark-energy candidate \cite{Padmanabhan:2002cp,Bagla:2002yn,Abramo:2003cp}.

 The complex tachyon field $\Phi=\phi e^{i\theta}$, where $\phi$ is the absolute value of the tachyon field and $\theta$ is a phase, is described by the Born-Infeld Lagrangian
 \begin{equation}\label{LBI1}
 \mathcal{L}_{BI}=-\sqrt{-g}V(|\Phi|)\sqrt{1+\partial^\mu\Phi^*\partial_\mu\Phi}\quad,
\end{equation} 

\noindent where $V(\phi)$ is the tachyon potential, which depends only on the absolute value of the scalar field $\phi\equiv|\Phi|$.  For a homogeneous field $\phi\equiv\phi(t)$ and $\theta\equiv\theta(t)$, in an expanding universe with Friedmann-Robertson-Walker metric, the Lagrangian becomes
\begin{equation}\label{LBIcosmo}
 \mathcal{L}_{BI}=-a^3V(\phi)\sqrt{1-\dot{\phi}^2-\phi^2\dot{\theta}^2}\quad,
\end{equation}

\noindent where $a\equiv a(t)$ is the scale factor. The equations of motion for $\phi$ and $\theta$ are, respectively,

\begin{equation}\label{eqmotion1}
 \frac{\ddot{\phi}}{1-\phi^2\dot{\theta}^2-\dot{\phi}^2}+\frac{\ddot{\theta}+\dot{\phi}\dot{\theta}\phi^{-1}}{(1-\phi^2\dot{\theta}^2-\dot{\phi}^2)}\frac{\phi^2\dot{\phi}\dot{\theta}}{(1-\phi^2\dot{\theta}^2)}+\frac{3H\dot{\phi}-\phi\dot{\theta}^2}{1-\phi^2\dot{\theta}^2}+\frac{V'(\phi)}{V(\phi)}=0\quad,
\end{equation}

\begin{equation}\label{eqmotion2}
 \frac{\ddot{\theta}}{1-\phi^2\dot{\theta}^2-\dot{\phi}^2}+3H\dot{\theta}+\frac{2\dot{\phi}\dot{\theta}}{\phi(1-\phi^2\dot{\theta}^2-\dot{\phi}^2)}=0\quad,
\end{equation}

\noindent where the prime denotes time derivative with respect to $\phi$. When the phase $\theta$ is zero, we recover the well-known equation of motion for the tachyon field. 
The interaction between the tachyon and the dark matter is driven by the phenomenological coupling $\mathcal{Q}=Q \rho_m\rho_\phi\dot{\phi}/H$, which in turn we consider that it modifies the right-hand side of Eq. (\ref{eqmotion1})

\begin{equation}\label{eqmotion4}
 \frac{\ddot{\phi}}{1-\phi^2\dot{\theta}^2-\dot{\phi}^2}+\frac{\ddot{\theta}+\dot{\phi}\dot{\theta}\phi^{-1}}{(1-\phi^2\dot{\theta}^2-\dot{\phi}^2)}\frac{\phi^2\dot{\phi}\dot{\theta}}{(1-\phi^2\dot{\theta}^2)}+\frac{3H\dot{\phi}-\phi\dot{\theta}^2}{1-\phi^2\dot{\theta}^2}+\frac{V'(\phi)}{V(\phi)}=\frac{Q \rho_m\rho_\phi}{H}\quad.
\end{equation}

 \noindent With this form of coupling, the time dependence of the coupling is implicit in the Hubble parameter $H$.

The Friedmann equations for the complex tachyon, in the presence of barotropic fluids,  are
\begin{equation}\label{eq:1stFEmatter3}
  H^2=\frac{1}{3}\left(\frac{V(\phi)}{\sqrt{1-\phi^2\dot{\theta}^2-\dot{\phi}^2}}+ \rho_m+\rho_r\right)\quad,
\end{equation}

\begin{equation}\label{eq:2ndFEmatter1}
  \dot{H}=-\frac{1}{2}\left(\frac{V(\phi)(\dot{\phi}^2+\phi^2\dot{\theta}^2)}{\sqrt{1-\phi^2\dot{\theta}^2-\dot{\phi}^2}} +\rho_m+\frac{4}{3}\rho_r\right)\quad,
\end{equation}

The equation of state for the tachyon field yields
\begin{equation}\label{eqstateT}
 w_\phi=\frac{p_\phi}{\rho_\phi}=\dot{\phi}^2+\phi^2\dot{\theta}^2-1\quad,
\end{equation}

\noindent thus the tachyon behavior is between the cosmological constant one ($w_\phi=-1$) and matter one ($w_\phi=0$), due to the square root in Eq. (\ref{LBIcosmo}). 

 We are now ready to proceed the dynamical analysis of the system.

\subsection{Autonomous system}\label{autosystem1}

The dimensionless variables for the case of tachyon field are

\begin{equation}\label{eq:dimensionlessXY}
\begin{aligned}
 x_1\equiv  &\dot{\phi}, \quad x_2\equiv \phi\dot{\theta}, \quad x_3\equiv \frac{1}{H\phi}, \quad y\equiv \frac{\sqrt{V(\phi)}}{\sqrt{3}H}\quad, \\ 
 &z\equiv\frac{\sqrt{\rho_r}}{\sqrt{3}H}, \quad \lambda\equiv -\frac{V'}{V^{3/2}}, \quad \Gamma\equiv \frac{VV''}{V'^2}\quad.\\
\end{aligned}
\end{equation}

\noindent Since $\dot{\phi}$ and $\theta$ are dimensionless variables, $\phi$ has dimension of time. For a real tachyon field neither the points $x_2$ nor $x_3$ exist. Thus, in what follows the real tachyon case is seen by just setting $x_2=x_3=0$.

The dark energy density parameter is written in terms of these new variables as
\begin{equation}\label{eq:densityparameterXY}
 \Omega_\phi \equiv \frac{\rho_\phi}{3H^2} = \frac{y^2}{\sqrt{1-x_1^2-x_2^2}}\quad,
 \end{equation}

\noindent so that Eq. (\ref{eq:1stFEmatter3}) can be written as 

\begin{equation}\label{eq:SomaOmegas}
\Omega_\phi+\Omega_m+\Omega_r=1\quad,
\end{equation}

\noindent where the matter and radiation density parameters are defined by $\Omega_i=\rho_i/(3H^2)$, with $i=m,r$. From Eqs. (\ref{eq:densityparameterXY}) and (\ref{eq:SomaOmegas}) we see that $x_1$, $x_2$ and $y$ are restricted in the phase plane by the relation
\begin{equation}\label{restriction}
0\leq x_1^2+x_2^2+y^4\leq 1\quad,
 \end{equation}
 
\noindent since $0\leq \Omega_\phi\leq 1$. In terms of these new variables the equation of state $w_\phi$  is
\begin{equation}\label{eq:equationStateXY}
 w_\phi =x_1^2+x_2^2-1\quad,
\end{equation}

\noindent which is clearly a trivial extension for the complex scalar field. The total effective equation of state is
\begin{equation}\label{eq:weff}
 w_{eff} = \frac{p_\phi+p_r}{\rho_\phi+\rho_m+\rho_r}=-y^2\sqrt{1-x_1^2-x_2^2}+\frac{z^2}{3}\quad,
\end{equation}

\noindent with an accelerated expansion for  $w_{eff} < -1/3$. The dynamical system for the variables  $x_1$, $x_2$, $x_3$, $y$, $z$  and $\lambda$ are

\begin{subequations}\label{dynsystem}\begin{align}\label{eq:dx1/dn}
\frac{dx_1}{dN}&=-(1-x_1^2-x_2^2)\left[3x_1-\sqrt{3}y\lambda+3Q\left(1-z^2-\frac{y^2}{\sqrt{1-x_1^2-x_2^2}}\right)\right]+x_2^2x_3\quad,\\
\frac{dx_2}{dN}&=-x_1x_2x_3-3x_2(1-x_1^2-x_2^2)\quad,\\
\frac{dx_3}{dN}&=-x_1x_3^2+\frac{x_3}{2}\left[3+z^2-\frac{3y^2}{\sqrt{1-x_1^2-x_2^2}}(1-x_1^2-x_2^2)\right]\quad,\\
\frac{dy}{dN}&=\frac{y}{2}\left[-\sqrt{3}x_1y\lambda+3+z^2- \frac{3y^2}{\sqrt{1-x_1^2-x_2^2}}(1-x_1^2-x_2^2)\right]\quad,\\
\frac{dz}{dN}&=-2z+\frac{z}{2}\left[3+z^2-\frac{3y^2}{\sqrt{1-x_1^2-x_2^2}}(1-x_1^2-x_2^2)\right]\quad,\\
\frac{d\lambda}{dN}&=-\sqrt{3}\lambda x_1y\left(\Gamma-\frac{3}{2}\right)\quad.
\end{align}
\end{subequations}

\subsubsection{Critical points}

 The fixed points of the system are obtained by setting $dx_1/dN=0$, $dx_2/dN=0$, $dx_3/dN=0$, $dy/dN=0$, $dz/dN$ and $d\lambda/dN=0$ in Eq. (\ref{dynsystem}). When $\Gamma=3/2$, $\lambda$ is constant the potential has the form found in Refs.~\cite{Aguirregabiria:2004xd,Copeland:2004hq} ($V(\phi)\propto \phi^{-2}$), known in the literature for both coupled \cite{Gumjudpai:2005ry,micheletti2009} and uncoupled \cite{Padmanabhan:2002cp,Bagla:2002yn} dark energy.\footnote{The equation for $\lambda$ is also equal zero when $x_1=0$, $y=0$ or $\lambda=0$, so that  $\lambda$ should not necessarily be constant, for the fixed points with these values of $x_1$ or $y$. However, for the case of dynamical $\lambda$, the correspondent eigenvalue is equal zero, indicating that the  fixed points is not hyperbolic.}  The fixed points are shown in Table \ref{criticalpoints}. Notice that $x_3$ and $y$ cannot be negative and recall that $\Omega_r=z^2$.

 \begin{table}\footnotesize
 \centering
\begin{tabular}{cccccccccc}
\hline\noalign{\smallskip}
Point & $x_1$ &$x_2$&$x_3$& $y$ & $z$& $w_\phi$ & $\Omega_\phi$& $w_{eff}$ \\
\noalign{\smallskip}\hline\noalign{\smallskip}
(a1)         & $\pm 1$ & 0& 0&0& 0& 0&0&0   \\
(a2)         & $0$ &$\pm 1$& 0& 0&0& 0& 0&0   \\
  (a3) 						& 1 &0&$ \frac{3}{2}$& 0&0& 0& 0&0  \\
   (a4) & $-Q$&$\pm\sqrt{1-Q^2}$ & 0 &0 &0 &0 &0  & 0\\
  (b) &any &$\pm\sqrt{1-x_1^2}$  & 0   & 0& $\pm 1$& 0& 0&  $\frac{1}{3}$  \\
     (c) &1 & 0  & 2   & 0& $\pm 1$& $0$& 0&  $\frac{1}{3}$  \\
      (d) &0 & 0  & any   & 0& $\pm 1$& $-1$& 0&  $\frac{1}{3}$ \\
           (e) & 0 &0&any & 1 & 0& $-1$ &1  & $-1$\\
  (f1) & $\frac{\lambda y_c}{\sqrt{3}}$ &0   &0& $y_c$ & 0& $\frac{\lambda^2 y_c^2}{3}-1$&1  &   $w_\phi$\\ 
  (f2) & $\frac{\lambda y_c}{\sqrt{3}}$ &0   &$\frac{\sqrt{3}\lambda y_c}{2}$& $y_c$ & 0& $\frac{\lambda^2 y_c^2}{3}-1$&1  &   $w_\phi$\\ 
  (g) & $-Q$ & 0 &0 &0&0   &$Q^2-1$  &0  & 0\\
   (h1) & $x_f$ &0&  0& $y_f$ &0 &$x_f^2-1$ &$\frac{w_{eff}}{w_\phi}$  &$\frac{x_fy_f\lambda}{\sqrt{3}}-1$ \\ 
 (h2) & $x_f$ &0&  $\frac{\sqrt{3\lambda y_f}}{2}$& $y_f$ &0 &$x_f^2-1$ &$\frac{w_{eff}}{w_\phi}$  &$\frac{x_fy_f\lambda}{\sqrt{3}}-1$ \\ 
  \noalign{\smallskip}\hline
\end{tabular}
\caption{\label{criticalpoints} Critical points ($x_1$, $x_2$, $x_3$, $y$ and $z$) of the Eq. (\ref{dynsystem}), for the tachyon field. The table shows the correspondent equation of state for the dark energy (\ref{eq:equationStateXY}), the effective equation of state (\ref{eq:weff}) and the density parameter for dark energy (\ref{eq:densityparameterXY}).}
\end{table}

The fixed points $y_c$, $x_f$ and $y_f$ are shown below
\begin{equation}\label{yc}
y_c=\sqrt{\frac{\sqrt{\lambda^4+36}-\lambda^2}{6}}\quad,
\end{equation}

\begin{equation}\label{xf}
x_f=-\frac{Q}{2}\pm \frac{\sqrt{Q^2+4}}{2}\quad,
\end{equation}

\begin{equation}\label{yf}
y_f=\frac{-\lambda x_f+ \sqrt{\lambda^2x_f^2+12\sqrt{1-x_f^2}}}{\sqrt{12(1-x_f^2)}}\quad.
\end{equation}

The eigenvalues of the Jacobian matrix were found for each fixed point in Table \ref{criticalpoints}.  The results are shown in Table \ref{stability}.

\begin{table}\footnotesize
\centering
\begin{tabular}{cccc}
\hline\noalign{\smallskip}
  Point & $\mu_1$ & $\mu_2$ & $\mu_3$ \\
\hline\noalign{\smallskip}
  (a1)        & $6(1\pm Q)$&0& $\frac{3}{2}$\\
  (a2)        & $0$&  $6$&$\frac{3}{2}$\\
   (a3)        & $6(1+ Q)$&$-\frac{3}{2}$& $-\frac{3}{2}$\\
  (a4)  & $0$&$6(1-Q^2)$  &$\frac{3}{2}$\\ 
  (b)    & $6x^2_1$&$6x_2^2$&2\\
  (c)    & $6$&$-2$&$-2$\\
   (d)   & $-3 $&$-3$&2\\
  (e)   & $-3 $&   $-3 $&0 \\
 (f1)   & $\sqrt{3}Q\lambda y_c -3\left(1-\frac{\lambda^2y_c^2}{3}\right)$&$-3\left(1-\frac{\lambda^2y_c^2}{3}\right)$   &   $\frac{\lambda^2y_c^2}{2}$      \\ 
 (f2)   & $\sqrt{3}Q\lambda y_c -3\left(1-\frac{\lambda^2y_c^2}{3}\right)$&$\frac{\lambda^2y_c^2}{2}-3$   &   $-\frac{\lambda^2y_c^2}{2}$      \\ 
  (g)  & $-3(1-Q^2)$&$-3(1-Q^2)$  &$\frac{3}{2}$\\ 
    (h1)   & $3\left(x_f^2-\frac{x_fy_f \lambda}{\sqrt{3}}\right)$& $-3(1-x_f)$ &$\frac{3}{2}$   \\
    (h2)   & $3\left(x_f^2-\frac{x_fy_f \lambda}{\sqrt{3}}\right)$& $-\frac{\sqrt{3}}{2}\lambda x_fy_f-3(1-x_f)$ &$-\frac{\sqrt{3}}{2}\lambda x_f y_f$  \\
 \hline
    &  $\mu_4$& $\mu_5$& Stability\\
\hline
  (a1)       & $\frac{3}{2}$&$-\frac{1}{2}$&saddle\\
  (a2)      & $\frac{3}{2}$&$-\frac{1}{2}$&saddle\\
   (a3)     & $\frac{3}{2}$&$-\frac{1}{2}$&saddle\\
  (a4)  & $\frac{3}{2}$& $-\frac{1}{2}$&saddle\\ 
  (b)    & 2& 1 &unstable\\
  (c)    & 2& 1&saddle\\
   (d)  & 4& 1&saddle\\
  (e)   &  $-3 $&$-2 $ &stable\\
 (f1)       &$\frac{\lambda^2y_c^2}{2}-3$ & $\frac{\lambda^2y_c^2}{2}-2$&saddle\\ 
 (f2)   &$\frac{\lambda^2y_c^2}{2}-3$ & $\frac{\lambda^2y_c^2}{2}-2$&stable for $\lambda<0$ or $Q=0$\\ 
  (g)  & $\frac{3}{2}$& $-\frac{1}{2}$&saddle\\ 
    (h1)   &$\frac{3}{2}\left(\frac{x_fy_f \lambda}{\sqrt{3}}-2\right)$& $\frac{3}{2}\left(\frac{x_fy_f \lambda}{\sqrt{3}}-\frac{4}{3}\right)$ &saddle \\
    (h2) &$\frac{3}{2}\left(\frac{x_fy_f \lambda}{\sqrt{3}}-2\right)$& $\frac{3}{2}\left(\frac{x_fy_f \lambda}{\sqrt{3}}-\frac{4}{3}\right)$  &stable\\
  
\noalign{\smallskip}\hline
\end{tabular}
\caption{\label{stability} Eigenvalues and stability of the fixed points for the tachyon field.}
\end{table} 

For the real tachyon field the eigenvalues $\mu_2$ and $\mu_3$ do not exist. The points (a1)--(a4) correspond to a matter-dominated solution, since $\Omega_m=1$ and $w_{eff}=0$.  They are saddle points because at least one eigenvalue has an opposite sign. The point (a4) is actually the point (a1), with $Q=1$. Points (b), (c) and (d) are radiation-dominated solutions, with $\Omega_r=1$ and $w_{eff}=1/3$. The difference between them is that (b) and (c) have $w_\phi=0$, while (d) has $w_\phi=-1$ and admits any value for $x_3$. They are unstable [(b)] or saddle points [(c) or (d)]. 

The point (e) is in principle a dark-energy-dominated solution with $\Omega_\phi=1$ and $w_{eff}=w_\phi=-1$, whose existence is restrict to $\lambda=0$. However, a careful analysis shows that the Jacobian matrix for this critical point has  zero eigenvector, thus the matrix cannot be diagonalized and the fixed point should not be considered. 

Points (f1) and (f2) are also a dark-energy-dominated solution ($\Omega_\phi=1$) whose equation of state depends on $\lambda$, which in turn can be either constant or zero. The cases with constant $\lambda$ are shown in  Table \ref{criticalpoints} and an accelerated expansion occurs for $\lambda^2<2/\sqrt{3}$.   For $\lambda=0$ we have $y_c=1$, $x_c=0$ and $w_\phi=-1$, thus we recover the point (e). All the points shown so far, but (e),  were also found in Refs. \cite{Gumjudpai:2005ry,Aguirregabiria:2004xd,Copeland:2004hq}. The  eigenvalues $\mu_2$, $\mu_4$ and $\mu_5$ of the fixed point (f1) and (f2) are always negative. For these points $\frac{\lambda^2y_c^2}{3}\leq 1$, then the first eigenvalue is also negative  if $Q=0$,  $\lambda<0$ or $\sqrt{3}Q\lambda y_c<3$. Therefore, the point (f2) describes a dark-energy-dominated universe and can lead to a late-time accelerated universe  if the requirement  $\mu_1<0$ is satisfied. On the other hand, (f1) is a saddle point. The effective equation of state depends only on $\lambda$, so the coupling $Q$ only changes the property of the fixed point.  

 The point (g) is also a saddle point with a matter-dominated solution, however, different from (a1)--(a4), the equation of state for the dark energy $w_\phi$ is no longer zero but depends on $Q$, leading to a universe with accelerated expansion for $Q^2<2/3$. For this point the coupling is restrict to values $0\leq Q^2\leq 1$.

\begin{figure}\center
\includegraphics[scale=0.55]{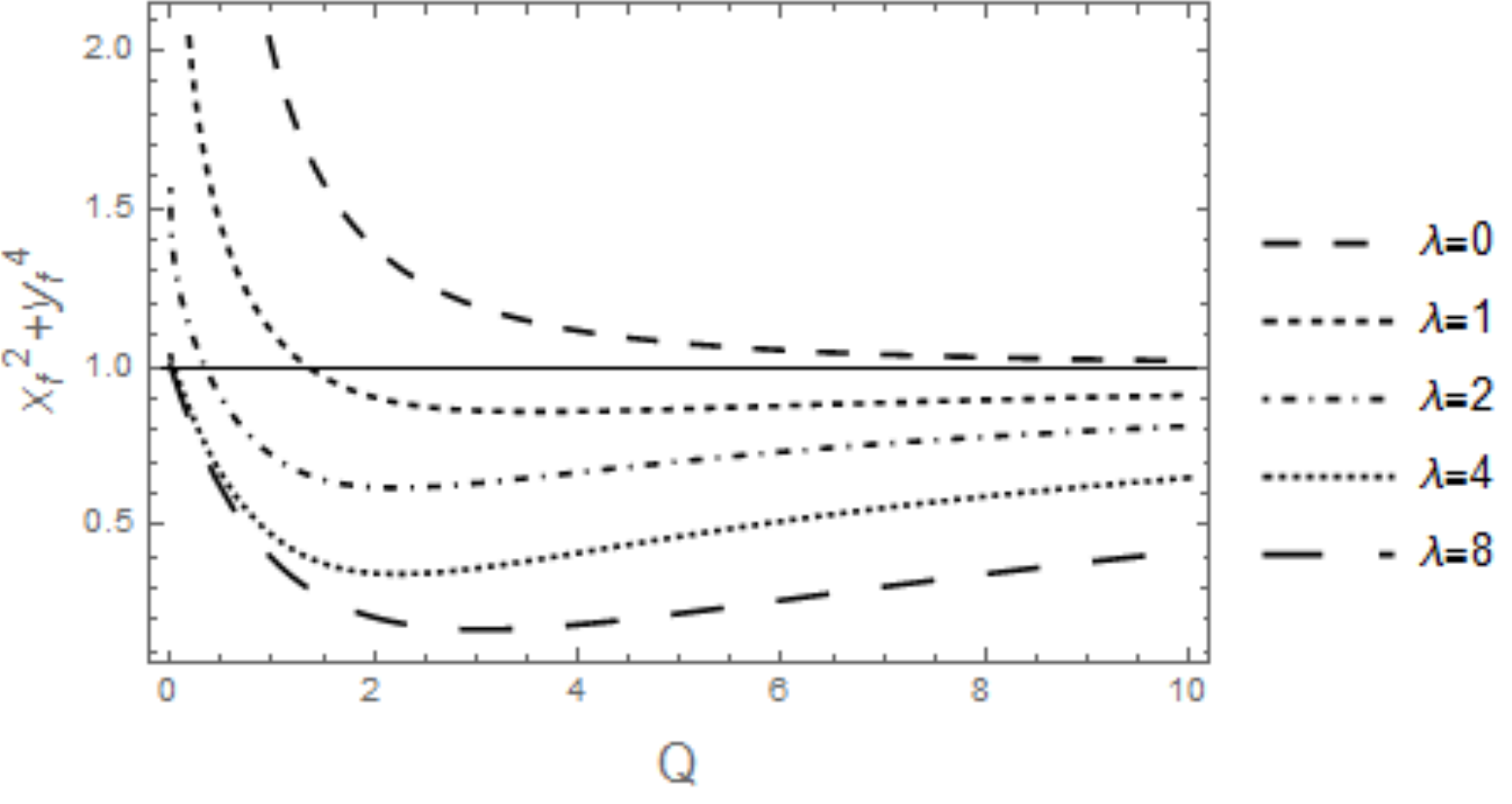}
\caption{Restriction $x_f^2+y_f^4$ as a function of the coupling $Q$, for $\lambda=0,1,2,4$ and $8$. The horizontal  line shows the upper bound for the restriction. Thus the fixed point does not exist for some values of $\lambda$ and $Q$. \label{f1}}
\end{figure}

\begin{figure}\center
\includegraphics[scale=0.55]{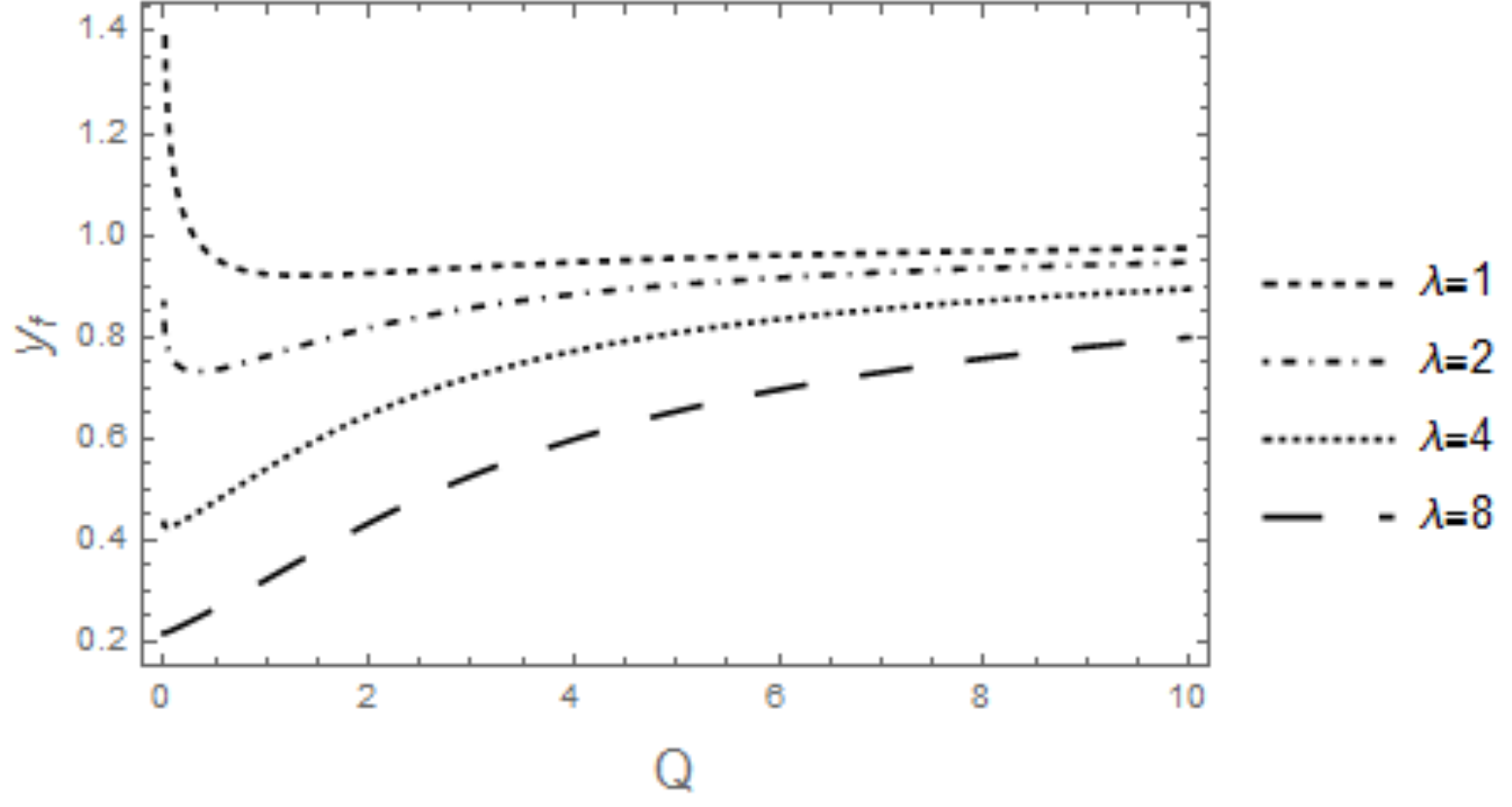}
\caption{Fixed point $y_f$ as a function of the coupling $Q$, for $\lambda=1,2,4$ and $8$. \label{f2}}
\end{figure}

The last fixed points (h1)  and (h2) require more attention. They are valid solutions for $x_f\neq 0$,\footnote{The case for $x_f=0$ is the fixed point (e).} for $Q\neq 0$ and for constant $\lambda$, and its behavior depends on $Q$. In order to have $x^2_f\leq1$, we must have $Q>0$ for the case with plus sign in $x_f$ (\ref{xf}), while we have $Q<0$ for the minus sign case. We restrict our attention for the plus sign case. When $Q\rightarrow |\infty|$,\footnote{$Q\rightarrow +\infty$ for the plus sign in $x_f$ and $Q\rightarrow -\infty$ for the minus sign in $x_f$.} $x_f\rightarrow 0$ and $y_f\rightarrow 1$, in agreement with the restriction (\ref{restriction}). Furthermore, the fixed points exist for some values of $\lambda>0$ and $Q$, due to Eq. (\ref{restriction}). In Fig. \ref{f1}, we show the restriction  $x_f^2+y_f^4$ as a function of $Q$, for some values of $\lambda$. As we see, for $\lambda=0$ we have $y_t=1$ only when $x_f\rightarrow 0$, but this value is never reached since the fixed points (h1) and (h2) are valid for $x_f\neq 0$. Thus for $\lambda\leq 0$ the restriction is not satisfied. From the figure we also notice that there is a lower bound for $Q$, for which the points (h1) and (h2) start existing. In Fig. \ref{f2} we plot $y_f$ as a function of $Q$, for the all the values of $\lambda$ but zero, shown in Fig. \ref{f1}.\footnote{We used the plus sign in $x_f$ (\ref{xf}) for Figs. \ref{f1} and \ref{f2}, but the minus sign shows similar behavior, with $Q<0$ due to Eq. (\ref{restriction}),  satisfied for  $\lambda<0$. The graphics for this case are reflections in the ordinate axis. } 

As in the case of quintessence and phantom, the fixed points that have $x_3=0$ [(a1), (a2), (a4), (b), (f1), (g) and (h1)] indicate that  $\phi\rightarrow \infty$ and therefore  $x_2\equiv  \phi\dot{\theta}\rightarrow \infty$ as well. However, this limit is in contradiction to Table \ref{criticalpoints} for $x_2$ showing that these seven critical points are not physically acceptable.

All fixed points reproduce the previous results in the literature \cite{Gumjudpai:2005ry,Aguirregabiria:2004xd,Copeland:2004hq} and they are generalizations of those analyses, with same stability behavior for the critical points. This indicates that the degree of freedom due to the complex scalar has no effect on the stability and on  the evolution of the system of equations, when compared with the case of real scalar field.

\begin{figure}\center
\includegraphics[scale=0.55]{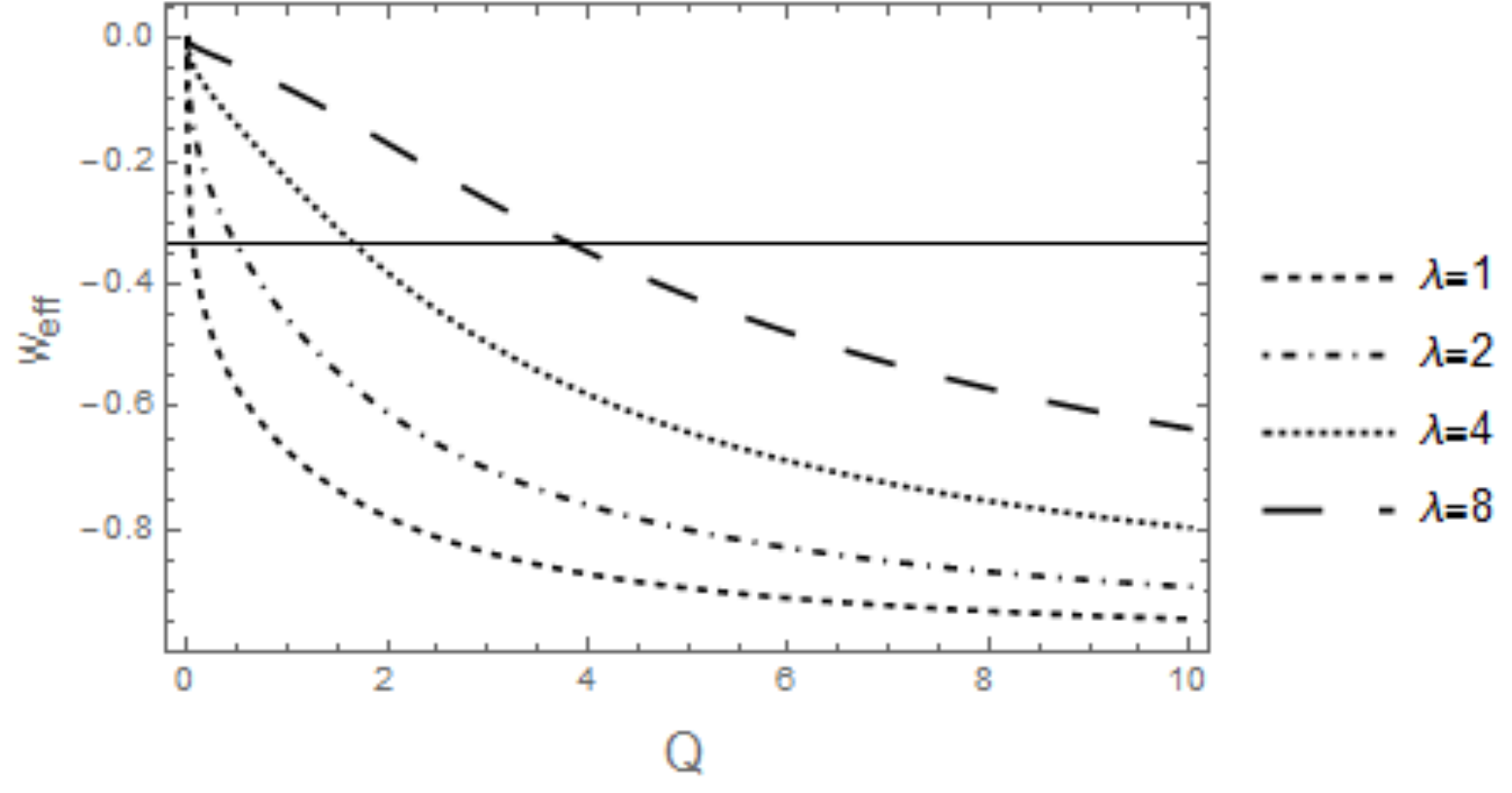}
\caption{Effective equation of state for the fixed point (h1) or (h2) as function of $Q$ for  $\lambda=1,2,4$ and $8$. The horizontal line ($w_{eff}=-1/3$) shows the border of acceleration and deceleration. \label{fig:weff}}
\end{figure}

\begin{figure}\center
\includegraphics[scale=0.55]{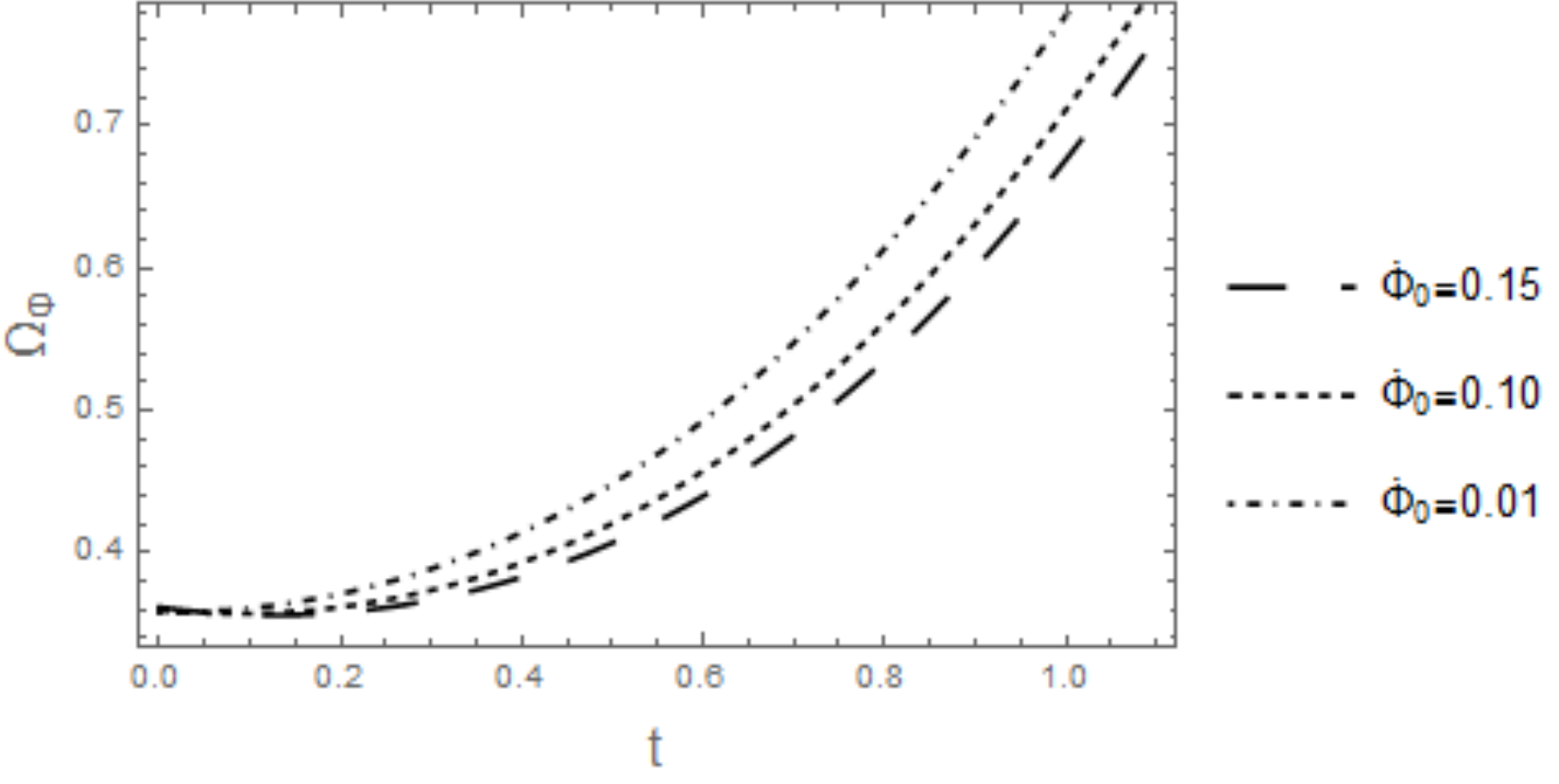}
\caption{Density parameter of the dark energy $\Omega_\phi$ as a function of $Q$, for $\lambda=1$, $2$ and $4$. We show in the horizontal line the limit $\Omega_\phi=1$, which does not allow a range of values of $Q$. \label{fig:omegade}}
\end{figure}

In Figs. \ref{fig:weff} and \ref{fig:omegade} we show the effective equation of state $w_{eff}$ and the density parameter for dark energy $\Omega_\phi$, respectively, as a function of $Q$ for the point (h2) (see Table \ref{criticalpoints}). Notice that depending on the value of $Q$, the universe does not exhibit an accelerated expansion for this fixed point, as we see from the horizontal line in Fig. \ref{fig:weff}. The values of $Q$, at which they reach the upper limits of Eq. (\ref{restriction}) and $\Omega_\phi$, are, of course, the same. This feature  can also be checked from  Figs. \ref{f1} and \ref{fig:omegade}. This fixed point is a tachyonic-dominated universe only for a range of values of $\lambda$ with specific $Q$. We have $\Omega_\phi=1$ for $\lambda=1$ and $Q\approx 1.35$, and for $\lambda=2$ and $Q\approx 0.35$, for instance. The effective equation of state for these two values  is $w_{eff}=-0.72$ and $-0.29$, respectively. Thus the former $w_{eff}$ implies an accelerated expansion, while the latter $w_{eff}$ does not.

\begin{figure}
\centering
	\includegraphics[scale=0.5]{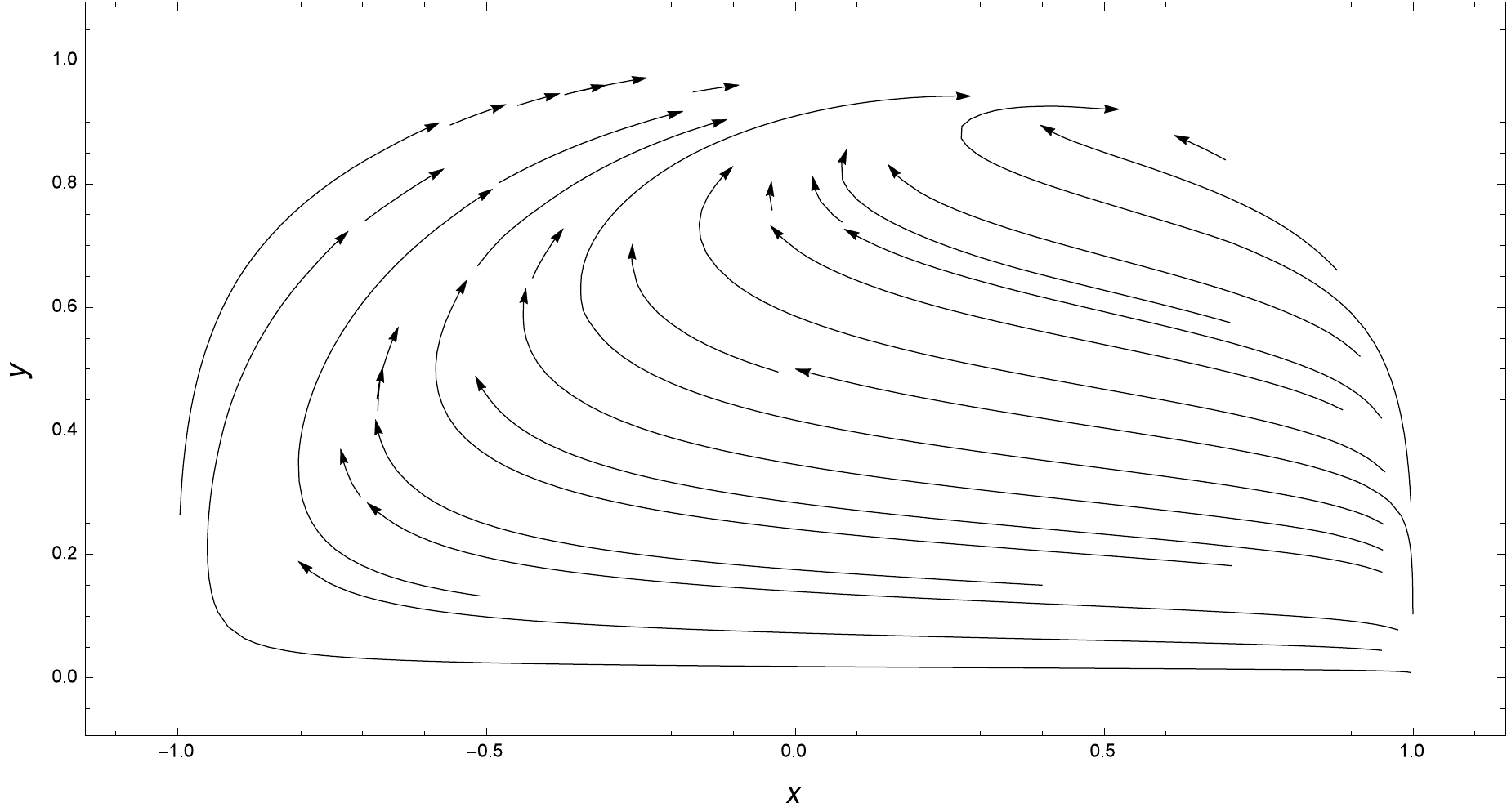}
	\caption{Phase plane $x_f--y_f$ for $\lambda=1$ and $Q=1.35$. The tachyonic-dominated solution (h2) is a stable point at $x\approx 0.53$ and $y\approx 0.92$.}
		\label{fig:PhasePlane}
\end{figure}

The fixed point (h2) is stable. The eigenvalues $\mu_4$ and $\mu_5$ are always negative because $w_{eff}$ is between $0$ and $-1$. The first eigenvalue is also negative because $\Omega_\phi\leq 1$, thus $x_f^2-1\leq\frac{x_fy_f \lambda}{\sqrt{3}}-1$, therefore $\frac{x_fy_f\lambda}{\sqrt{3}}\geq x_f^2$, since $x_f$ is always positive. Therefore, the point (h2) can lead to a late-time accelerated universe, depending on the value of $\lambda$ and $Q$. Its behavior is shown in Fig. \ref{fig:PhasePlane}, for $\lambda=1$ and $Q=1.35$.

\subsubsection{Summary}\label{conclu1}

The critical points shown in the tachyonic case describe the three phases of the universe: the radiation-dominated era, the matter-dominated era, and the present dark-energy-dominated universe. The matter-dominated universe can be described by the saddle point (a3). There are two points that  can represent the radiation-dominated era: (c) and (d). The two  are saddle points, with the additional difference that the point (d) has an  equation of state for dark energy equal to $-1$. 

A tachyonic-dominated universe is described by  the points  (f2) and (h2).  The point (f2) can be stable only if the coupling is zero or $\lambda<0$. The last fixed point (h2)  is stable and can describe an accelerated universe depending on the value of $\lambda$ and $Q$. 

%Despite of the similarity of some fixed points ((h1) and (h2) for intance), some of them does not reproduce exactly the case of the real tachyon field, because they have different stabilities. Thus, the dynamical analysis of the complex tachyon field provides additional fixed points, in comparison with the real field.

From all the critical points,  the cosmological transition radiation $\rightarrow$ matter $\rightarrow$ dark energy is achieved considering the following sequence of fixed points:  (c) or (d) $\rightarrow$ (a3) $\rightarrow$ (f2) [$\lambda$ dependent] or (h2) [$Q$ and $\lambda$ dependent]. Although the sequence is viable, the form of the potential dictates whether the fixed points are allowed or not.  Among several possibilities in the literature, the potential $V(\phi)\propto \phi^{-n}$, for instance, leads to a dynamically changing $\lambda$ (either if $\lambda\rightarrow 0$ for $0<n<2$, or $\lambda\rightarrow \infty$ for  $n>2$) \cite{Abramo:2003cp}. A dynamically changing $\lambda$ is allowed for the fixed points (a3), (c) and (d). On the other hand, points (f2) and (h2) require a constant $\lambda$, implying $V(\phi)\propto \phi^{-2}$ \cite{Gumjudpai:2005ry,micheletti2009,Padmanabhan:2002cp,Bagla:2002yn}.

 \subsection{Overall summary}\label{summary}
 
 In this section we presented the coupled dark energy using a complex scalar field, in the light of the dynamical system theory. There were analyzed three possibilities: quintessence, phantom and tachyon field.  All three possibilities are known  in the literature for the real field \cite{Amendola:1999er,Gumjudpai:2005ry,ChenPhantom}, and for uncoupled and complex quintessence field \cite{Zhai2005}. Thus a natural question that arises is  how a complex scalar field changes the previous results  and if there are new fixed points due to the complex field. Although some equations for the dimensionless variables are trivial extensions of the real field case (e.g. the equation of state for the scalar field), the differential equations were generalized. All fixed points found here are in agreement with the previous results, with no new fixed points, however, there are some crucial differences. For the quintessence and the phantom there is a contradiction between the fixed points $x_2$ and $x_3$ when the latter is zero.  This situation occurs for almost all fixed points and the only two exceptions are unstable points that  represent respectively the radiation and matter era, so the dark-energy-dominated universe is absent. Therefore the extra degree of freedom spoils the results known in the case of real scalar field. For the tachyon field all the critical points are also similar to the real field case, with same stabilities.  Therefore, the extra degree of freedom due to the complex tachyon field plays no role on the stability of the critical points. Although the results presented here enlarge the previous results found in the literature, with the generalization of the equations of motion,  the dynamical system theory does not provide further information in what is already known for the case of real scalar field, letting open the possibility of studying complex scalar fields by other ways of analysis.

\section{Dynamical analysis for a vector-like dark energy}\label{sec:vecDE}

\subsection{Vector-like dark energy dynamics}\label{VDE}

The  Lagrangian for three identical copies of an abelian field (called cosmic triad in  \cite{ArmendarizPicon:2004pm}), here uncoupled to matter, is given by
\begin{equation}\label{LVDE}
 \mathcal{L}_A=-\sqrt{-g}\sum_{a=1}^3\left(\frac{1}{4}F^{a \mu\nu}F_{\mu\nu}^a+V(A^{a2})\right)\quad,
\end{equation} 

\noindent where $F_{\mu\nu}^a=\partial_\mu A^a_\nu -\partial_\nu A_\mu^a$ and $V(A^2)$ is the potential for the vector field, which breaks gauge invariance, with $A^{a2}\equiv A_\mu^a A^{a\mu}$. The energy-momentum tensor of the field is obtained varying the Lagrangian (\ref{LVDE}) with respect to the metric and it is $T_{\mu\nu}^A=\sum_{a=1}^3T^{a}_{\mu\nu}$, where
\begin{equation}
T^a_{\mu\nu}=\left[F_{\mu\rho}^aF_\nu^{a\rho}+2\frac{dV}{dA^{a2}}A^a_\mu A^a_\nu-g_{\mu\nu}\left(\frac{1}{4}F_{\rho\sigma}^{a\rho\sigma}+V(A^2) \right)\right]\quad.
\label{eq:Tmunu}
\end{equation}

\noindent Varying Eq. (\ref{LVDE}) with respect to the fields $A_\mu^a$ gives the equations of motion 
 \begin{equation}
\partial_\mu\left(\sqrt{-g}F^{a\mu\nu}\right)=2\sqrt{-g}V'A^{a\nu}\quad, 
\label{eq:eom}
\end{equation}
\noindent  where from now on we use $V'\equiv\frac{dV}{dA^{a2}} $. 

	In an expanding universe, with FLRW metric and scale factor $a$, each one of the three vectors should be along a coordinate axis with same magnitude.  An ansatz for the $i$ components  of the vector $A^a_\mu$ compatible with homogeneity and isotropy is
\begin{equation}
A_i^a=\delta^a_i A(t)\quad,
\label{eq:ansatzA}
\end{equation}

\noindent where a scalar product with an unit vector is implicit. From Eq. (\ref{eq:eom}) the component $A_0^a$ is zero and using Eqs.  (\ref{eq:ansatzA}) into (\ref{eq:eom}) the equation of motion becomes
\begin{equation}\label{eqmotionvec}
  \ddot{A}+H\dot{A}+2V'A=0\quad.
\end{equation} 

The pressure and energy density for the cosmic triad is obtained from Eq. (\ref{eq:Tmunu})
\begin{equation}\label{rhoA}
\rho_A=\frac{3\dot{A}^2}{2a^2}+3V\quad,
\end{equation} 

 \begin{equation}\label{pA}
 p_A=\frac{\dot{A}^2}{2a^2}-3V+2V'\frac{A^2}{a^2}\quad.
\end{equation} 

\noindent With this ansatz\footnote{In \cite{ArmendarizPicon:2004pm} the author used a comoving vector ansatz: $A_\mu^a=\delta_\mu^a A(t)\cdot a$. This choice leads, of course, to a different equation of motion, energy density and pressure. However, the effect due to the scale factor that here appears in the denominator of $\rho_A$ and $p_A$, for instance, appears as a Hubble friction term ($H\dot{A}$) in the same expressions.} the potential depends now on $V(3A^2/a^2)$ and the prime is the derivative with respect to $3A^2/a^2$. We assume that the potential is given by $V=V_0 e^{-\frac{3\lambda A^2}{a^2}}$, where $V_0$ is a constant. With this form the quantity $-V'/V$ will be constant, as we will see soon. Thus, for the comoving vector $A_{ic}^a=A^a_i\cdot a$ (as used in \cite{ArmendarizPicon:2004pm}), the potential does not have an explicit dependence on the scale factor. If the cosmic triad were massless, we would have $\dot{A}\propto a^{-1}$, thus $\rho_A \propto a^{-4}$, as it should be for relativistic matter.

Inspired by the quintessence case \cite{Wetterich:1994bg,Amendola:1999er}, where the coupling is $Q\rho_m \dot{\phi}$, we assume the phenomenological interaction $\mathcal{Q}=3 Q \rho_m\dot{A}/a$ for the vector dark energy, where $Q$ is a positive constant. The coupling has this form in order to the right-hand side of the Proca-like equation (\ref{eqmotionvec}) to be no longer zero but to equal $Q\rho_m$. The case with negative $Q$ is the same of the case with $Q>0$ but with negative fixed point $x$, defined soon.  In the presence of a barotropic fluid, the Friedmann equations are
\begin{align}\label{eq:1stFEmatterS3V}
  H^2&=\frac{1}{3}\left(\frac{3\dot{A}^2}{2a^2}+3V+ \rho_m\right)\quad,\\
  \dot{H}&=-\frac{1}{2}\left(\frac{2\dot{A}^2}{a^2}+2V'\frac{A^2}{a^2}+(1+w_m)\rho_m\right)\quad.
\end{align}

We now proceed to the dynamical analysis of the system.

\subsubsection{Autonomous system} 

The dimensionless variables are defined as

\begin{eqnarray}\label{eq:dimensionlessXYS3V}
 x\equiv  \frac{\dot{A}}{\sqrt{2}Ha},  \quad y\equiv \frac{\sqrt{V(\phi)}}{H},\quad
   z\equiv \frac{A}{a}, \quad \lambda\equiv -\frac{V'}{V}, \quad \Gamma\equiv \frac{VV''}{V'^2}\quad.
\end{eqnarray}

The dark energy density parameter is written in terms of these new variables as
\begin{equation}\label{eq:densityparameterXYS3V}
 \Omega_A \equiv \frac{\rho_A}{3H^2} =x^2+y^2\quad,
 \end{equation}

\noindent so that Eq. (\ref{eq:1stFEmatterS3V}) can be written as 
\begin{equation}\label{eq:SomaOmegasS3V}
\Omega_A+\Omega_m=1\quad,
\end{equation}

\noindent where the density parameter of the barotropic fluid is defined by $\Omega_m=\rho_m/(3H^2)$. From Eqs. (\ref{eq:densityparameterXYS3V}) and (\ref{eq:SomaOmegasS3V}) $x$ and $y$ are restricted in the phase plane by the relation
\begin{equation}\label{restrictionS3V}
0\leq  x^2+y^2\leq 1\quad,
 \end{equation}
 
\noindent due to $0\leq \Omega_A\leq 1$. 
The equation of state $w_A=p_A/\rho_A$  becomes
\begin{equation}\label{eq:equationStateXYS3V}
 w_A =\frac{ x^2-3y^2-2\lambda z^2}{3x^2+3y^2}\quad.
\end{equation}

\noindent Depending on the value of $\lambda$ the equation of state can be less than minus one. 

The total effective equation of state is
\begin{eqnarray}\label{eq:weffS3V}
 w_{eff} = \frac{p_A+p_m}{\rho_A+\rho_m}=&w_m+x^2(\frac{1}{3}-w_m)-y^2(1+w_m)-\frac{2}{3}\lambda y^2z^2\quad,
\end{eqnarray}

\noindent with an accelerated expansion for  $w_{eff} < -1/3$.  The dynamical system for the variables  $x$, $y$, $z$  and $\lambda$ are

\begin{align}\label{dynsystemS3V}
x'&=-x+\sqrt{2} y^2z\lambda-\frac{3Q}{\sqrt{2}} (1-x^2-y^2)-x\left[yz^2\lambda-x^2+y^2-\frac{1+3w_m}{2}(1-x^2-y^2)\right],\\
y'&=-3 yz\lambda(\sqrt{2}x-z)-y\left[y^2z^2\lambda-2x^2-\frac{3}{2}(1+w_m)(1-x^2-y^2)\right]\quad,\\
z'&=2x-z\quad,\\
\lambda'&=-6\lambda^2 z\left(\Gamma-1\right)(\sqrt{2}x-z)\quad,
\end{align}

\noindent where the prime is the derivative with respect to $N$.

\subsubsection{Critical points}

 The fixed points of the system are obtained by setting $x'=0$,  $y'=0$, $z'=0$ and $\lambda'=0$ in Eq. (\ref{dynsystemS3V}). When $\Gamma=1$, $\lambda$ is constant the potential is $V(A^2)=V_0e^{\frac{-3\lambda A^2}{a^2}}$.\footnote{The equation for $\lambda$ is also equal zero when $z=0$ or $\lambda=0$, so that  $\lambda$ should not necessarily be constant, for the fixed point with this value of $z$. However, for the case of dynamical $\lambda$, the correspondent eigenvalue is equal zero, indicating that the  fixed points are not hyperbolic.} Different from the scalar field case, where $V=V_0 e^{-\lambda\phi}$, the exponent of the potential also depends on the scale factor $a$. The fixed points are shown in Table \ref{criticalpointsS3V} with the eigenvalues of the Jacobian matrix. Notice that $y$ cannot be negative.

 \begin{table*}\centering\footnotesize
\begin{tabular}{lllllll}
\hline\noalign{\smallskip}
Point  & $x$& $y$  &$z$& $w_A$ & $\Omega_A$& $w_{eff}$\\
\noalign{\smallskip}\hline\noalign{\smallskip}
(a)  &$\pm 1$   &$0 $  &$\pm 2$  &$\frac{1}{3}(1-8\lambda)$&$1$ &$\frac{1}{3}$ \\
								(b) &$\frac{\sqrt{2}Q}{w_m-1/3} $ & $0$& $2x$ &$ \frac{1}{3}(1-8\lambda)$& $\frac{2Q^2}{(w_m-1/3)^2}$ &$6Q^2$\\
		(c) &$ 0$ &$1$ &$0$ & $-1$ &$1$ &$-1$\\
   \noalign{\smallskip}\hline
	 &  $\mu_1$ & $\mu_2$ & $\mu_3$ \\
\noalign{\smallskip}\hline\noalign{\smallskip}
(a) & $-1$ & $1\pm 3(\sqrt{2}Q-w_m)$&	$2+6(2-\sqrt{2})\lambda$   \\
								(b) &$-1$ &$-\frac{1}{2}+9Q^2$ &$ \frac{3}{2}+Q^2(9+108(2-\sqrt{2})\lambda)$\\
		(c)  &$-3(1+w_m)$& 	$-\frac{3}{2}(1+\sqrt{1+8\sqrt{2}\lambda})$&	$-\frac{3}{2}(1-\sqrt{1+8\sqrt{2}\lambda})$\\
   \noalign{\smallskip}\hline
\end{tabular}
\caption{\label{criticalpointsS3V} Critical points ($x$, $y$ and $z$) of  Eq. (\ref{dynsystemS3V}) for the vector-like dark energy.  The table shows the correspondent equation of state for the dark energy (\ref{eq:equationStateXYS3V}), the effective equation of state (\ref{eq:weffS3V}), the density parameter for dark energy (\ref{eq:densityparameterXYS3V}) and the eigenvalues of the Jacobian matrix.}
\end{table*}

The point (a) corresponds to a radiation solution, once $w_{eff}=1/3$. It can be a saddle or a stable point, depending on the value of $Q$ and $\lambda$. However, the universe is dominated by the cosmic triad, as indicated by $\Omega_A=1$, and therefore the fixed point does not describe a radiation-dominated universe, since $\Omega_m=0$. The point (b) is valid only for $w_m\neq 1/3$ and it is a saddle point,  since two eigenvalues are negative and one is positive. Since $y=0$ for this critical point, $x^2$ should be less than or equal to one (since $\Omega_A\leq 1$), so the coupling should be  $Q\leq 1/\sqrt{2}$. However, this critical point can describe a matter-dominated universe only if $Q=0$ or sufficiently small $Q\ll 1$, so that  $w_{eff} \approx 0$, as so for $\Omega_A$. The last fixed point (c) is an attractor and describes a dark-energy dominated universe ($\Omega_A=1$) that leads to an accelerated expansion of the universe, since $w_A=w_{eff}=-1$. It is a stable spiral if $\lambda<-1/(8\sqrt{2})$, otherwise it is a saddle point. The potential for this condition for $\lambda$ is $V=V_0 e^{3|\lambda|z^2}$ and it behaves as the cosmological constant at the fixed point, since $z\equiv A/a=0$ for (c). Once the coupling is constant and sufficiently small (to the fixed point (b) describe the matter-dominated universe), it has the same value, of course, for the point (c). 

\begin{figure}\centering
\includegraphics[scale=0.6]{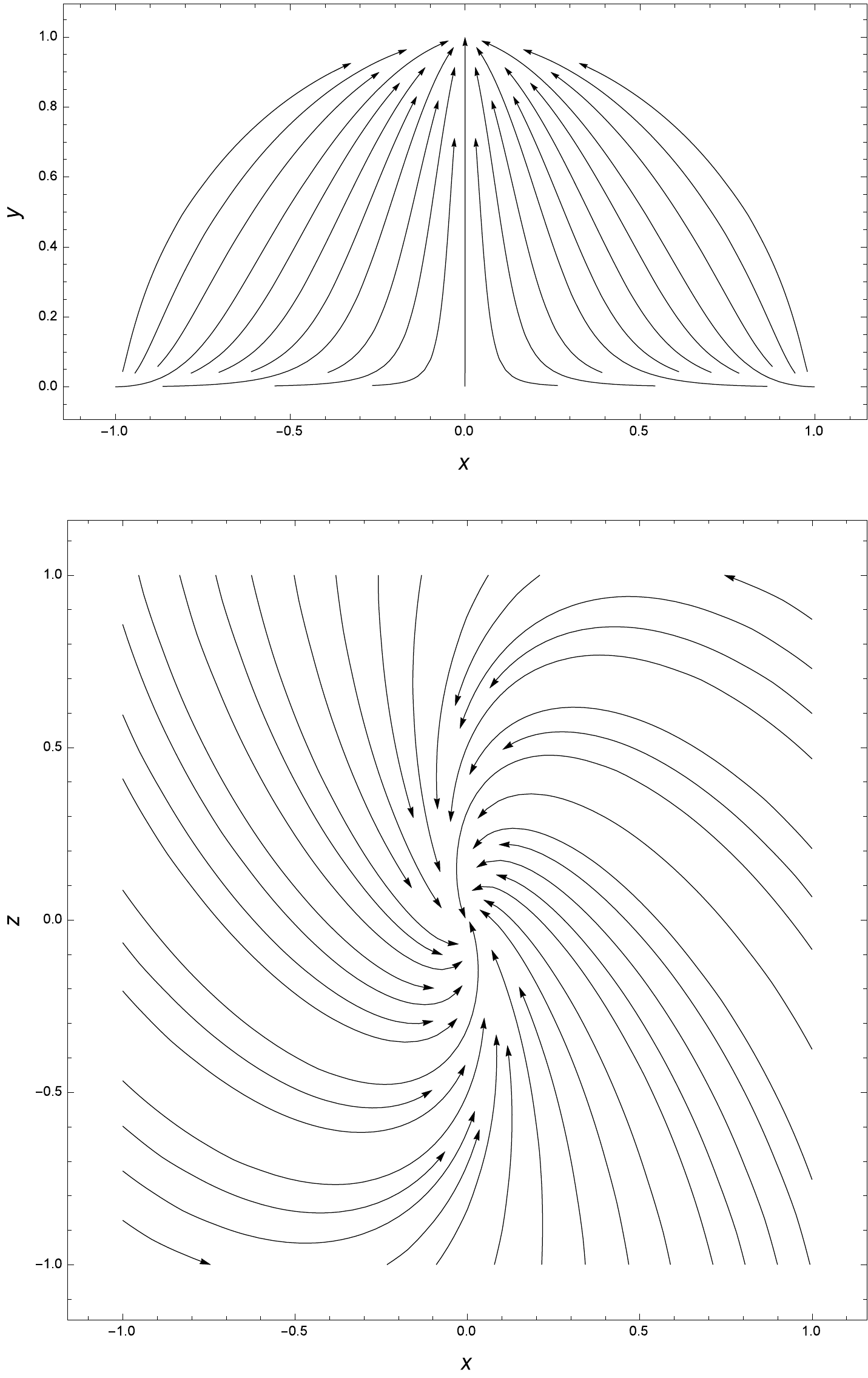}%
\caption{Phase portrait of the system, with $Q=0$ and $\lambda=-0.3$. All trajectories converge to the attractor (c) at $x=0$, $y=1$ and $z=0$, which is a stable spiral that describes the dark-energy dominated universe. The top panel shows the slice $z=0$, while the bottom panel shows the phase plane at $y=1$. }%
\label{phaseportV}%
\end{figure}

\begin{figure}\centering
\includegraphics[scale=0.7]{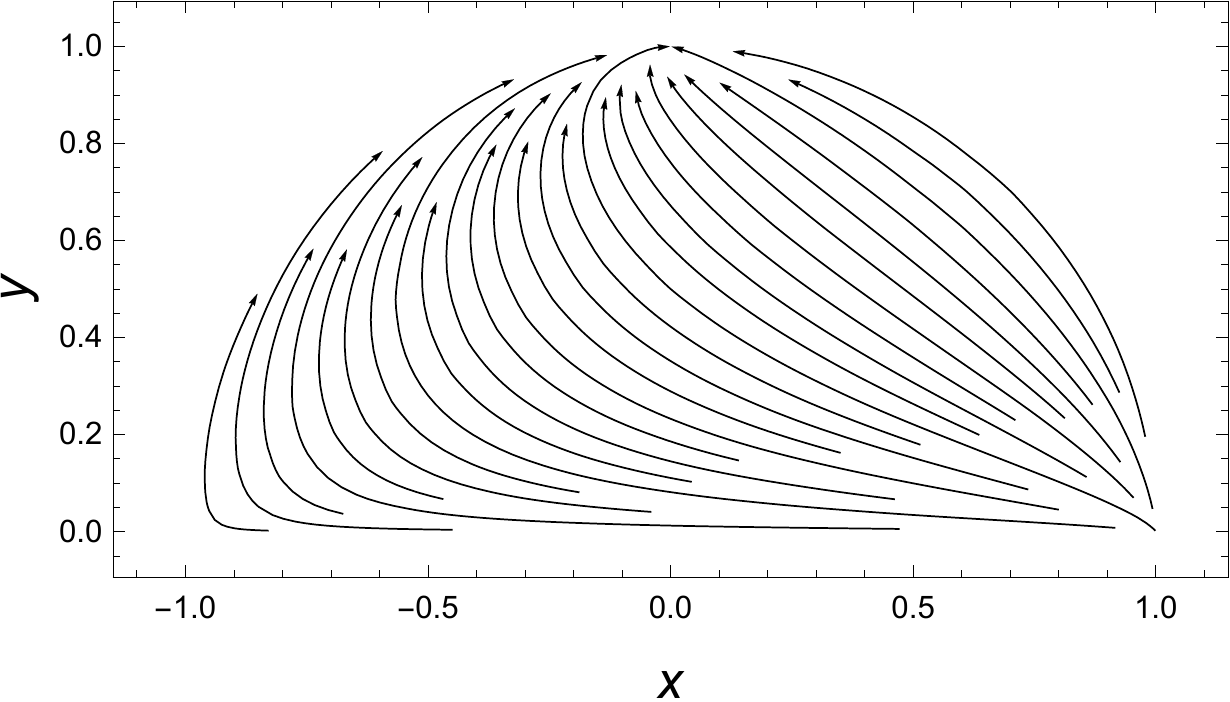}%
\caption{Phase portrait of the system, with $Q=1/\sqrt{6}$ and $\lambda=-0.3$. All trajectories converge to the attractor (c) at $x=0$, $y=1$ and $z=0$, which is a stable spiral that describes the dark-energy dominated universe. The panel shows only the slice $z=0$ because the phase plane for $y=1$ is similar to that one showed in Fig. \ref{phaseportV}. }%
\label{phaseport2V}%
\end{figure}

We show the phase portrait of the system in Figs. \ref{phaseportV} ($Q=0$) and \ref{phaseport2V} ($Q=1/\sqrt{6}$). The latter case is shown just to illustrate how the interaction affects the phase portrait, although we expect a very small $Q$, as discussed for the fixed point (b). We see that all trajectories converge to the attractor point.

\begin{figure}\centering
\includegraphics[scale=0.8]{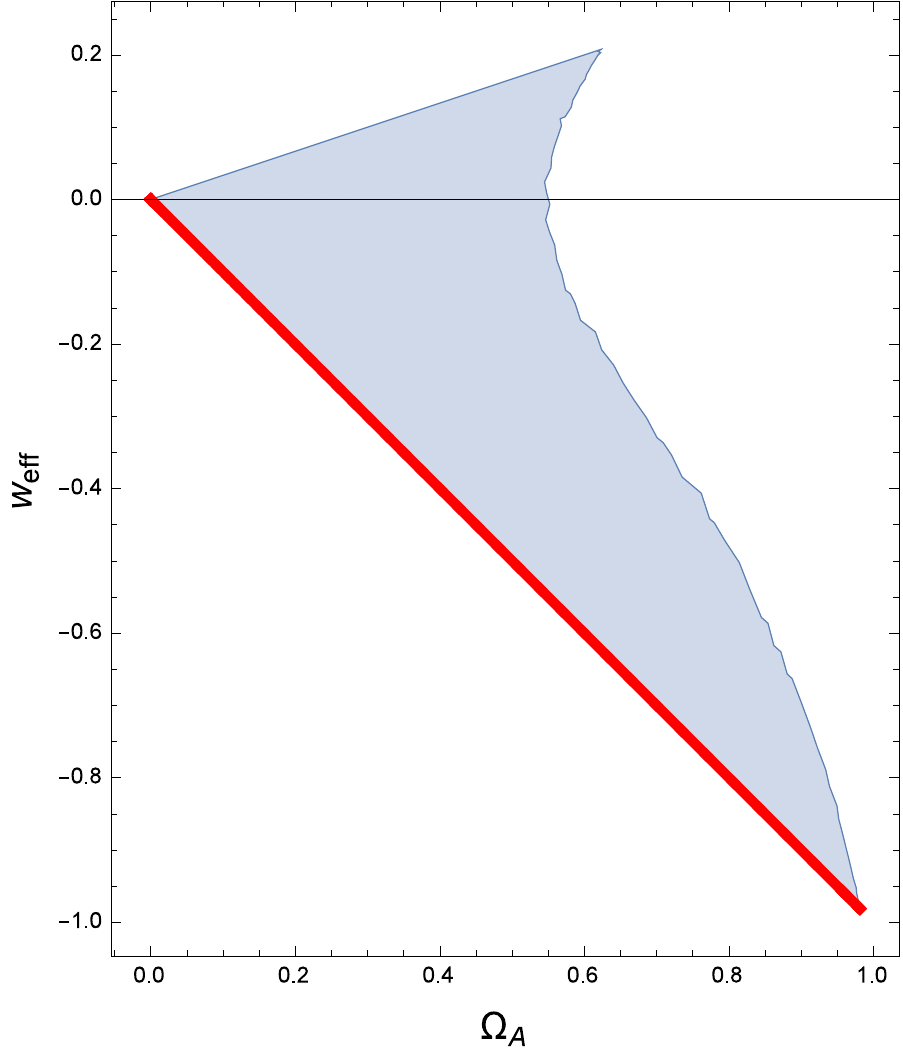}%
\caption{Effective equation of state  $w_{eff}$ (\ref{eq:weffS3V}) as a function of the dark energy density parameter $\Omega_A$ (\ref{eq:densityparameterXYS3V}). This parametric plot is independent of $Q$, once both $w_{eff}$ and $\Omega_A$  have no explicit dependence on the interaction. The blue shaded region represents the allowed values of  $w_{eff}$ and $\Omega_A$. We used $w_m=0$ since this is the only allowed value for the fixed point (b). The red line shows the transition from the fixed point (b) ($\Omega_A=0$) to the fixed point (c)  ($\Omega_A=1$). }%
\label{parametricV}%
\end{figure}

In Fig. \ref{parametricV} we show the effective equation of state  $w_{eff}$ (\ref{eq:weffS3V}) as a function of the dark energy density parameter $\Omega_A$ (\ref{eq:densityparameterXYS3V}), where the blue shaded region represents the allowed values of  $w_{eff}$ and $\Omega_A$. The red line shows the transition from the fixed point (b) ($\Omega_A=0$) to the fixed point (c)  ($\Omega_A=1$).   

 \subsection{Summary}\label{conclu2}
 
 In this section we used the dynamical system theory to investigate if a vector-like dark energy, similar to \cite{ArmendarizPicon:2004pm}, in the presence of a barotropic fluid can lead to the three cosmological eras, namely, radiation, matter and dark energy. The analysis was generalized for the case of coupled dark energy, with a phenomenological interaction $3Q\dot{A}\rho_m/a$. There are fixed points that successfully describe the matter-dominated and the dark-energy-dominated universe. Only the radiation era was not cosmologically viable, however, if one is interested in the last two periods of the evolution of the universe, the dynamical system theory provides a good	tool to analyze asymptotic states of such cosmological models.        

%% file: sugra.tex
%%%%%%%%%%%%%%%%%%%%%%%%%%%%%%%%%%%%%%%%%%%%%%%%%%%%%%%%%%%%%%%%%%%%%%%%%%%%%%%%%%%%
\chapter{Supergravity-based models}\label{sugrachapter}

 In this chapter we use supergravity to build dark energy models. In Sect. \ref{chapter:sugra} we review the basics of minimal supergravity, including the mechanism of spontaneous (local) supersymmetry breaking. We follow the references \cite{wess1992,Freedman2012,MoroiThesis} in the first section. The new results presented in Sects. \ref{modelSugra1} and \ref{HDEsugra} are based on \cite{Landim:2015hqa, Landim:2015upa}.

\section{Basics of Minimal Supergravity}\label{chapter:sugra}

%%%%%%%%%%%%%%%%%%%%%%%%%%%%%%%%%%%%%%%%%%%%%%%%%%%%%%%%%%%%%%%%%%%%%%%%%%%%%%%%%%%%
\subsection{From Wess-Zumino to minimal Supergravity model}\label{section:msugra}

Supergravity is the supersymmetric theory of gravity, understood also as the theory with local supersymmetry. In order to see this latter point, which is deeply related with the former one, we start with the Wess-Zumino model. A complex scalar field $\phi$ and its supersymmetric fermionic field $\chi$ are described by the following Lagrangian \cite{wess1992} in on-shell representation\footnote{The on-shell representation is written for the sake of simplicity, in the off-shell case it is necessary to introduce auxiliary fields, in order for the bosonic degrees of freedom to match the fermionic ones.}   
\begin{equation}\label{Lwz}
\mathcal{L}_{WZ}=\partial_{\mu}\phi\partial^{\mu}\phi^*+i\partial_\mu\bar{\chi}\bar{\sigma}^\mu\chi\quad,
\end{equation}

\noindent which is invariant (up to a total derivative) under the supersymmetric transformations
\begin{subequations}\label{susytransf}
\begin{align}
\delta\phi&= \sqrt{2}\xi\chi\quad,\\
\delta\chi&=i\sqrt{2}\sigma^\mu\bar{\xi}\partial_\mu\phi\quad, \end{align}
\end{subequations}

\noindent where $\xi$ is a grassmannian parameter associated with the global SUSY transformations. The generator of these transformations (\ref{susytransf}) obeys the relations 

\begin{subequations}\label{commutators}
\begin{eqnarray}
\{Q_\alpha\!^A,\bar{Q}_{\dot{\beta}B}\}=2\sigma_{\alpha\dot{\beta}}^\mu P_\mu\delta^A\!_B\end{eqnarray}
\begin{eqnarray}
\{Q_\alpha\!^A,Q_\beta\!^B\}=\{\bar{Q}_{\dot{\alpha}A},\bar{Q}_{\dot{\beta}B}\}=0\end{eqnarray}
\begin{eqnarray}
[P_\mu,Q_\alpha\!^A]=[P_\mu,\bar{Q}_{\dot{\alpha}A}]=0\end{eqnarray}
\begin{eqnarray}
[P_\mu,P_\nu]=0\label{4pcommut}\end{eqnarray}
\end{subequations}

\noindent where Eq. (\ref{4pcommut}) is indicated to complete the SUSY algebra. Capital indices ($A, B, \ldots$) refer to an internal space and run from 1 to $\mathcal{N}>1$, but here we consider $\mathcal{N}=1$. Indices with the first part of the Greek symbols ($\alpha$, $\beta$, $\ldots$, $\dot{\alpha}$, $\dot{\beta}$, $\ldots$) run from one to two and denote two-component Weyl spinors, with the dotted indices for the complex conjugate components. 

When the SUSY tranformations (\ref{susytransf}) are promoted to be local ($\xi=\xi(x)$), variation of Eq.  (\ref{Lwz}) yields the extra terms
\begin{align}\label{Lwzlocal}
\delta\mathcal{L}_{WZ\text{local}}&=\sqrt{2}[(\partial_{\mu}\xi)\sigma^\nu\bar{\sigma}^\mu\chi\partial_\nu\phi^*+\bar{\chi}\bar{\sigma}^\mu\sigma^\nu(\partial_\mu\bar{\xi})\partial_\nu\phi] + \text{total deriv.}\notag\\
&\equiv (\partial_\mu\xi)J^\mu+h.c.+\text{total deriv.}\quad,
\end{align}

\noindent which are those ones that survive after using Euler-Lagrange equations, where
\begin{equation}\label{susycurrent}
J^\mu\equiv\sqrt{2}\sigma^\nu\bar{\sigma}^\mu\chi\partial_\nu\phi^*
\end{equation}

\noindent is the Noether current associated with the invariance of the Lagrangian (\ref{Lwz}) under SUSY transformations (\ref{susytransf}). 

As usual in quantum field theory, we coupled the current with a gauge field $\psi_\mu$, then this contribution is
\begin{equation}\label{gravCurrent}
\mathcal{L}_{\psi J}=-\frac{i}{2}\kappa\psi_\mu J^\mu+h.c\quad.
\end{equation}

\noindent where $\kappa$ is a generic coupling constant. The field $\psi_\mu$ has both spinor and vector indices, and it is build from the representation
\begin{equation}\label{repgrav}
\left(\frac{1}{2},\frac{1}{2}\right)\otimes\left(\frac{1}{2},0\right)=\left(1,\frac{1}{2}\right)\oplus\left(0,\frac{1}{2}\right)\quad, 
 \end{equation}

 \noindent where $\left(\frac{1}{2},\frac{1}{2}\right)$ represents a vector with total spin 1 and $\left(0,\frac{1}{2}\right)$ represents a right-handed spinor with spin $\frac{1}{2}$. The result $\left(1,\frac{1}{2}\right)$ has total spin $\frac{3}{2}$ and besides the  helicities $\pm\frac{3}{2}$, it has helicity $\pm\frac{1}{2}$ from the spin $\frac{3}{2}$, and helicities $\pm\frac{1}{2}$ from the right-handed spinor in the right hand side. The same occurs for the right-handed spinor. As we will see soon, all helicities $\pm\frac{1}{2}$ vanish with spinor equations.
 
 The variation of $\mathcal{L}_{\psi J}$ gives 
 \begin{equation}\label{vargravCurrent}
\delta\mathcal{L}_{\psi J}=-\frac{i}{2}\kappa[(\delta\psi_\mu) J^\mu+\psi_\mu (\delta J^\mu)]+h.c\quad.
\end{equation}

\noindent Then, if the gauge field $\psi_\mu$ transforms as 
 \begin{equation}\label{vargravCurrent2}
\delta\psi_\mu=\frac{2}{\kappa}\partial_\mu\xi
\end{equation}

\noindent the first term in right-hand side of Eq. (\ref{vargravCurrent}) cancels the same terms in Eq.  (\ref{Lwzlocal}).

The second term will be canceled out as follows. The SUSY algebra (\ref{commutators}) yields the energy-momentum tensor $T_\mu\!^\nu$ as an anticommutator 
 \begin{equation}\label{tminianticom}
\{\bar{Q}_{\dot{\alpha}},J_\alpha\!^\nu \}=2\sigma_{\alpha\dot{\alpha}}\!^{\mu} T_\mu\!^{\nu}+\text{S.T.}\quad,
\end{equation}

\noindent where the Schwinger terms (S.T.) have zero vacuum expectation value \cite{wess1992}. With this relation, we  find
 \begin{equation}\label{vargravCurrent3}
-\frac{i}{2}\kappa\psi_\mu (\delta J^\mu)+h.c.=\frac{i}{2}k(\psi_\mu\sigma_\nu\bar{\xi}+\psi_\nu\sigma_\mu\bar{\xi}+\bar{\psi}_\mu\bar{\sigma}_\nu\xi+\bar{\psi}_\nu\bar{\sigma}_\mu\xi)T^{\mu\nu}\quad,
\end{equation}

\noindent which can be canceled out if we change the Lagrangian (\ref{Lwz}) by its version in a curved spacetime
 \begin{equation}\label{Lwzlocalgravity}
\mathcal{L}_{WZ}\rightarrow \sqrt{-g}(\partial_{\mu}\phi\partial^{\mu}\phi^*+i\partial_\mu\bar{\chi}\bar{\sigma}^\mu\chi)\equiv \mathcal{L}_{local}\quad.
\end{equation}

\noindent With this change, the variation of the Lagrangian has now a component due to the metric 
 \begin{equation}\label{}
 \frac{\delta\mathcal{L}_{local}}{\delta g_{\mu\nu}}\delta g_{\mu\nu}+\sqrt{-g}\delta\tilde{\mathcal{L}}_{WZ} =0\quad.
\end{equation}

But the second term is what we have found so far, that is, the variation of $\mathcal{L}_{WZ}+\mathcal{L}_{\psi J}$, which is Eq. (\ref{vargravCurrent3}). The energy-momentum tensor can be defined as 
 \begin{equation}\label{}
 T^{\mu\nu}\equiv \frac{2}{\sqrt{-g} }\frac{\delta\mathcal{L}_{local}}{\delta g_{\mu\nu}}\quad,
\end{equation}

\noindent thus
 \begin{align}\label{}
 \frac{1}{2}T^{\mu\nu}\delta g_{\mu\nu}=&\! -\delta(\mathcal{L}_{WZ\text{local}}+\mathcal{L}_{\psi J})\\\notag =&-\frac{i}{2}k(\psi_\mu\sigma_\nu\bar{\xi}+\psi_\nu\sigma_\mu\bar{\xi}+\bar{\psi}_\mu\bar{\sigma}_\nu\xi+\bar{\psi}_\nu\bar{\sigma}_\mu\xi)T^{\mu\nu}\quad,
\end{align}

\noindent which implies 
\begin{align}\label{}
 \delta g_{\mu\nu}= -ik(\psi_\mu\sigma_\nu\bar{\xi}+\psi_\nu\sigma_\mu\bar{\xi}+\bar{\psi}_\mu\bar{\sigma}_\nu\xi+\bar{\psi}_\nu\bar{\sigma}_\mu\xi)\quad.
\end{align}

The overall result is that when we promote the supersymmetry from a global symmetry to a local one, we need to introduce the gauge field $\psi_\mu$ and curvature $\eta^{\mu\nu}\rightarrow \sqrt{-g}g^{\mu\nu}$ in order to cancel out the extra contribution due to $\xi(x)$ in the variation of the Lagrangian.  A more convenient treatment is done, actually, in terms of the vielbein $e_\mu\!^{ a}$ instead of the metric, whose variation is
 \begin{align}\label{}
 \delta e_\mu\!^{a}= -ik(\xi\sigma^a\bar{\psi}_\mu+\bar{\xi}\bar{\sigma}^a\psi_\mu)\quad.
\end{align}

Since gravity is introduced in a ``natural'' way, the coupling is expected to be $k=M_{pl}^{-1}$, and the theory of local supersymmetry is called supergravity, whose multiplet is the vielbein $ e_\mu\!^{ a}$ and the gravitino $\psi_\mu$.

%========================================================================================
\subsection{Minimal Supergravity}

The on-shell Supergravity Lagrangian is given by the sum of the Einstein-Hilbert Lagrangian with the Rarita-Schwinger Lagrangian\footnote{In four dimensions $\gamma^{\mu\nu\rho}=\frac{1}{2}\epsilon^{\mu\nu\rho\sigma}\gamma_\sigma\gamma_5$ for Majorana spinors.} \cite{Freedman2012}
\begin{equation}\label{Lsugra}
e^{-1}\mathcal{L}_{Sugra}=\frac{R}{2}-\frac{1}{2}\bar{\Psi}_\mu\gamma^{\mu\nu\rho}D_\nu\Psi_\rho\quad,
\end{equation}

\noindent where $e$ means the determinant of $e_\mu\!^{a}$. Here, gravitino covariant derivative is given by
\begin{equation}\label{Dgravitino}
D_\nu\Psi_\rho\equiv\partial_\nu\Psi_\rho+\frac{1}{4}\omega_{\nu ab}\gamma^{ab}\Psi_\rho\quad,
\end{equation}

\noindent where $\omega_{\nu ab}$ is the spin connection, $\gamma^{ab}\equiv \frac{1}{2}[\gamma^a,\gamma^b]$ and in Eq. (\ref{Lsugra})  $\gamma^{\mu\nu\rho}\equiv\frac{1}{2}\{\gamma^\mu,\gamma^{\nu\rho}\}$. The Christoffel connection term $\Gamma_{\nu\rho}^\sigma\Psi_\sigma$ is not included in Eq. (\ref{Dgravitino}) because $\Gamma_{\nu\rho}^\sigma$ is symmetric while $\gamma^{\mu\nu\rho}$ is antisymmetric, in such a way that this term vanishes in the Lagrangian $\mathcal{L}_{Sugra}$. 

The supersymmetric transformations of $\mathcal{L}_{Sugra}$ are \cite{Freedman2012}
\begin{subequations}\label{sugratransf}
\begin{align}
\delta e_\mu\!^a&= \frac{1}{2}\bar{\xi}\gamma^a\Psi_\mu\quad,\\
\delta \Psi_\mu&=D_\mu\xi\equiv\partial_\mu\xi+\frac{1}{4}\omega_{\nu ab}\gamma^{ab}\xi\quad, \end{align}
\end{subequations}

\noindent where obviously $\xi=\xi(x)$. Thus, $\mathcal{L}_{Sugra}$ is invariant under (\ref{sugratransf}). In the presence of matter, the  gravitino transformation above contains more terms due to the extra fields.

%==================================================================================%
%==================================================================================%
\subsection{Super-Higgs mechanism}\label{massivegravitino}

In this section we present the mechanism of Super-Higgs, in which gravitino acquires mass by absorving the goldstone fermion. This mechanism of spontaneous symmetry breaking occurs in much the same way as the bosons $W^{\pm}$ and $Z^0$ obtain mass in the breakdown of the gauge symmetry of the electroweak interactions \cite{weinberg2000}.
	 In the presence of matter fields, supergravity Lagrangian has the form shown below \cite{wess1992}, where we consider the supergravity multiplet coupled with ``several'' Wess-Zumino multiplet $\{\phi^i,\chi^i\}$
\begin{align}\label{LsugraInter}
e^{-1}\mathcal{L}=&-\frac{R}{2}+\epsilon^{\mu\nu\rho\sigma}\bar{\psi}_\mu\bar{\sigma}_\nu\mathcal{D}_\rho\psi_\sigma-g_{ij^*}\partial_{\mu}\phi^i\partial^{\mu}\phi^{*j}-ig_{ij^*}\bar{\chi}^j\bar{\sigma}^\mu\mathcal{D}_\mu\chi^i
-\frac{1}{\sqrt{2}}g_{ij^*}\partial_\mu\phi^{*j}\chi^i\sigma^\mu\bar{\sigma^\nu}\psi_\mu\notag\\
&-\frac{1}{\sqrt{2}}g_{ij^*}\partial_\mu\phi^{j}\bar{\chi}^i\bar{\sigma^\mu}\sigma^\nu\bar{\psi}_\mu+e^{-1}\mathcal{L}^{(2f)}+e^{-1}\mathcal{L}^{(4f)}-\mathcal{V}\quad,
\end{align}

\noindent where 
\begin{subequations}\label{DcovLsugraInter}
\begin{align}
\mathcal{D}_\mu\chi=&\partial_\mu\chi^i+\chi^i\omega_\mu+\Gamma^i_{jk}\partial_\mu\phi^j\chi^k-\frac{1}{4}(K_j\partial_\mu\phi^j-K_{j*}\partial_\mu\phi^{j*})\chi^i\quad.\\
\mathcal{D}_\mu\psi_\nu=&\partial_\mu\psi_\nu+\psi_\nu\omega_\mu+\frac{1}{4}(K_j\partial_\mu\phi^j-K_{j*}\partial_\mu\phi^{j*})\psi_\nu\quad,\\
e^{-1}\mathcal{L}^{(2f)}= &e^{K/2}\left[W^*\psi_\mu\sigma^{\mu\nu}\psi_\nu+W\bar{\psi}_\mu\bar{\sigma}^{\mu\nu}\bar{\psi}_\nu+\frac{i}{\sqrt{2}}D_iW\chi^i\sigma^\mu\bar{\psi}_\mu+ \frac{i}{\sqrt{2}}D_{i*}W^*\bar{\chi}^i\bar{\sigma}^\mu\psi_\mu\right.\notag\\ &\left.+\frac{1}{2}\mathcal{D}_iD_jW\chi^i\chi^j+\frac{1}{2}\mathcal{D}_{i*}D_{j*}W^*\bar{\chi}^i\bar{\chi}^j \right]\quad,\\
\mathcal{V}=&e^{K}\left[g^{ij*}(D_iW)(D_iW)^*-3WW^*\right]\quad,\\
D_iW=&W_i+K_iW\quad,\\
\mathcal{D}_{i}D_{j}W=&W_{ij}+K_{ij}W+K_iD_jW-K_iK_jW-\Gamma^k_{ij}D_kW\quad.
\end{align}\end{subequations}

\noindent with $\Gamma^k_{ij}=K^{kl^*}K_{ijl^*}$ and lower indices indicate derivatives with respect to $\phi^i$. The $\mathcal{L}^{(4f)}$ term means the coupling of four fermions, which does not play a relevant role for gravitino mass. In the expressions above $K\equiv K(\phi^i,\phi^{i*})$ is the K\"ahler potential, a real function of the scalar and its conjugate, and $W\equiv W(\phi^i)$ is the superpotential, a holomorphic function of $\phi^i$. It is interesting to note that when supergravity is unbroken $D_iW=0$, therefore the mixing between gravitino and fermions vanishes. 

To realize what is the gravitino mass, the  following K\"ahler transformation is done
\begin{align}\label{kahlertransf}
K\rightarrow K+ \log W +\log W^*\equiv G\quad.
\end{align}

This transformation can be done only if $\langle W\rangle\neq 0$, and this assumption has a deep meaning. The potential in Eq. (\ref{DcovLsugraInter}) is zero if supergravity is unbroken and  $\langle W\rangle= 0$, what does not allow the K\"ahler transformation (\ref{kahlertransf}) due to the logarithm, therefore the gravitino remains massless. In the case of broken supergravity a vanishing cosmological constant is only achieved if there is a contribution of the negative part in the potential, therefore the transformation (\ref{kahlertransf}) can be done  and we go further. 

The kinetic terms in the supergravity Lagrangian and the K\"ahler metric are left invariant under a K\"ahler transformation. The new quadratic part in Eq. (\ref{DcovLsugraInter}) is found performing the following replacements: 
\begin{equation}\label{L2fchange}
\begin{aligned}
K \rightarrow G, \quad W\rightarrow 1\\
D_iW\rightarrow G_i, \quad K_{ij}\rightarrow G_{ij}\\
\mathcal{D}_iD_jW\rightarrow G_{ij}+G_iG_j-\Gamma^k_{ij}G_k\quad.
\end{aligned}
\end{equation}
\noindent The Christoffel symbol $\Gamma^k_{ij}$ also does not change. Then $\mathcal{L}^{(2f)}$ yields

\begin{align}\label{L2fchanged}
e^{-1}\mathcal{L}^{(2f)}= &e^{G/2}\left[\psi_\mu\sigma^{\mu\nu}\psi_\nu+\bar{\psi}_\mu\bar{\sigma}^{\mu\nu}\bar{\psi}_\nu+\frac{i}{\sqrt{2}}G_i\chi^i\sigma^\mu\bar{\psi}_\mu+ \frac{i}{\sqrt{2}}G_{i*}\bar{\chi}^i\bar{\sigma}^\mu\psi_\mu\right.\notag\\ &\left.+\frac{1}{2}(\nabla_iG_j+G_iG_j)\chi^i\chi^j+\frac{1}{2}(\nabla_{i^*}G_{j^*}+G_{i^*}G_{j^*})\bar{\chi}^i\bar{\chi}^j \right]\quad,
\end{align}

\noindent where the covariant derivative is $\nabla_iG_j\equiv G_{ij}-\Gamma^k_{ij}G_k$. The potential $\mathcal {V}$ becomes
\begin{align}\label{vsugrachanged}
\mathcal{V}=e^{G}\left[g^{ij*}G_iG_{j^*}-3\right]\quad.
\end{align}

We can see that the mixing still exist, thus, in order to get the mass of the fermions, the spinor mass matrix can be  diagonalized to get rid of the mixing. A field redefinition play the necessary role as well, and an easy way to see this is transform $\mathcal{L}^{(2f)}$ in the Majorana basis
\begin{align}\label{L2fchangedMajbasis}
e^{-1}\mathcal{L}^{(2f)}= \frac{e^{G/2}}{2}\left[\bar{\Psi}_\mu\gamma^{\mu\nu}\Psi_\nu+i\sqrt{2}\bar{\eta}\gamma^\mu\Psi_\mu +\bar{\eta}\eta+(\nabla_iG_j\chi^i\chi^j+h.c.)\right]\quad,
\end{align}

\noindent where $\eta\equiv G_i\chi^i$ is the Goldstone fermion, called goldstino. Using $\gamma^{\mu\nu}\gamma_\nu=3\gamma^\mu$ and the field redefinition \cite{wess1992}
\begin{equation}\label{gravitinoredef}
\tilde{\Psi}_\mu=\Psi_\mu+\frac{i}{6}\sqrt{2}\gamma_\mu\eta
\end{equation}

\noindent  the mixing is eliminated and the Lagrangian (\ref{L2fchangedMajbasis}) has now only diagonal terms
\begin{align}\label{L2fchangedMajbasisNew}
e^{-1}\mathcal{L}^{(2f)}= \frac{e^{G/2}}{2}\bar{\Psi}_\mu\gamma^{\mu\nu}\Psi_\nu+\frac{e^{G/2}}{2}\left[\left(\nabla_iG_j+\frac{1}{3}G_iG_j\right)\chi^i\chi^j+h.c.\right]\quad.
\end{align}

Finally, the mass of the gravitino and the spinor mass are 
\begin{equation}\label{gravitinomass}
m_{3/2}\equiv\langle e^{G/2}\rangle\quad,
\end{equation}

\begin{equation}\label{fermionmass}
m_{ij}\equiv\left\langle e^{G/2}\left(\nabla_iG_j+\frac{1}{3}G_iG_j\right)\right\rangle\quad.
\end{equation}

We have seen that gravitino acquires mass by absorbing the goldstino, with their masses given above. This process is also a gauge choice, in such a way that goldstino can be eliminated and the gauge invariance no longer keeps the gravitino massless,  bringing back the helicity $\pm\frac{1}{2}$ of the gravitino, which is predominant for light gravitino. If the potential $\mathcal{V}$ has an extremum value $\langle \partial \mathcal{V}/\partial \phi^k\rangle=0$ and the resulting cosmological constant is zero,  the following conditions come from Eq. (\ref{vsugrachanged}) 

\begin{align}
\langle G_iG^i\rangle&=3\\
\langle G^j\nabla_iG_j+G_i\rangle&=0\quad,
\end{align}

\noindent where $G^i\equiv g^{ij^*}G_{j^*}$, which leads to a zero spinor matrix mass (\ref{fermionmass}), therefore a massless goldstino.

\subsection{Rarita-Schwinger field equations}

\subsubsection{Massless field}

For the massless spin-$3/2$ particle, Rarita-Schwinger Lagrangian  is
\begin{equation}\label{}
\mathcal{L}_{RS}=-\frac{1}{2}\bar{\Psi}_\mu\gamma^{\mu\nu\rho}\partial_\nu\Psi_\rho\quad,
\end{equation}

\noindent with the equation of motion
\begin{equation}\label{elgravmassless}
\gamma^{\mu\nu\rho}\partial_\nu\Psi_\rho=0\quad.
\end{equation}

\noindent Using $\gamma_\mu\gamma^{\mu\nu\rho}=(D-2)\gamma^{\nu\rho}$, yields $\gamma^{\nu\rho}\partial_\nu\Psi_\rho=0$, and noting that $\gamma^{\mu\nu\rho}=\gamma^\mu\gamma^{\nu\rho}-2\eta^{\mu[\nu}\gamma^{\rho]}$, Eq. (\ref{elgravmassless})  implies 
\begin{equation}\label{elgravmassless2}
\gamma^{\mu}(\partial_\mu\Psi_\nu-\partial_\nu\Psi_\mu)=0\quad.
\end{equation}

From the discussion of Eq. (\ref{repgrav}) we see that since the free spin-$3/2$ particle is massless, it has only two degrees of freedom, thus we can choose a gauge, as the following
\begin{equation}\label{gravgauge}
\gamma^i\Psi_i=0\quad.
\end{equation}

\noindent The $\nu=0$ and $\nu=i$ components of Eq. (\ref{elgravmassless2}) are
\begin{align}\label{}
\gamma^i\partial_i\Psi_0-\partial_0\gamma^i\Psi_i=0\quad,\\
\gamma\cdot\partial\Psi_i-\partial_i\gamma\cdot\Psi=0\quad.
\end{align}

\noindent With this gauge choice, the first equation yields $\nabla^2\Psi_0=0$, so $\Psi_0=0$ and the second one leads to the Dirac equation for the spatial components $\Psi_i$
\begin{align}\label{}
\gamma\cdot\partial\Psi_i=0\quad,
\end{align}

\noindent which is a time evolution equation. However, contracting the equation above with $\gamma^i$ yields another constraint $\partial^i\Psi_i=0$.

Then, from the gauge condition (\ref{gravgauge}) and the equation of motion we get $3\times2^{D/2}$ independent constraints
\begin{equation}\label{2ndConstr}
\begin{aligned}
\gamma^i\Psi_i=0\quad,\\
\Psi_0=0\quad,\\
\partial^i\Psi_i=0\quad.
\end{aligned}
\end{equation}
 
\noindent Since $\Psi_\mu$ has $D\times2^{D/2}$ components, the number of  on-shell degrees of freedom of the massless Rarita-Schwinger field is $\frac{1}{2}(D-3)2^{D/2}$. Thus,  in $D=4$ the massless field with the gauge condition has only the helicitites $\pm3/2$.

\subsubsection{Massive field}
For the massive spin-$3/2$ particle, Rarita-Schwinger Lagrangian  is
\begin{equation}\label{}
\mathcal{L}_{RS}=-\frac{1}{2}\bar{\Psi}_\mu\gamma^{\mu\nu\rho}\partial_\nu\Psi_\rho+\frac{m_{3/2}}{2}\bar{\Psi}_\mu\gamma^{\mu\nu}\Psi_\nu\quad.
\end{equation}

\noindent Using the Euler-Lagrange equation, we get the equation 
\begin{equation}\label{elgrav}
\gamma^{\mu\nu\rho}\partial_\nu\Psi_\rho-m_{3/2}\gamma^{\mu\nu}\Psi_\nu=0\quad.
\end{equation}

\noindent Operating $\partial_\mu$ in the  equation above yields 
\begin{equation}\label{none}
m_{3/2}(\slashed{\partial}\gamma^{\nu}\Psi_\nu-\gamma^\nu\slashed{\partial}\Psi_\nu)=0 \quad,
\end{equation}

\noindent and if $\Psi_\nu$ is massive ($m_{3/2}\neq 0$)
\begin{equation}\label{0steqgrav}
\slashed{\partial}\gamma^{\nu}\Psi_\nu-\gamma^\nu\slashed{\partial}\Psi_\nu=[\slashed{\partial},\gamma^{\nu}]\Psi_\nu=0\quad.
\end{equation}

\noindent Applying $\gamma_\mu$ on Eq. (\ref{elgrav}) yields
\begin{equation}\label{1steqgrav}
\slashed{\partial}\gamma^{\nu}\Psi_\nu-\gamma^\nu\slashed{\partial}\Psi_\nu+3m_{3/2}\gamma^\nu\Psi_\nu=0\quad,
\end{equation}

\noindent then, using Eq.  (\ref{1steqgrav}), for a massive field
\begin{equation}\label{2steqgrav}
\gamma^\nu\Psi_\nu=0\quad.
\end{equation}

\noindent With this results is possible to write other condition
\begin{equation}\label{3steqgrav}
\partial^\nu\Psi_\nu=0\quad.
\end{equation}

Therefore, using Eqs.  (\ref{0steqgrav}), (\ref{2steqgrav}) and (\ref{3steqgrav}), Eq. (\ref{elgrav}) can be written in the form of a Dirac equation
 \begin{equation}\label{diracgravitino}
(\slashed{\partial}+m_{3/2})\Psi^\mu=0\quad,
\end{equation}

\noindent which shows that this is the free field of a particle of mass $m_{3/2}$ and spin-$3/2$. 

We see that the equation of motion (\ref{elgrav}) splits into three equations: two  spinor constraints  (\ref{2steqgrav}) and (\ref{3steqgrav}),  and  the Dirac equation version for spin-$3/2$ particle (\ref{diracgravitino}).

\section{Cosmological tracking solution and the Super-Higgs mechanism}\label{modelSugra1}

Supergravity with four supercharges  exists in four dimensions at most (that is, $\mathcal{N}\times 2^{D/2}=4$ for  $\mathcal{N}=1$ in $D=4$), so in higher dimensions (such as $D=10$ in superstring theory) one needs more supersymmetries.  Thus minimal supergravity can be seen as an effective theory in four dimensions, and can be applied to cosmology at least as a first approximation or a toy model.\footnote{For some models of extended supergravities in cosmology, see \cite{kallosh2002supergravity,kallosh2002gauged}.}
 In the framework of minimal supergravity, Refs. \cite{Brax1999,Copeland2000} were the first attempts to describe dark energy through quintessence. Moreover,  models of holographic dark energy were embedded in minimal supergravity  in \cite{Landim:2015hqa}. 

As usual in minimal supergravity, the scalar potential can be negative, so some effort should be made in order to avoid this negative contribution. In \cite{Brax1999}  one possibility was to require $\langle W \rangle=0$, for instance.\footnote{The recent work \cite{Bergshoeff:2015tra} shows  the explicit de Sitter supergravity action, where the negative contribution of the scalar potential is avoided in another way.}

Supergravity can provide a candidate for dark matter as well, the gravitino \cite{Pagels1982}. Once local supersymmetry is broken, the gravitino acquires a mass by absorbing the goldstino, but its mass is severely constrained when considering standard cosmology. The gravitino may be the lightest superparticle (LSP) being either stable, in a scenario that preserves R-parity \cite{Pagels1982}, or unstable, but long-lived and with a  small R-parity violation \cite{Takayama2000, Buchmuller:2007ui}. Another possibility is the gravitino to be the next-to-lightest superparticle (NLSP), so that it decays into standard model particles or into  LSP.  

From the cosmological point-of-view, if the gravitino is stable its mass should be $m_{3/2}\leq 1$ keV in order not to overclose the universe \cite{Pagels1982}. Thus the gravitino may be considered as dark matter; however such value is not what is expected to solve the hierarchy problem and the gravitino describes only hot/warm dark matter; then another candidate  is needed. If the gravitino is unstable it should have $m_{3/2}\geq 10$ TeV to decay before the Big Bang Nucleosynthesis (BBN) and therefore not to conflict its results \cite{Weinberg1982}.

 These problems may be circumvented if the initial abundance of the gravitino is diluted by inflation, but since the gravitino can be produced afterward through  scattering processes when the universe  is reheated, these problems can still exist.  Furthermore, the thermal gravitino production provides an upper bound for the reheating temperature ($T_R$), in such a way that if the gravitino mass is, for instance, in the range $10$ TeV $\leq m_{3/2}\leq100$ TeV, the reheating temperature should be $T_R<10^{10}-10^{11}$ GeV \cite{Kawasaki:1994af,Kawasaki:1994bs}. Depending on which is the LSP, the gravitino mass should be even bigger than $100$ TeV \cite{nakamura2006}. On the other hand, thermal leptogenesis requires $T_R\geq 10^8-10^{10}$ GeV \cite{Davidson2002,Buchmuller2002}, which leads also to strict values for the gravitino mass. All these issues regarding gravitino mass are refered sometimes as `gravitino problem'.

In this section we expand the previous results in the literature, taking the spontaneous breaking of local supersymmetry (SUSY) into account and regarding quintessence in minimal supergravity. We consider the  scalar potential defined by the usual flat K\"ahler potential $K$ (which leads to a flat  K\"ahler metric, thus to a canonical kinetic term)  and a power-law superpotential $W$,  whose scalar field $\varphi$ leads to the local SUSY breaking. After this mechanism (known also as Super-Higgs mechanism), the massive gravitino can decay into another scalar field $\Phi$. This scalar will play the role of the dark energy and the potential $V(\Phi)$ will be deduced from the original  potential $V(\varphi)$. It turns out that the potential $V(\Phi)$  corresponds to the so-called tracker behavior, whose initial conditions for the scalar field do not change the attractor solution. 

The rest of the section is organized in the following manner. In Sect. \ref{model} we present the local supersymmetry breaking process, assuming the flat K\"ahler potential $K$ and a power-law superpotential $W$. In  Sect. \ref{DE} we relate the previous scalar potential to a new one, responsible for the accelerated expansion of the universe. Sect. \ref{conclus} is reserved for a summary.

\subsection{The Super-Higgs mechanism}\label{model}
 
 We start our discussion using a complex scalar field $\phi=(\phi_R+i\phi_I)/\sqrt{2}$. We set $8\pi G=M_{pl}^{-1}=1$ in the following steps for simplicity. The scalar potential in $\mathcal{N}=1$ supergravity  depends on a real function $K\equiv K(\phi^i,\phi^{i*})$, called K\"ahler potential, and a holomorphic function $W\equiv W(\phi^i)$, the superpotential. We do not consider D-terms, so the potential for one scalar field  is given by
   \begin{equation}V= e^{K}\left[ K_{\phi\phi^*}^{-1}\left|W_\phi+K_\phi W\right|^2-3|W|^2\right]\quad,\label{potential}
   \end{equation}

 \noindent where $K_{\phi\phi^*}\equiv \frac{\partial^2 K}{\partial \phi\partial\phi^*}$ is the K\"ahler metric, $W_\phi\equiv\frac{\partial W}{\partial \phi}$, $K_\phi\equiv\frac{\partial K}{\partial \phi}$. When supergravity is spontaneously broken, the gravitino acquires a mass given by
  \begin{equation}m_{3/2}=\left\langle e^{K/2}|W|\right\rangle\quad,\end{equation}
 
 \noindent where $\langle ...\rangle$  means the vacuum expectation value. 
 
\begin{figure}
\centering
\includegraphics[scale=0.55]{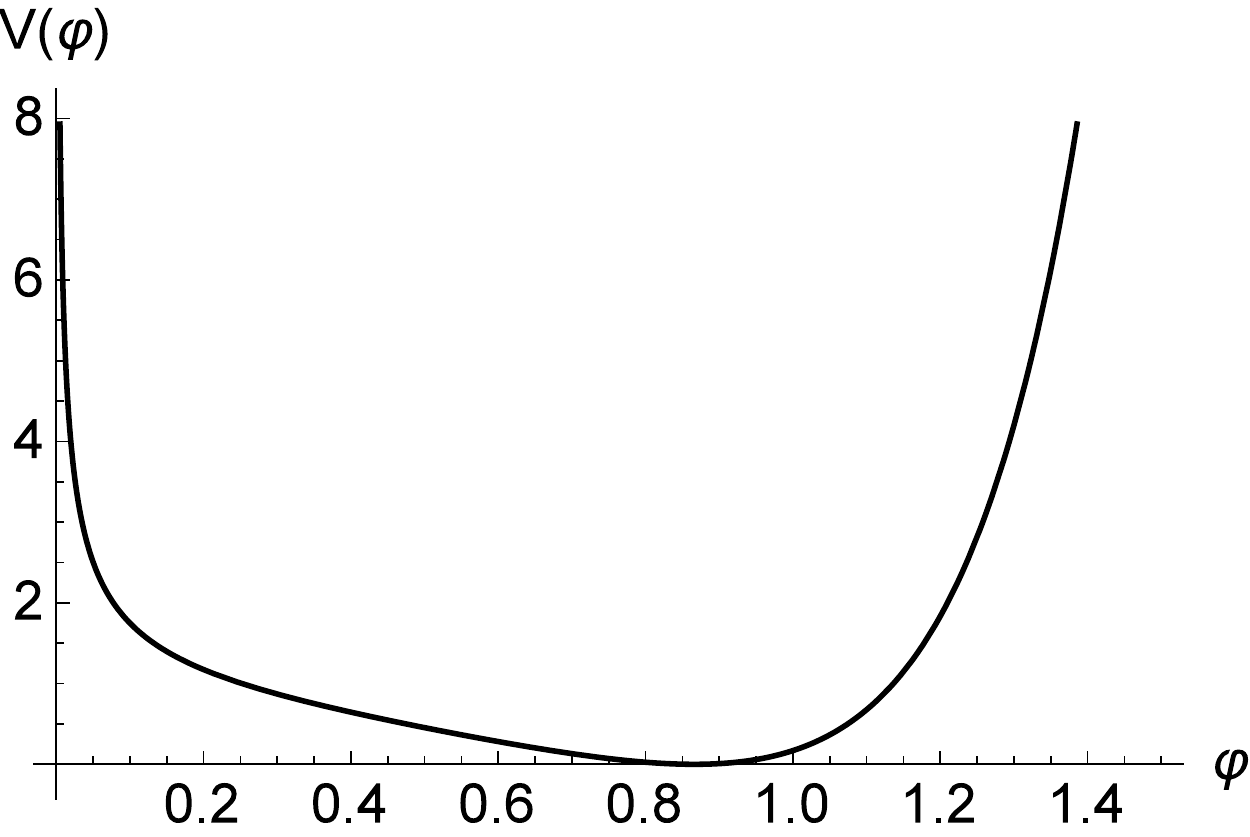}
\caption{\label{figPotCaseA} Potential $V(\varphi)$ as a function of the field $\varphi $, with $\lambda=1$ (in Planck units), for $n=3/4$.}
\end{figure}

 We use the K\"ahler potential $K_{f}=\phi\phi^*$, which leads to a flat K\"ahler metric and the superpotential  $W=\lambda^2\phi^n$, for real $n$, with $\lambda$ being a free parameter. For these choices of $K$ and $W$  the potential (\ref{potential}) is
   \begin{equation}
 V= \lambda^4e^{\varphi^ 2}[n^2\varphi^{2(n - 1)} + (2n - 3)\varphi^{2n} + \varphi^{2(n + 1)}]\quad,\label{potcaseA}
 \end{equation}

\noindent where  $\varphi\equiv(\sqrt{\phi_R^2+\phi_I^2})/\sqrt{2}$ is the absolute value of the complex scalar field.
We found that Eq.  (\ref{potcaseA}) has extremal points at 
\begin{align}
\varphi_1&=\sqrt{1-n-\sqrt{1-n}}\quad,\\
\varphi_2&=\sqrt{-n}\quad,\\
 \varphi_3&=\sqrt{1-n+\sqrt{1-n}}\quad,
 \end{align}

\noindent for $n<0$, with $\varphi_3$ being the global minimum, which is also  valid for the case $0<n<1$. For $n<0$ there are two local minima points ($\varphi_1$ and $\varphi_3$) which correspond to $V<0$ at these values. For negative $n$ the potential at $\varphi_3$  corresponds to the true vacuum, while the potential at $\varphi_1$ corresponds to the false vacuum. For $n>1$ $\varphi^2$ at the minimum point is negative, but this is  not allowed since $\varphi$ is  an absolute value. We consider the case of vanishing cosmological constant, so Eq. (\ref{potcaseA}) is zero at the global minimum $\varphi_3$ for $n=3/4$. This fractional number is the only possibility of the potential given by Eq (\ref{potcaseA}), which has $V=0$ at the minimum point. This case can be seen in Fig. \ref{figPotCaseA}. Due to its steep shape, the potential cannot drive the slow-roll inflation.

 SUSY is spontaneously broken at  the minimum point $\varphi_0\equiv \varphi_3$  and the gravitino becomes massive by absorbing the massless goldstino.  We estimate the gravitino mass for different values of $\lambda$, which are shown in Table \ref{tab:lambda}. The scalar mass has the same order of magnitude as the gravitino mass, therefore it may also lead to cosmological difficulties  in much the same way as the Polonyi field does \cite{Coughlan1983}. However, such a problem can be alleviated as the gravitino problem is, i.e., if the scalar mass is larger than $\mathcal{O}$(10TeV)  BBN starts after the decay of the scalar $\varphi$  has finished. From Table \ref{tab:lambda} we see that $\lambda\geq 10^{11}$ GeV does not spoil BBN.  Since the gravitino interactions are suppressed by the Planck mass, the dominant interaction is the one proportional to $M_{pl}^{-1}$ \cite{Freedman2012}, which is, for our purposes, the gravitino decay only into another complex scalar field ($\Phi$) plus its spin-1/2 superpartner ($\Psi\rightarrow \Phi+\chi$).\footnote{Although the interaction with a gauge field is also $\propto M_{pl}^{-1}$ we do not consider it for simplicity, since we have also not considered D-terms in the scalar potential. Gravitino decay into MSSM particles, such as photon and photino \cite{Kawasaki:1994af} or neutrino and sneutrino \cite{Kawasaki:1994bs}, is not considered here.}  The decay rate for the gravitino  is $\Gamma_{3/2} \sim m_{3/2}^3/M_{pl}^2$, thus for $\lambda\geq10^{12}$ GeV, the  gravitino time decay is $\leq10^{-3}$s, hence it does not conflict BBN results. The temperature due to the decay is $T_{3/2}\sim \sqrt{\Gamma_{3/2}}\leq 10^{7}$ GeV for $\lambda\leq10^{12}$ GeV, in agreement with \cite{Kawasaki:1994af,Kawasaki:1994bs}, but with a gravitino mass one order greater than the range showed in these references. For $\lambda\geq10^{13}$ GeV we get $T_{3/2}\geq 10^{10}$ GeV.

	As a result of the gravitino decay, it also appears a massless spin-1/2 fermion $\chi$, which behaves like radiation. The $\chi$ interactions are also suppressed  by the Planck scale because they are due to four fermions terms ($\propto M_{pl}^{-2}$) or through its covariant derivative, whose interaction is with $\partial \Phi$. Since the fermion $\chi$ has a radiation behavior, its cosmological contribution is diluted as the universe expands by $a^{-4}$. When the universe is radiation-dominated, $\Omega_{\Phi}$ is small, as so $\dot{\Phi}$, which implies that the interaction due to the covariant derivative is also small. Therefore, the contribution of $\chi$ can be ignored and  the gravitino can decay into dark energy, whose associated potential for the scalar is going to be deduced from Eq. (\ref{potcaseA}).

   \subsection{Dark energy} 	\label{DE}
  
 We first review the complex quintessence as a dark energy candidate, presented in \cite{Gu2001}. In this section the quintessence field will be related with the results of the last section and the  corresponding quintessence potential will be deduced from Eq. (\ref{potcaseA}). 

As seen in Eq. (\ref{potcaseA}) the scalar potential depends on the absolute value of the scalar field, as it is in fact for a complex quintessence field. As in Ref. \cite{Gu2001}, the complex quintessence can be written as $S=\Phi e^{i\theta}$, where $\Phi\equiv|S|$ is the absolute value of the scalar and $\theta$ is a phase, both depending only on time. The equations of motion for the complex scalar field in an expanding universe with FLRW metric and with a potential that depends only on the absolute value $\Phi$ (as  in our case) has been discussed in Sect. \ref{quintphantom}  \cite{Gu2001} 
  \begin{equation}\label{eq:KG}
  \ddot{\Phi}+3H\dot{\Phi}+\frac{d}{d\Phi}\left(\frac{\omega^2}{2a^6}\frac{1}{\Phi^2}+V(\Phi)\right)=0\quad,
 \end{equation}

  \noindent where $a$ is the scale factor and the first term in the brackets comes from the equation of motion for $\theta$ (\ref{eqmotionscalar2}), with $\omega$ being an integration constant interpreted as angular velocity \cite{Gu2001}. This term drives $\Phi$ away  from zero  and the factor $a^{-6}$ may make the term decrease very fast, provided that $\Phi$ does not decrease faster than $a^{-3/2}$. This situation never happens for our model, so that the contribution due to the phase component $\theta$ decreases faster than the matter density  $\rho_m$ (proportional to $a^{-3}$). Furthermore, if $\omega$ is small the complex scalar field behaves like a real scalar. From all these arguments, we neglect the complex contribution in the following.

 \begin{table}\centering
\begin{tabular}{ll}
\hline\noalign{\smallskip}
$\lambda$ (GeV)&$m_{3/2}$ (GeV)\\

\noalign{\smallskip}\hline\noalign{\smallskip}
$10^{14}$  &  $10^{10}$  \\
$10^{13}$  &  $10^{8}$  \\
$10^{12}$  &  $10^{6}$  \\
$10^{11}$  &  $10^{4}$  \\
$10^{10}$  &  $10^{2}$  \\
$10^{9}$  &  $1$  \\
$10^{8}$  &  $10^{-2}$  \\
$10^{7}$  &  $10^{-4}$  \\
$10^{6}$  &  $10^{-6}$  \\ 
  
 \noalign{\smallskip}\hline
\end{tabular}
\caption{\label{tab:lambda} Gravitino masses for different values of $\lambda$.}
\end{table}

After supergravity is spontaneously broken, the scalar field $\varphi$ may oscillate around $V(\varphi_0)=0$ and the massive gravitino can decay into the scalar field $\Phi$.  Since the  scalar $\varphi$ oscillates around $V(\varphi_0)=0$ before decay,  we perform the following steps. We assume the dark energy, given by the field $\Phi$, has a similar potential to the one in Eq. (\ref{potcaseA}). Then, we just shift the scalar potential from $V(\Phi)$ to $V(\Phi-\Phi_0)$, thus the minimum point goes from $\Phi_0\neq 0$ to $\Phi_0=0$. The Taylor expansion of $V(\Phi)$ around the new minimum $\Phi=\Phi_0=0$ (for $n=3/4$) leads to a natural exponential term $e^{\Phi^2}\approx 1$. All these procedures give rise to the potential $V(\Phi)$ 
    \begin{equation}\label{potexpanded}
  V(\Phi)=\frac{M^{9/2}}{\sqrt{\Phi}} +\mathcal{O}(\Phi^{1/2})\quad.
   \end{equation}
 
  \noindent The constant $M^{9/2}$ is written in this form for convenience. The leading-order term of this potential (\ref{potexpanded}) is an example of the well-known tracking behavior \cite{Zlatev:1998tr,Steinhardt:1999nw}. 
	
	The parameter $\lambda$ does not have the same order of $M^{9/2}$, as long as the original shifted potential was expanded around zero. The order of magnitude of the  potential $V(\Phi)$ ($M^{9/2}$) is determined phenomenologically and can be as $10^{-47} $ GeV$^4$.  The tracker condition $\Gamma\equiv VV''/V'^2=(\alpha+1)/\alpha>1$ (for a generic potential $V(\phi)\sim \phi^{-\alpha}$ for $\alpha>0$) is satisfied in our case, where $\alpha=1/2$. 
	
	The dynamical system analysis of the quintessence field shows that the potential (\ref{potexpanded}) leads to a fixed point in the phase plane, which is a stable attractor \cite{copeland2006dynamics, Landim:2015uda}. The equation of state for the quintessence field is given by $w_\Phi=-1+\tilde{\lambda}^2/3$, where $\tilde{\lambda}\equiv -V'(\Phi)/V((\Phi)$ decreases to zero for the tracking solution. The potential Eq. (\ref{potexpanded}) gives an equation of state $w_\Phi=-0.96$, in agreement with the Planck results \cite{Planck2013cosmological}, for $\tilde{\lambda}=0.35$, which in turn is easily achieved since $\tilde{\lambda}$ decreases to zero.
 
 As is usual in quintessence models in supergravity, such as in \cite{Brax1999,Copeland2000},  different energy scales are needed to take both dynamical dark energy and local SUSY breaking into account. This  unavoidable feature is present in our model in the fact that we have needed two different free parameters $\lambda$ and $M^{9/2}$ and they are determined by phenomenological arguments.

 \subsection{Summary} \label{conclus}
 In this section we have analyzed the scalar potential in minimal supergravity using the traditional flat K\"ahler potential and the power-law superpotential, whose  single scalar field $\varphi$  is responsible for the Super-Higgs mechanism. The gravitino decays before BBN and originates other scalar $\Phi$, which is regarded as the dark energy. The scalar $\Phi$ had  its potential derived from the potential $V(\varphi)$, where the leading-order term of the expanded  potential  is a well-known example that satisfies the tracker condition.

%%%%%%%%%%%%%%%%%%%%%%%%%%%%%%%%%%%%%%%%%%%%%%%%%%%%%%%%%%%%%%%%%%%%%%%%%%%%%%%%%%%%%%%%%%%%%%%%%%%%%%%%%%%%%%%%%
 \section{Holographic dark energy from minimal supergravity}\label{HDEsugra}
\subsection{Holographic dark energy}

Another striking attempt  to explain the acceleration comes from holography. An effective quantum field theory in a box of size $L$ with ultraviolet (UV) cutoff $\Lambda$ has an entropy that scale extensively, given by $S\sim L^3\Lambda^3$ \cite{Cohen:1998zx}. Based on the peculiar thermodynamics of black holes \cite{Bekenstein:1973ur,Bekenstein:1974ax,Bekenstein:1980jp,Hawking:1974sw,Bekenstein:1993dz}, Bekeinstein proposed that the maximum entropy in a box of volume $L^3$ behaves non-extensively \cite{Bekenstein:1973ur,Bekenstein:1974ax,Bekenstein:1980jp,Bekenstein:1993dz}, growing with area of the box. Moreover, the degrees of freedom of a physical system also scales with its boundary area  rather than its volume \cite{'tHooft:1993gx,Susskind:1994vu}. A system satisfies the Bekenstein entropy bound if its volume is limited according to
\begin{equation}\label{BHentropy}
L^3\Lambda^3 \lesssim L^2M_{pl}^2\quad,
\end{equation}

\noindent where the RHS is the entropy of a black hole of radius $L$. The length $L$ acts as an infrared (IR) cutoff and from Eq. (\ref{BHentropy}) it is straightforward that $L\lesssim M_{pl}^2 \Lambda^{-3}$. 

Field theories appear not to be a good effective low energy description of any system containing a black hole, since infalling particles experience Planck scale interactions with outgoing Hawking radiation near the horizon. Therefore, they should probably not attempt to describe particle states whose volume is smaller than their corresponding Schwarzschild radius. But an effective theory that saturates (\ref{BHentropy}) includes states with the Schwarzschild radius much larger than the box size. This is seen if we recall that effective quantum field theory is expected to describe a system at a temperature $T$, provided that $T\leq\Lambda$. Since the thermal energy is an extensive quantity, it is $M\sim L^3T^4$  with an entropy $S\sim L^3T^3$. Eq. (\ref{BHentropy}) is saturated at a temperature $T\sim (M_{pl}^2/L)^{1/3}$ and the corresponding Schwarzschild radius is $L_S\sim M/M_{pl}^2=L(LM_{pl}^2)^{2/3}\gg L$.   

To avoid the difficulties pointed above, Cohen \textit{et al.} \cite{Cohen:1998zx}  proposed a stronger constraint on the IR cutoff. Since the
maximum energy density in the effective theory is $\Lambda^4$, the new constraint is
   \begin{equation}\label{BHentropy2}
L^3\Lambda^4 \lesssim LM_{pl}^2\quad.
\end{equation}

\noindent A system at temperature $T$ saturates (\ref{BHentropy2}) at $T\sim (M_{pl}/L)^{1/2}$ and the corresponding Schwarzschild radius is now $L_S\sim L$. 

Once a relation between the UV and IR cutoff was established, if $\rho_D$ is the vacuum energy density, Eq. (\ref{BHentropy2}) becomes $L^3\rho_D \lesssim LM_{pl}^2$. This last inequality is saturated at \cite{Hsu:2004ri,Li:2004rb}
   \begin{equation}\label{li2}
\rho_D=3c^2M_{pl}^2L^{-2}\quad,
\end{equation}

\noindent where $3c^2$ is a constant introduced for convenience.

 The first choice for $L$ was  the Hubble radius $H^{-1}$. With this choice the Friedmann equation is $3 H^2 = \rho_D+\rho_m=3c^2H^2+\rho_m$, so $\rho_m=3H^2(1-c^2)$ and $\rho_m$  and $\rho_D$ behave as $H^2$. The $\rho_m$ scales with $a^{-3}$, as so $\rho_D$, thus the dark energy is pressureless and its equation of state is $w_m=0$ \cite{Hsu:2004ri}. The correct equation of state for dark energy was obtained by \cite{Li:2004rb}, when he chose the future event horizon as the IR cutoff. The future event horizon is
\begin{equation}\label{re}
R_E=a\int_t^\infty\frac{dt}{a}=a\int_a^\infty\frac{da}{Ha^2}\quad.
\end{equation}

 Assuming that the dark energy $\rho_D$ dominates the Friendmann equation  $3H^2=\rho_D=3c^2R_E^{-2}$ simplifies to $HR_E=c$. Using Eq. \ref{re} into $HR_E=c$ we can write it as \cite{Li:2004rb}
\begin{equation}\label{re2}
\int_a^\infty\frac{da}{Ha^2}=\frac{c}{Ha}\quad,
\end{equation}

\noindent which has the solution $H^{-1}=\alpha a^{1-1/c}$, with a constant $\alpha$. The density energy for HDE becomes
\begin{equation}\label{re3}
\rho_D=3\alpha^2a^{-2(1-1/c)}\quad,
\end{equation}

\noindent hence the equation of state for HDE is 
\begin{equation}\label{li1}
w_D=-\frac{1}{3}-\frac{2}{3c}\end{equation}

\noindent and can describe the accelerated expansion of the universe. For a IR cutoff with the particle horizon 
\begin{equation}\label{re4}
R_H=a\int_0^t\frac{dt}{a}=a\int_0^a\frac{da}{Ha^2}
\end{equation}

\noindent the equation of state would be $w_D=-\frac{1}{3}+\frac{2}{3c}>-\frac{1}{3}$.

The problem with the Hubble radius as the IR cutoff could be avoided assuming the interaction between dark energy and dark matter (DM) \cite{Pavon:2005yx}. Such interaction was first proposed in the context of quintessence\cite{Wetterich:1994bg,Amendola:1999er}, and since the energy  densities of the DE and DM are comparable, the interaction can alleviate the coincidence problem\cite{Zimdahl:2001ar,Chimento:2003iea}. Taking the coupling $\mathcal{Q}= 3b^2H \rho_D$ into account, with constant $b$, the HDE with Hubble radius as IR cutoff could lead to an accelerated expansion of the universe and also solve the coincidence problem \cite{Pavon:2005yx}. To see this fact, consider the ratio of the energy densities 
\begin{equation}\label{pavon1}
r\equiv \frac{\rho_M}{\rho_D}=\frac{1-c^2}{c^2}\quad,
\end{equation}

\noindent 	whose evolution equation is
\begin{equation}\label{pavon2}
\dot{r}=3Hr\left[w+\left(1+\frac{1}{r}\right)b^2 \right]\quad.
\end{equation}

The equation of state for dark energy with this interaction is written in terms of the ratio $r$ as 
  \begin{equation}w_D=-\left(1+\frac{1}{r}\right)b^2=-\frac{b^2}{1-c^2}
   \label{eqPavon}
   \end{equation}
\noindent and when Eq. (\ref{eqPavon}) is used in Eq. (\ref{pavon1}), $\dot{r}=0$, alleviating the coincidence problem.

Li's proposal could also be generalized assuming the interaction between the two components of the dark sector as $\mathcal{Q}\propto H(\rho_M+\rho_D)$ \cite{Wang:2005jx,Wang:2005pk,Wang:2005ph,Wang:2007ak}. 

Due to the prominent role of the AdS/CFT correspondence\cite{Maldacena:1997re} to relate both supergravity and holography concepts, it is natural to ask if there is any connection between supergravity (even for $\mathcal{N}=1$) and HDE. As showed in \cite{Sheykhi:2011cn}, it is possible to establish a connection between HDE and different kind of scalar fields, such as quintessence, tachyon and K-essence, through reconstructed scalar potentials. In Sects. \ref{HDEHUbble} and \ref{HDEfuture} we show that the HDE in interaction with DM can be embedded in minimal supergravity plus matter with a single chiral superfield. The interaction between HDE and DM is taken into account because they are generalizations of some uncoupled cases, namely, the models presented in \cite{Hsu:2004ri,Li:2004rb}, thus the correspondent models are good choices to analyze whether they can be embedded in minimal supergravity or not.

\subsection{HDE with Hubble radius as IR cutoff} \label{HDEHUbble}

 We use $M_{pl}=1$ from now on for simplicity. First we consider the size $L$  as the Hubble radius $H^{-1}$, thus the energy density for the dark energy becomes
   \begin{equation}\rho_D=3c^2H^2\quad, \label{eqH}
   \end{equation} 
  
  \noindent which describes a pressureless fluid \cite{Hsu:2004ri} in the absence of interaction with dark matter. When one considers such interaction \cite{Pavon:2005yx}, the fluid  has an equation of state that can describe the dark energy. We consider the interaction $Q=3b^2H\rho_D$, where $b$ is a constant which measures the strength of the interaction.    
  
In order to reconstruct a holographic scalar field model we should relate Eq. (\ref{eqPavon}) with the scalar field $\phi$. Using the energy density and the pressure for the scalar field $\rho_\phi=\dot{\phi}^2/2+V(\phi)$ and $p_\phi=\dot{\phi}^2/2-V(\phi)$, we have
   \begin{equation}\dot{\phi}^2=(1+w_\phi)\rho_\phi, \quad V(\phi)=(1-w_\phi)\frac{\rho_\phi}{2}\quad.
   \label{scalarV}
   \end{equation}

Using Eqs. (\ref{eqH}) and (\ref{eqPavon}) into Eq. (\ref{scalarV}) we have the potential \cite{Sheykhi:2011cn}
 \begin{equation}
 V(\phi)=B^{-2}e^{-\sqrt{2}B\phi}\quad,
   \label{V1}
   \end{equation}
   
   \noindent where $B=\frac{3k}{2\sqrt{2}c}\left(3-\frac{3b^2}{1-c^2}\right)^{-1/2}$ and $k=1-b^2c^2/(1-c^2)$.

On the other hand, the scalar potential in minimal supergravity with no D-terms is given by 
  \begin{equation}V= e^{K}\left( K^{\Phi\bar{\Phi}}\left|W_\Phi+K_\Phi W\right|^2-3|W|^2\right)\quad,\label{potentialHDE}
   \end{equation} 
   
    \noindent for a single complex scalar field.  We will use the same choices of Ref. \cite{Copeland2000} for $K$ and $W$, namely, the string-inspired K\"ahler potential $K=-\ln(\Phi+\bar{\Phi})$, which is present at the tree level for axion-dilaton field in string theory, and the superpotential $W=\Lambda^2\Phi^{-\alpha}$, where $\Lambda$ is a constant. With these choices, with the imaginary part of the scalar field stabilized at zero $\langle$Im $\Phi\rangle=0$ and the field redefinition Re $\Phi=e^{\sqrt{2}\phi}$, we get the scalar potential expressed in terms of the canonical normalized field $\phi$ 
    \begin{equation}V(\phi)= \frac{\Lambda^4}{2}(\beta^2-3)e^{-\sqrt{2}\beta\phi}\quad,\label{potentialsugra}
   \end{equation} 
   
   \noindent where  $\beta=2\alpha+1$. Comparing Eq. (\ref{V1}) with Eq.  (\ref{potentialsugra}) we have $B=\beta$ and $\Lambda^2=\sqrt{2}/(\beta\sqrt{\beta^2-3})$, with $\beta^2\geq 3$. Thus, the two parameters $b$ and $c$ determine the exponent $\alpha$ of the superpotential and the parameter $\Lambda$. We notice in this case that $\Lambda$ is not an independent parameter, but it depends on $b$ and $c$ as well. Even if we would have a way to know what is the value of $\alpha$ or rather, $\beta$, we could not know the specific values of $b$ and $c$. The opposite direction is favored, because once one finds out observationally $b$ and $c$, the HDE can be embedded in a specific supergravity model. If there is no interaction ($b=0$) $\beta=\sqrt{3}/(2c\sqrt{2})$, but $w_D=0$.

   The kernel used so far simplifies $w_D$ in such a way that Eq. (\ref{V1}) has been written with no need of any approximation. When one considers the other possibilities for the interaction $Q$ ($\propto \rho_M$ or $\propto\rho_M+\rho_D$), the scalar potential cannot be written as easy as it was in Eq. (\ref{V1}). We illustrate this possibility now, but with other IR cutoff.  
   
  \subsection{HDE with future event horizon as IR cutoff}\label{HDEfuture}

   We will analyze another possibility of HDE with the kernel $\mathcal{Q}=3b^2H(\rho_M+\rho_D)$ and with the future event horizon $R_E=a\int_t^\infty{dt/a}=c\sqrt{1+r}/H$ as the IR cutoff.  The choice of $L$ is done because when $b=0$ the original Li's model of HDE \cite{Li:2004rb} is recovered. The energy density for the HDE is 
       \begin{equation}\label{rho2}
   \rho_D=\frac{3H^2}{1+r}
   \end{equation}
   
\noindent and the equation of state for HDE becomes
        \begin{equation}
     w_D=-\frac{1}{3}\left(1+\frac{2\sqrt{\Omega_D}}{c}+\frac{3b^2}{\Omega_D}\right)\quad,
   \label{eqabdalla}
   \end{equation}
   
  \noindent where $\Omega_D\equiv\rho_D/(\rho_D+\rho_M)=(1+r)^{-1}$ is the density parameter for DE. Using Eqs. (\ref{rho2}) and (\ref{eqabdalla}) into Eq.  (\ref{scalarV}) we have
          \begin{equation}
    \dot{\phi}^2=2H^2\Omega_D\left(1-\frac{\sqrt{\Omega_D}}{c}-\frac{3b^2}{2\Omega_D}\right)\quad,
   \label{phidot}
   \end{equation}
   
      \begin{equation}
    V(\phi)=H^2\Omega_D\left(2+\frac{3\sqrt{\Omega_D}}{c}+\frac{9b^2}{2\Omega_D}\right)\quad.
   \label{V}
   \end{equation}
   
 \noindent From the equations above we see that we have to know the evolution of $\Omega_D$ in order to fully determine  $V(\phi)$ as a function of $\phi$. The evolution equation  for $\Omega_D$ was found in \cite{Wang:2005jx} and it is
       \begin{equation}
   \frac{\Omega_D'}{\Omega_D^2}=(1-\Omega_D)\left(\frac{1}{\Omega_D}+\frac{2}{c\sqrt{\Omega_D}}-\frac{3b^2}{\Omega_D(1-\Omega_D)}\right)\quad,
   \label{omegaevolut}
   \end{equation}

   \noindent where the prime is the derivative with respect to $\ln a$. To solve this differential equation we let $y=1/\sqrt{\Omega_D}$, thus  Eq.  (\ref{omegaevolut}) becomes
    \begin{equation}
   y^2y'=(1-y^2)\left(\frac{1}{c}+\frac{y}{2}+\frac{3b^2y^3}{2(1-y^2)}\right)\quad.
   \label{yevolut}
   \end{equation}

 In order to have an analytic solution of Eq. (\ref{yevolut}) we will investigate the asymptotic behavior of $\Omega_D$, for small and large $a$. For very small $a$ we have $\Omega_D\rightarrow 0$ and $y\rightarrow \infty$, therefore Eq. (\ref{yevolut}) is approximately
     \begin{equation}
   y'\approx(1-y^2)\left(\frac{3b^2}{2}-\frac{1}{2}\right)y\quad,
   \label{yevolut2}
   \end{equation}
  
  \noindent which leads to the solution $\Omega_D=\Omega_0 a^{-(3b^2-1)}\approx \Omega_0 a $, provided that $b$ should be small \cite{Wang:2005jx}. Since  $\Omega_D$ scales with $a$, the contribution of the dark energy in the early universe is negligible. We also see from Eq. (\ref{phidot}) that the term between brackets should be positive and small, then $ \dot{\phi}\propto H\sqrt{2\Omega_D} $ is small. Therefore, we have $\phi(a)\sim a^{1/2}$. Due to the small contribution of  $\Omega_D$, we neglect it at this limit and we will focus on large $a$, where HDE is dominant. 
  
  For large $a$ we have $\Omega_D\rightarrow 1$, and $y\rightarrow 1$. Thus, the last term in Eq.  (\ref{yevolut}) is the dominant one, so $y'\approx\frac{3b^2}{2}y$, and $\Omega_D=\Omega_0a^{-3b^2}$, where $\Omega_0$ is the value of $\Omega_D$ at $a_0=1$. Since $\Omega_D=\Omega_0a^{-3b^2}=\Omega_0a^{-3(1+w_D)}$, we have $1+w_D=b^2$ which is very small if we consider the present value of $w_D$ given by Planck \cite{Planck2013cosmological}. Thus we neglect the last term in Eq. (\ref{phidot}) and we assume the dark energy dominance. We have  $\Omega_D=\Omega_0\approx 1$ at large $a$ and Eq. (\ref{phidot}) yields
     \begin{equation}
    \dot{\phi}^2\approx 2H^2\Omega_D\left(1-\frac{1}{c}\right)\quad.
   \label{phidot2}
   \end{equation}

  \noindent The equation above implies that $w_D\approx -\frac{1}{3}(1+2/c)$, which is the Li's proposal \cite{Li:2004rb}. Eq. (\ref{phidot2}) has the solution
     \begin{equation}
   \phi(a)=\left(1-\frac{1}{c}\right)^{1/2}\sqrt{2\Omega_D}\ln a\quad.
   \label{phidota}
   \end{equation}
   
   The second Friedmann equation at large $a$ leads to  $H=\frac{3c}{(c-1)t}$ and $a=t^{3/(c-1)}$. The scalar potential (\ref{V}) becomes
       \begin{equation}
V(\phi)\approx\left(2+\frac{3}{c}\right)H^2=\left(2+\frac{1}{c}\right)\frac{9c^2}{(c-1)^2}e^{-\frac{1}{3}\left(\frac{c-1}{c}\right)^{1/2}\phi}\quad.\label{V2}
   \end{equation}
   
   Similarly to before, comparing Eq. (\ref{V2}) with Eq. (\ref{potentialsugra}) we see that both $\beta$ and $\Lambda$ are determined by the constant $c$. Eq. (\ref{V2}) was deduced for a dark-energy-dominated universe, in such a way that the interaction with dark matter is absent, as it should be in this limit.

   A more realistic scenario could not be found analytically, in opposite to the previous case, where the scalar potential was written with no approximations.  Different kernels for the coupling $Q$ may be tried, although the main features of the method were expressed in these two cases. The other alternatives are similar.

  \subsection{Summary}
 In this section we embedded two models of HDE in the minimal supergravity with one single chiral superfield. In the first model we used the Hubble radius as the IR cutoff and an interaction proportional to $\rho_D$, while in the second one we used the future event horizon as IR cutoff and the kernel proportional to $\rho_D+\rho_M$. In both cases the free parameters  of our superpotential could be expressed in terms of the constants $c$ and $b$, depending on the model. The second case is embedded only for a dark-energy-dominated universe, while the first one is more general. There are other ways to embed HDE in minimal supergravity, as for instance the choices of K\"ahler potential and superpotential made in \cite{Nastase:2015pua} for the inflationary scenario. However, the main results are the same, that is, a way to relate HDE and supergravity. Due to the nature of the holographic principle and recalling that extended supergravity is the low-energy limit of string theory, the relation presented here has perhaps a deeper meaning, when one takes a quantum gravity theory into account.  Such idea was explored in \cite{Nastase:2016sji}.

%% file: MDE.tex
\chapter{Metastable dark energy}\label{MDE}

The chapter is structured as follows. In Sect. \ref{MSDE} we present a model of metastable dark energy. It is embedded into a dark $SU(2)_R$ model in Sect. \ref{darkSU2} and we summarize our results in Sect. \ref{concluSU2}.  This chapter is based on our paper \cite{Landim:2016isc}.

\section{A model of metastable dark energy}\label{MSDE}

The current stage of accelerated expansion of the universe will be described by a canonical scalar field $\varphi$ at a local minimum $\varphi_0$ of its potential $V(\varphi)$, while the true minimum of $V(\varphi)$ is at $\varphi_{\pm}=\langle\varphi\rangle$. The energy of the true vacuum is below the zero energy of the false vacuum, so that this difference is interpreted as the observed value of the vacuum energy ($10^{-47}$ GeV$^4$). 

We assume that by some mechanism the scalar potential is positive definite (as e.g. in supersymmetric models) and the true vacuum lies at zero energy. As we will see below this value is adjusted by the mass of the scalar field and the coefficient of the quartic and sixth-order interaction. The rate at which the false vacuum decays into the true vacuum state will be calculated.

The process of barrier penetration in which the metastable false vacuum decays into the stable true vacuum is similar to the  old inflationary scenario and it occurs through the formation of bubbles of true vacuum in a false vacuum background. After the barrier penetration the bubbles grow at the speed of light and eventually collide with other bubbles until all  space is in the lowest energy state. The energy release in the process can produce new particles and  a Yukawa interaction $g \varphi\bar{\psi}\psi$ can account for the production of a fermionic field which can be the pressureless fermionic dark matter.  However, as we will see, the vacuum time decay is of the order of the age of the universe, so another dominant process for the production of cold dark matter should be invoked in order to recapture the standard cosmology.

If one considers a scalar field $\varphi$ with the even self-interactions up to order six, one gets

\begin{equation}\label{VScalar}
 V(\varphi)=\frac{m^2}{2}\varphi^2-\frac{\lambda}{4}\varphi^4+\frac{\lambda^2}{32 m^2}\varphi^6\quad,
\end{equation} 

\noindent where $m$ and $\lambda$ are positive free parameters of the theory and the coefficient of the $\varphi^6$ interaction is chosen in such a way that the potential (\ref{VScalar}) is a perfect square. This choice will be useful to calculate the false vacuum decay rate.

The potential (\ref{VScalar}) has  extrema  at $\varphi_0=0$, $\varphi_{\pm}=\pm\frac{ 2m}{\sqrt{\lambda}}$ and $\varphi_1=\frac{\varphi_{\pm}}{\sqrt{3}}$, but it is zero in all of the minima ($\varphi_0$ and $\varphi_{\pm}$). In order to have a cosmological constant, the potential should  deviate slightly from the perfect square (\ref{VScalar}).  Once the coupling present in GR is the Planck mass $M_{pl}$ it is natural to expect that the deviation from the Minkowski vacuum is due to a term proportional to $M^{-2}_{pl}$. Thus we assume that the potential (\ref{VScalar}) has a small deviation given by $\frac{\varphi^6}{M_{pl}^2}$. Although the value of the scalar field at the minimum point $\varphi_\pm$ also changes, the change is very small and we can consider that the scalar field at the true vacuum is still $\pm\frac{ 2m}{\sqrt{\lambda}}$. The difference between the true vacuum and the false one is 
\begin{equation}
V(\varphi_0)-V(\varphi_\pm)\approx\frac{64m^6}{\lambda^3M_{pl}^2}\quad.
\label{eq:cosmoconst}
\end{equation}

As usual in quantum field theory it is expected that the parameter $\lambda$ is smaller than one, thus, if we assume $\lambda \sim 10^{-1}$, the Eq. (\ref{eq:cosmoconst}) gives $\sim 10^{-47}$ GeV$^4$ for $m\sim \mathcal {O}(\text{MeV})$. Bigger values of $\lambda$ imply smaller values of $m$. Therefore, the cosmological constant is determined by the mass parameter and the coupling of the quartic interaction.

\begin{figure}\centering
\includegraphics[scale=0.6]{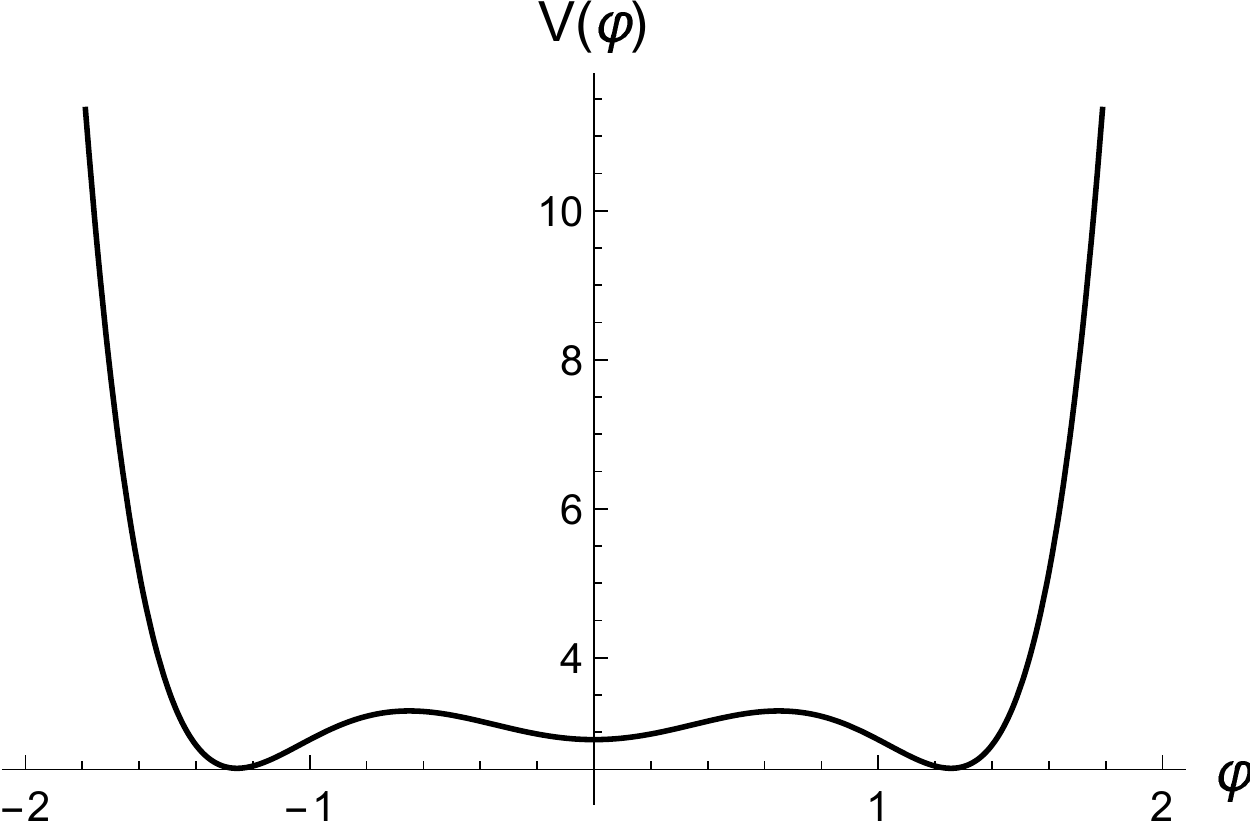}%
\caption{Scalar potential (\ref{VScalar}) with arbitrary parameters and values. The difference between the true vacuum at $\varphi_\pm\approx \pm 1.2 $   and  the false vacuum at $\varphi_0=0$ is $\sim 10^{-47}$ GeV$^4$.}%
\label{potentialFig}%
\end{figure}

 The potential (\ref{VScalar}) with the term $\frac{\varphi^6}{M_{pl}^2}$ is shown in Fig. \ref{potentialFig}.

\subsection{Decay rate}\label{sec:Decay}

The computation of the decay rate is based on the semi-classical theory presented in \cite{Coleman:1977py}. The energy of the false vacuum state at which $\langle \varphi\rangle=0$ is given by \cite{Weinberg:1996kr}
\begin{equation}
E_0=-\lim_{T\rightarrow\infty}\frac{1}{T}\ln\left[\int{\exp\left(-S_E[\varphi;T]\right)\prod_{\vec{x},t}\,d\varphi(\vec{x},t)}\right]\quad ,
\label{eq:E0}
\end{equation}
where $S_E[\varphi;T]$ is the Euclidean action,
\begin{equation}
S_E=\int\,d^3x\int_{-\frac{T}{2}}^{+\frac{T}{2}}\,dt\left[\frac{1}{2}\left(\frac{\partial\varphi}{\partial t}\right)^2+\frac{1}{2}\left(\nabla\varphi\right)^2+V(\varphi)\right]\quad.
\label{eq:S}
\end{equation}

 The imaginary part of $E_0$ gives the decay rate\footnote{As we know from Quantum Mechanics. See for instance the Problem 1.15 from \cite{griffiths2005introduction}. } and all the fields $\varphi(\vec{x},t)$ integrated in Eq. (\ref{eq:E0})  satisfy the boundary conditions 
\begin{equation}\label{boundarycond}\varphi(\vec{x},+T/2)=\varphi(\vec{x},-T/2)=0\quad .
\end{equation} 

The action (\ref{eq:S}) is stationary under variation of the fields that satisfy the equations
\begin{equation}
\frac{\delta S_E}{\delta\varphi}=-\frac{\partial^2\varphi}{\partial t^2}-\nabla^2\varphi+V'(\varphi)=0
\label{eq:Seq}
\end{equation}
and  are subject to the boundary conditions (\ref{boundarycond}). In order to get the solution of Eq. (\ref{eq:Seq}) we make an ansatz that the field $\varphi(\vec{x},t)$ is invariant under rotations around $\vec{x}_0,t_0$ in four dimensions, which in turn is valid for large $T$ \cite{Coleman:1977th}. The ansatz is 
\begin{equation}
\varphi(\vec{x},t)=\varphi(\rho) \quad \text{with } \quad \rho\equiv\sqrt{(\vec{x}-\vec{x}_0)^2+(t-t_0)^2}\quad .
\label{eq:ansatz}
\end{equation}

In terms of the Eq. (\ref{eq:ansatz}), the field equations (\ref{eq:Seq}) becomes
\begin{equation}
\frac{d^2\varphi}{d \rho^2}+\frac{3}{\rho}\frac{d\varphi}{d\rho}=V'(\varphi)\quad .
\label{eq:Seq2}
\end{equation}

The above equation of motion  is analogous to that of a particle at position $\varphi$  moving in a time $\rho$, under the influence of a potential $-V(\varphi)$ and a viscous force $-\frac{3}{\rho}\frac{d\varphi}{d\rho}$. This particle travels from an initial value $\varphi_i$ and $\rho=0$ and reaches $\varphi=0$ at $\rho\rightarrow\infty$. The Euclidean action (\ref{eq:S}) for the rotation invariant solution becomes
\begin{equation}
S_E=\int_{0}^{\infty}2\pi^2 \rho^3\,d\rho\left[\frac{1}{2}\left(\frac{\partial\varphi}{\partial \rho}\right)^2+V(\varphi)\right]\quad .
\label{eq:Srot}
\end{equation}

The metastable vacuum decay into the true vacuum is seen as the formation of bubbles of true vacuum surrounded by the false vacuum outside. The friction term $\frac{d\varphi}{d\rho}$ is different from zero only at the bubble wall, since the field is at rest inside and outside. The decay rate per volume of the false vacuum, in the semi-classical approach, is of order 
\begin{equation}
\frac{\Gamma}{V}\approx M^{-4}\exp(-S_E)\quad ,
\label{eq:decayrate}
\end{equation}
where $M$ is some mass scale. When $S_E$ is large the barrier penetration is suppressed and the mass scale is not important. This is the case when the energy of the true vacuum is slightly below the energy of the false vacuum, by an amount $\epsilon$, considered here as small as $\epsilon\sim 10^{-47}$ GeV$^4$. On the other hand, the potential $V(\varphi)$ is not small between $\varphi_0$ and $\varphi_\pm$.

We will use the so-called `thin wall approximation', in which $\varphi$ is taken to be inside of a four-dimensional sphere of large radius $R$. For a thin wall we can consider $\rho\approx R$  in this region and since $R$ is large we can neglect the viscous term, which is proportional to $3/R$ at the wall. The action (\ref{eq:Srot}) in this approximation is 
\begin{equation}
S_E\simeq -\frac{\pi^2}{2}R^4\epsilon+2\pi^2R^3S_1\quad ,
\label{eq:Sthin}
\end{equation}
where $S_1$ is a surface tension, given by
\begin{equation}
S_1= \sqrt{2}\int_{\varphi_0}^{\varphi_+}\,d\varphi{ \sqrt{V}}\quad ,
\label{eq:S1}
\end{equation}
for small $\epsilon$.  The action (\ref{eq:Sthin}) is stationary at the radius 
\begin{equation}
R\simeq \frac{3S_1}{\epsilon}\quad ,
\label{eq:Rmin}
\end{equation}
and at the stationary point the action (\ref{eq:Sthin}) becomes
\begin{equation}
S_E\simeq \frac{27\pi^2S_1^4}{2\epsilon^3}\quad .
\label{eq:SthinA}
\end{equation}

Using the potential (\ref{VScalar}) into Eq. (\ref{eq:S1}) we obtain\footnote{The term $\frac{\varphi^6}{M^2_{pl}}$ is very small and can be ignored. }
\begin{equation}
S_1= \frac{m^3}{\lambda}\quad ,
\label{eq:S1Pot}
\end{equation}
which in turn gives the Euclidean action at the stationary point in the thin wall approximation (\ref{eq:Sthin}) 
\begin{equation}
S_E\simeq \frac{27\pi^2m^{12}}{2\lambda^4\epsilon^3}\quad.
\label{eq:SthinPot}
\end{equation}

Substituting the action (\ref{eq:SthinPot}) into the decay rate (\ref{eq:decayrate}) with $\epsilon\sim 10^{-47}$ GeV$^4$ and the mass scale being $M\sim 1$ GeV for simplicity, we have
\begin{equation}
\frac{\Gamma}{V}\approx \exp\left[-10^{143}\left(\frac{m}{\text{GeV}}\right)^{12}\lambda^{-4}\right] \text{GeV}^4
\quad .
\label{eq:decayrate2}
\end{equation}

The decay time is obtained inverting the above expression,
\begin{equation}
t_{decay}\approx 10^{-25}\left\{\exp\left[10^{143}\left(\frac{m}{\text{GeV}}\right)^{12}\lambda^{-4}\right] \right\}^{1/4}\text{s}\quad.
\label{eq:tdecayrate}
\end{equation}

The expression for the decay time gives the lowest value of the mass parameter $m$ for which (\ref{eq:tdecayrate}) has at least the age of the universe ($10^{17}$ s). Therefore the mass parameter should be
\begin{equation}
m\gtrsim 10^{-12} \text{GeV}\quad ,
\label{eq:m}
\end{equation}
for $\lambda\sim 10^{-1}$. Thus, it is in agreement with  the values for $m$ at which the scalar potential describes the observed vacuum energy, as discussed in the last section.  The mass of the scalar field can be smaller if the coupling $\lambda$ is also smaller than $10^{-1}$. The decay rate (\ref{eq:decayrate2}) is strongly suppressed for larger values of $m$. The bubble radius given in Eq. (\ref{eq:Rmin}) for the mass parameter (\ref{eq:m}) is $R\gtrsim 0.03$ cm.

Notice that the axion would still be a possibility, although it arises in a quite different context. We can also consider the gravitational effect in the computation of the decay rate. In this case the new action $\bar{S}$ has the Einstein-Hilbert term $\frac{M_{pl}^2}{2}\mathcal{R}$, where  $\mathcal{R}$ is the Ricci scalar. The relation between the new action  $\bar{S}$ and the old one $S_E$ can be deduced using the thin wall approximation and it gives \cite{Coleman:1980aw}
\begin{equation}
\bar{S}=\frac{S_E}{\left(1+\left(\frac{R}{2\Delta}\right)^2\right)^2}\quad ,
\label{Sbar}
\end{equation}
where $S_E$ and $R$ are given by Eqs. (\ref{eq:SthinA}) and (\ref{eq:Rmin}), respectively, in the absence of gravity, and $\Delta=\frac{\sqrt{3}M_{pl}}{\sqrt{\epsilon}}$ is the value of the bubble radius when it is equal to the   Schwarzschild radius associated with the energy released by the conversion of false vacuum to true one. 

 For $\epsilon\sim 10^{-47}$ GeV$^4$ we get $\Delta\sim 10^{27}$ cm, thus the gravitational correction $R/\Delta$ is very small. Larger values of $m$ give larger $R$, implying that  the gravitational effect should be taken into account. Even so, the decay rate is still highly suppressed.

\section{A dark $SU(2)_R$ model}\label{darkSU2}

As an example of how the metastable dark energy can be embedded into a dark sector model we restrict our attention to  a model with  $SU(2)_R$ symmetry. Both dark energy and dark matter are doublets under $SU(2)_R$ and singlets under any other symmetry.\footnote{We assume here only interactions among the components of the dark sector, leading the possible interactions with the standard model particles to a future work.} After the spontaneous symmetry breaking by the dark Higgs field $\phi$, the gauge bosons $W_d^+$, $W_d^-$ and $Z_d$ acquire the same mass given by $m_W=m_Z=g v/2$, where $v$ is the VEV of the dark Higgs. The dark $SU(2)_R$ model contains a  dark matter candidate $\psi$,  a dark neutrino $\nu_d$ (which can be much lighter than $\psi$), and the dark energy doublet $\varphi$, which contains $\varphi^0$ and $\varphi^+$,  the latter being  the heaviest particle. After  symmetry breaking $\varphi^0$ and $\varphi^+$ have different masses and both have a potential given by Eq. (\ref{VScalar}) plus the deviation $\frac{(\varphi^\dagger\varphi)^3}{M_{pl}^2}$  The interaction between the fields are given by the Lagrangian
%\bar{\psi}_R i \slashed{\partial}\psi_R +\bar{\nu}_{dR} i \slashed{\partial}\nu_{dR}+\partial\varphi\dagger\partial{\varphi}+
\begin{eqnarray}
 \mathcal{L}_{int}=
g\left(W_{d\mu}^+J_{dW}^{+\mu} +W_{d\mu}^-J_{dW}^{-\mu}+Z_{d\mu}^0J_{dZ}^{0\mu}\right)\quad,
\label{eq:Lint}
\end{eqnarray}
where the currents are 
\begin{eqnarray}
&&\quad J_{dW}^{+\mu} =\frac{1}{\sqrt{2}}[\bar{\nu}_{dR}\gamma^\mu \psi_R+i(\varphi^0\partial^\mu\bar{\varphi}^+-\bar{\varphi}^+\partial^\mu\varphi^0)]\quad , \label{eq:JW+}\\
&&\quad J_{dW}^{-\mu} =\frac{1}{\sqrt{2}}[\bar{\psi}_{R}\gamma^\mu \nu_{dR}+i(\varphi^+\partial^\mu\bar{\varphi}^0-\bar{\varphi}^0\partial^\mu\varphi^+)]
\quad ,\label{eq:JW-}\\
J_{dZ}^{0\mu} &=&\frac{1}{2}[\bar{\nu}_{dR}\gamma^\mu \nu_{dR}-\bar{\psi}_{R}\gamma^\mu \psi_{R}+
i(\varphi^+\partial^\mu\bar{\varphi}^+-\bar{\varphi}^+\partial^\mu\varphi^+)-i(\varphi^0\partial^\mu\bar{\varphi}^0-\bar{\varphi}^0\partial^\mu\varphi^0)]\quad .
\label{eq:JZ0}
\end{eqnarray}

 The currents above are very similar to the ones in the electroweak theory. The main differences are that there is no hypercharge due to $U(1)_Y$ and there is a new doublet, given by $\varphi^+$ and $\varphi^0$. 

Among the interactions shown in Eq. (\ref{eq:Lint}), it is of interest to calculate the decay rate due to the process $\varphi^+\rightarrow \varphi^0+\psi+\nu_d$. The three-body decay leads to a cold dark matter particle whose mass can be accommodated to give the correct relic abundance, to a dark neutrino which is a hot/warm dark matter particle, and to a scalar field $\varphi^0$. Similar to the weak interactions, we assume that the energy involved in the decay is much lower than the mass of the gauge fields, thus the propagator of $W$ is proportional to $g^2/m_W^2$ and the currents interact at a point. We can also define 
\begin{equation}
 \frac{g^2}{8m_W^2}\equiv \frac{G_d}{\sqrt{2}}\quad ,
\label{eq:darkcoupling}
\end{equation}
where $G_d$ is the dark coupling.

		\begin{figure}
		\centering\includegraphics[scale=1]{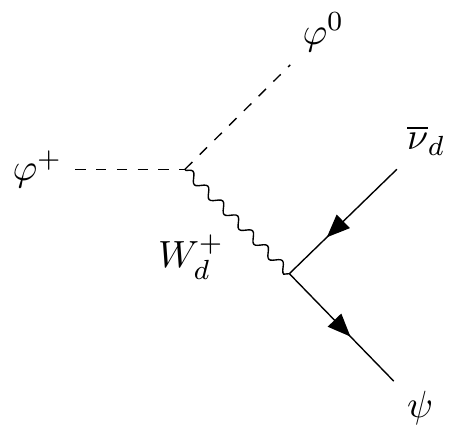}
%\begin{tikzpicture}
%\begin{feynman}
%\vertex (a) {\(\varphi^{+}\)};
%\vertex [right=of a] (b);
%\vertex [above right=of b] (f1) {\(\varphi^0\)};
%\vertex [below right=of b] (c);
%\vertex [above right=of c] (f2) {\(\overline \nu_{d}\)};
%\vertex [below right=of c] (f3) {\(\psi\)};
%\diagram* {
%(a) -- [scalar] (b) -- [scalar] (f1),
%(b) -- [boson, edge label'=\(W^{+}_d\)] (c),
%%(c) -- [anti fermion] (f2),
%(c) -- [fermion] (f3),
%};
%\end{feynman}\end{tikzpicture}
\caption{Feynman diagram for the decay $\varphi^+\rightarrow \varphi^0+\psi+\nu_d$.}\label{feyndiagram}
\end{figure}

The Feynman diagram for the decay is shown in Fig. \ref{feyndiagram} and the amplitude for the decay is 
\begin{eqnarray}
\mathcal{M}=\frac{g^2}{4m_W^2}(P+p_1)_\mu\overline{v}(p_3)\gamma^\mu(1+\gamma^5)u(p_2)
\quad ,
\label{eq:M}
\end{eqnarray}
where the labels $1$, $2$ and $3$ are used, respectively, for the particles $\varphi^0$, $\psi$ and $\nu_d$. The energy-momentum conservation implies that $P=p_1+p_2+p_3$, where $P$ is the four-momentum of the field $\varphi^+$ and $M$ will be its mass. 

The averaged amplitude squared for the decay $\varphi^+\rightarrow \varphi^0+\psi+\nu_d$ is
\begin{eqnarray}
\overline{ |\mathcal{M}|^2}= 16 G_d^2\left\{2[(P+p_1)\cdot p_2][(P+p_1)\cdot p_3]-(P+p_1)^2(p_2\cdot p_3+m_2m_3)\right\}\quad.
\label{eq:M12}
\end{eqnarray}

Using the energy-momentum conservation  and defining the invariants $s_{ij}$ as $s_{ij}\equiv (p_i+p_j)^2=(P-p_k)^2$, we can reorganize the amplitude squared. The three invariants are not independent, obeying $s_{12}+s_{23}+s_{13}=M^2+m_1^2+m_2^2+m_3^2$ from their definitions and the energy-momentum conservation. With all these steps we  eliminate $s_{13}$ and get
\begin{eqnarray}
\overline{ |\mathcal{M}|^2}= &16 G_d^2[ -2 s_{12}^2-2s_{12}s_{23}+2(M^2+m_1^2+m_2^2+m_3^3)s_{12}+\frac{(m_2+m_3)^2}{2}s_{23}\nonumber\\
&-2m_2m_3(M^2+m_1^2)-2m_1^2M^2-2m_2^2(m_1^2+m_2^2)-\frac{(m_2+m_3)^2}{2} ]\quad .
\label{eq:M2}
\end{eqnarray}

The decay rate can be evaluated from \cite{Agashe:2014kda}
\begin{equation}
d\Gamma= \frac{ 1}{(2\pi)^3}\frac{1}{32M^3}\overline{|\mathcal{M}|^2}d s_{12}d s_{23}\quad , \label{eq:dgamma}
\end{equation}
where for a given value of $s_{12}$, the range of $s_{23}$ is determined by its
values when $\vec{p_2}$ is parallel or antiparallel to $\vec{p_3}$ 
\begin{eqnarray}
(s_{23})_{max}&=& (E_2^*+E_3^*)^2-\left(\sqrt{E_2^{*2}-m_2^2}-\sqrt{E_3^{*2}-m_3^2}\right)^2
\quad ,\label{s23max}\\
(s_{23})_{min}&=& (E_2^*+E_3^*)^2-\left(\sqrt{E_2^{*2}-m_2^2}+\sqrt{E_3^{*2}-m_3^2}\right)^2\quad . 
\label{s23min}
\end{eqnarray}

The energies $E_2^*=(s_{12}	-m_1^2+m_2^2)/(2\sqrt{s_{12}})$ and $E_3^*=(M	-s_{12}-m_3^2)/(2\sqrt{s_{12}})$ are the energies
of particles 2 and 3 in the $s_{12}$ rest frame \cite{Agashe:2014kda}. The invariant $s_{12}$, in turn, has the limits 
\begin{equation}
\label{limits12}
(s_{12})_{max}=(M-m_3)^2, \qquad (s_{12})_{min}=(m_1+m_2)^2\quad . 
\end{equation}

\begin{figure}\centering
\includegraphics[scale=0.55]{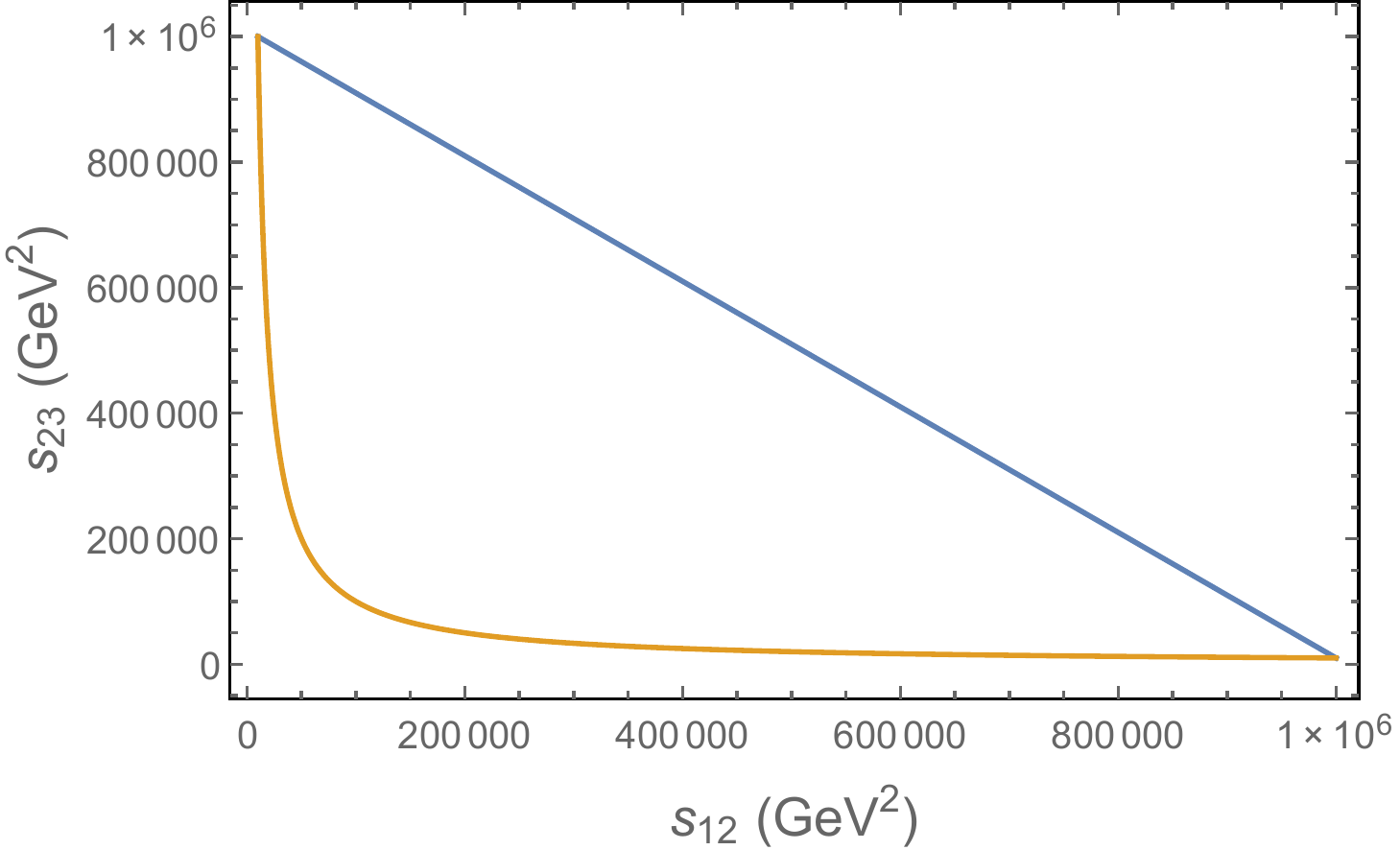}%
\caption{Dalitz plot for the $s_{12}$ and $s_{23}$ limits (\ref{s23max}), (\ref{s23min}) and (\ref{limits12}). The region limited by the curves is kinetically allowed . We used $M=1000$ GeV, $m_1=1$ MeV, $m_2=100$ GeV, $m_3=0$ and $G_d\sim 10^{-27} $ GeV$^{-2}$.  }%
\label{dalitzplot}%
\end{figure}

The plot of the kinetic limits of $s_{12}$ and $s_{23}$ is shown in Fig. \ref{dalitzplot} and it is called Dalitz plot. 

\begin{figure}\centering
\includegraphics[scale=0.6]{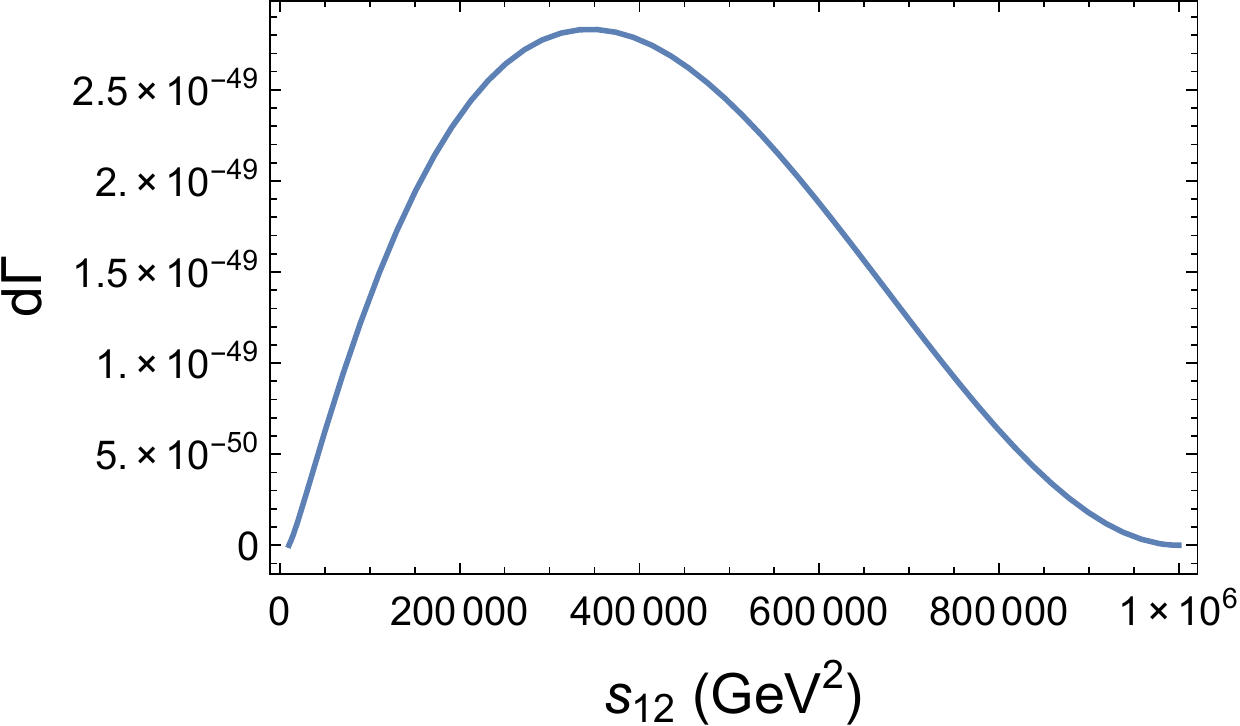}%
\caption{Differential decay rate $d\Gamma$ (\ref{eq:dgamma}) as a function of $s_{12}$ for $M=1000$ GeV, $m_1=1$ MeV, $m_2=100$ GeV, $m_3=0$ and $G_d\sim 10^{-27} $ GeV$^{-2}$.}%
\label{decaywidth}%
\end{figure}

With the limits for $s_{12}$ (\ref{limits12}) and for $s_{23}$ (\ref{s23max})--(\ref{s23min}) and with the amplitude squared  (\ref{eq:M2})  we can integrate Eq. (\ref{eq:dgamma}) for different values of masses, in order to get the decay time $t_{dec}$.

The plot of $d\Gamma$ as a function of $s_{12}$ is shown in Fig. \ref{decaywidth} and the decay rate $\Gamma$ is the area under the curve. For illustrative purposes, we set the mass of the particles as being $M=1000$ GeV, $m_1=1$ MeV, $m_2=100$ GeV and $m_3=0$ GeV. With these values of masses,  the decay time is of the order of the age of the universe ($10^{17}$s) with $G_d\sim 10^{-27}$ GeV$^{-2}$, while  with $G_d\sim 10^{-26}$ GeV$^{-2}$ the decay time is $t_{dec}\sim 10^{15}$s. If $g$ is for instance of the same order of the fine-structure constant,  the gauge bosons $W_d^\pm$ and $Z_d$ have masses  around $10^{11}$ GeV in order to the decay time to be $10^{15}$s. Such decay times are compatible with  phenomenological models of interacting dark energy,  where the coupling is proportional to the Hubble rate  \cite{Costa:2016tpb,Wang:2016lxa}. In addition, depending on the values of the free parameters, the mass of the gauge bosons can be of the same order of the grand unified theories scale.

 \section{Summary}\label{concluSU2}

In this chapter we presented a model of metastable dark energy, in which the dark energy is a scalar field with a potential given by a sum of even self-interactions up to order six. The parameters of the model can be adjusted in such a way that the difference between the energy of the true vacuum and the energy of the false one is around $10^{-47}$ GeV$^4$. The decay of the false vacuum to the true one is highly suppressed, thus the metastable dark energy can explain the current accelerated expansion of the universe. We do not need a very tiny mass for the scalar field (as it happens for some models of quintessence), in order to get the observed value of the vacuum energy.

The metastable dark energy can be inserted into a more sophisticated model for the dark sector. In this paper we restricted our attention to a Lagrangian invariant under $SU(2)_R$ (before the spontaneous symmetry breaking by the dark Higgs), in which the dark energy doublet and the dark matter doublet naturally interact with each other. The decay of the heaviest particle of the dark energy doublet into the three daughters (dark energy particle, cold and hot dark matter) was calculated and the decay time can be as long as the age of the universe, if the mediator is massive enough. Such a decay shows a different form of interaction between dark matter and dark energy, and the model opens a new window to investigate the dark sector from the point-of-view of particle physics.

%% file: conclusions.tex
\chapter{Final remarks}\label{conclusions}

This thesis explored models of the dark sector, in which the dark energy is described by some candidates other than the conventional cosmological constant. Scalar fields (real or complex) were mostly used in  the models, although a vector field was taken into account as well. 
The two components of the dark sector were assumed to interact with each other, in different contexts. Equipped with the dynamical system theory we analyzed the asymptotic states of some cosmological models, in which the dark energy was either scalar fields or a vector field. Scalar fields were also used as candidates to explain the accelerated expansion of the universe using minimal supergravity. Finally, we proposed a model of metastable dark energy and we inserted it in a dark $SU(2)_R$ model, in which the dark energy and dark matter doublets naturally interact with each other. 

The results presented here can be extended in several different directions. The dynamical system theory has shown itself to be a good tool to analyze asymptotic states of cosmological models and we  have studied other couplings in \cite{Landim:2016gpz}. In addition, we obtained an analytic formula for the growth rate of structures in a coupled dark energy model in which the exchange of energy-momentum is proportional to the dark energy density \cite{Marcondes:2016reb}.   

The dark $SU(2)_R$ model  opens a new window to investigate the whole dark sector from the point-of-view of particle physics. We started attacking this issue here and a natural evolution is to build extensions of the Standard Model. Furthermore, as part of an extended Standard Model, it may be expected that a grand unified theory contains the dark $SU(2)_R$.

%% file: notations.tex
\chapter{Notations and Conventions}

We use Planck units $\hbar=c=8\pi G=k_B=1$ unless where is explicitly written. The FLRW metric has signature ($-,+,\ldots, +$) and the universe is flat. Weyl spinor indices are the first part of Greek symbols ($\alpha$, $\beta$, $\ldots$, $\dot{\alpha}$, $\dot{\beta}$, $\ldots$), run from one to two and are implicit. Second part of Greek symbols ($\mu, \nu, \rho, \ldots$) indicate Lorentz elements, and in four dimension they run from 0 to 3. 

Generalized Pauli matrices are 

\begin{equation}\label{paulimatrices}
 \begin{aligned}
       \sigma^0&= \begin{pmatrix}
-1 & 0 \\
0 & -1
 \end{pmatrix} 
              \\
        \sigma^2&= \begin{pmatrix}
0 & -i \\
i & 0
\end{pmatrix} 
       \end{aligned}
 \qquad 
   \begin{aligned}
       \sigma^1&= \begin{pmatrix}
0 & 1 \\
1 & 0
 \end{pmatrix} 
              \\
        \sigma^3&= \begin{pmatrix}
1 & 0 \\
0 & -1
 \end{pmatrix} 
       \end{aligned}
\end{equation}

\noindent and the generators of the Lorentz group are written in terms of \eq{paulimatrices} by

\begin{equation}\label{genlorentz}
\begin{aligned}
\sigma^{\mu\nu}\!_{\alpha}\!^\beta\equiv\frac{1}{4}(\sigma_{\alpha\dot{\alpha}}\!^\mu\bar{\sigma}^{\nu\dot{\alpha}\beta}-\sigma_{\alpha\dot{\alpha}}\!^\nu\bar{\sigma}^{\mu\dot{\alpha}\beta})\\
\bar{\sigma}^{\mu\nu\dot{\alpha}}\!_{\dot{\beta}}\equiv\frac{1}{4}(\bar{\sigma}^{\mu\alpha\dot{\alpha}}\sigma_{\alpha\dot{\beta}}\!^\nu-\bar{\sigma}^{\nu\alpha\dot{\alpha}}\sigma_{\alpha\dot{\beta}}\!^\mu)\quad,
\end{aligned}
\end{equation}

\noindent where $\bar{\sigma}^\mu=(\sigma^0, -\sigma^i)$. Gamma matrices in the Weyl basis are

\begin{equation}\label{gammamatrix}
\gamma^\mu=\begin{pmatrix}
0 & \sigma^\mu \\
\bar{\sigma}^\mu& 0  \\
\end{pmatrix}\quad,
\end{equation}

\noindent and $\gamma_5$ is defined as

\begin{equation}\label{gammamatrix5}
\gamma_5=i\gamma^0\gamma^1\gamma^2\gamma^3\quad.
\end{equation}

\noindent Majorana spinors contain one Weyl spinor

\begin{equation}\label{majoranaspinor}
\Psi_M=\begin{pmatrix}
\chi_\alpha \\
\bar{\chi}^{\dot{\alpha}} 
\end{pmatrix}\quad.
\end{equation}

Gravitino is written in the following way

\begin{equation}\label{gravitinospinor}
\Psi_\mu=\begin{pmatrix}
\psi_{\mu\alpha} \\
\bar{\psi}_\mu\!^{\dot{\alpha}} 
\end{pmatrix}\quad,
\qquad
\bar{\Psi}_\mu=\begin{pmatrix}
\psi_{\mu\alpha} &\bar{\psi}_\mu\!^{\dot{\alpha}}
\end{pmatrix} \quad.
\end{equation}

%% file: main.bbl
\providecommand{\noopsort}[1]{}\providecommand{\singleletter}[1]{#1}%
\begin{thebibliography}{100}

\bibitem{reiss1998}
A.~G. Riess et~al.
\newblock {Observational evidence from supernovae for an accelerating universe
  and a cosmological constant}.
\newblock {\em Astron.J.}, 116:1009--1038, 1998.

\bibitem{perlmutter1999}
S.~Perlmutter et~al.
\newblock {Measurements of Omega and Lambda from 42 high redshift supernovae}.
\newblock {\em Astrophys.J.}, 517:565--586, 1999.

\bibitem{Planck2013cosmological}
P.~A.~R. Ade et~al.
\newblock {Planck 2013 results. XVI. Cosmological parameters}.
\newblock {\em Astron.Astrophys.}, 571:A16, 2014.

\bibitem{Weinberg:1988cp}
S.~Weinberg.
\newblock {The Cosmological Constant Problem}.
\newblock {\em Rev. Mod. Phys.}, 61:1--23, 1989.

\bibitem{copeland2006dynamics}
E.~J. Copeland, M.~Sami, and S.~Tsujikawa.
\newblock {Dynamics of dark energy}.
\newblock {\em Int. J. Mod. Phys.}, D15:1753--1936, 2006.

\bibitem{dvali2000}
G.~Dvali, G.~Gabadadze, and M.~Porrati.
\newblock {4D Gravity on a Brane in 5D Minkowski Space}.
\newblock {\em Phys. Lett. B}, 485:208, 2000.

\bibitem{yin2005}
S.~Yin, B.~Wang, E.~Abdalla, and C.~Lin.
\newblock {Transition of equation of state of effective dark energy in the
  Dvali-Gabadadze-Porrati model with bulk contents}.
\newblock {\em Phys.Rev. D}, 76:124026, 2007.

\bibitem{Dymnikova:2001ga}
I.~Dymnikova and M.~Khlopov.
\newblock {Decay of cosmological constant as Bose condensate evaporation}.
\newblock {\em Mod. Phys. Lett.}, A15:2305--2314, 2000.

\bibitem{Dymnikova:2001jy}
I.~Dymnikova and M.~Khlopov.
\newblock {Decay of cosmological constant in selfconsistent inflation}.
\newblock {\em Eur. Phys. J.}, C20:139--146, 2001.

\bibitem{Mukhopadhyay:2007ed}
U.~Mukhopadhyay, P.~P. Ghosh, M.~Khlopov, and S.~Ray.
\newblock {Phenomenology of $\Lambda$-CDM model: A Possibility of accelerating
  universe with positive pressure}.
\newblock {\em Int. J. Theor. Phys.}, 50:939--951, 2011.

\bibitem{peebles1988}
P.~J.~E. Peebles and B.~Ratra.
\newblock {Cosmology with a Time Variable Cosmological Constant}.
\newblock {\em Astrophys.J.}, 325:L17, 1988.

\bibitem{ratra1988}
B.~Ratra and P.~J.~E. Peebles.
\newblock {Cosmological Consequences of a Rolling Homogeneous Scalar Field}.
\newblock {\em Phys.Rev.}, D37:3406, 1988.

\bibitem{Frieman1992}
J.~A. Frieman, C.~T. Hill, and R.~Watkins.
\newblock {Late time cosmological phase transitions. 1. Particle physics models
  and cosmic evolution}.
\newblock {\em Phys.Rev.}, D46:1226--1238, 1992.

\bibitem{Frieman1995}
J.~A. Frieman, C.~T. Hill, A.~Stebbins, and I.~Waga.
\newblock {Cosmology with ultralight pseudo Nambu-Goldstone bosons}.
\newblock {\em Phys. Rev. Lett.}, 75:2077, 1995.

\bibitem{Caldwell:1997ii}
R.~R. Caldwell, R.~Dave, and P.~J. Steinhardt.
\newblock {Cosmological imprint of an energy component with general equation of
  state}.
\newblock {\em Phys. Rev. Lett.}, 80:1582, 1998.

\bibitem{Caldwell:1999ew}
R.~R. Caldwell.
\newblock {A Phantom Menace?}
\newblock {\em Phys. Lett. B}, 545:23--29, 2002.

\bibitem{Caldwell:2003vq}
R.~R. Caldwell, M.~Kamionkowski, and N.~N. Weinberg.
\newblock {Phantom energy and cosmic doomsday}.
\newblock {\em Phys.Rev.Lett.}, 91:071301, 2003.

\bibitem{Zlatev:1998tr}
I.~Zlatev, L.-M. Wang, and P.~J. Steinhardt.
\newblock {Quintessence, cosmic coincidence, and the cosmological constant}.
\newblock {\em Phys. Rev. Lett.}, 82:896--899, 1999.

\bibitem{Steinhardt:1999nw}
P.~J. Steinhardt, L.-M. Wang, and I.~Zlatev.
\newblock {Cosmological tracking solutions}.
\newblock {\em Phys. Rev.}, D59:123504, 1999.

\bibitem{Gu2001}
J.-A. Gu and W-Y.~P. Hwang.
\newblock {Can the quintessence be a complex scalar field?}
\newblock {\em Phys.Lett.}, B517:1--6, 2001.

\bibitem{Essig:2013lka}
R.~Essig et~al.
\newblock {Working Group Report: New Light Weakly Coupled Particles}.
\newblock In {\em {Community Summer Study 2013: Snowmass on the Mississippi
  (CSS2013) Minneapolis, MN, USA, July 29-August 6, 2013}}, 2013.

\bibitem{Brax1999}
P.~Brax and J.~Martin.
\newblock {Quintessence and supergravity}.
\newblock {\em Phys. Lett.}, B468:40--45, 1999.

\bibitem{Copeland2000}
E.~J. Copeland, N.~J. Nunes, and F.~Rosati.
\newblock {Quintessence models in supergravity}.
\newblock {\em Phys. Rev.}, D62:123503, 2000.

\bibitem{ArmendarizPicon:2004pm}
C.~Armendariz-Picon.
\newblock {Could dark energy be vector-like?}
\newblock {\em JCAP}, 0407:007, 2004.

\bibitem{Koivisto:2008xf}
T.~Koivisto and D.~F. Mota.
\newblock {Vector Field Models of Inflation and Dark Energy}.
\newblock {\em JCAP}, 0808:021, 2008.

\bibitem{Bamba:2008ja}
K.~Bamba and S.~D. Odintsov.
\newblock {Inflation and late-time cosmic acceleration in non-minimal
  Maxwell-$F(R)$ gravity and the generation of large-scale magnetic fields}.
\newblock {\em JCAP}, 0804:024, 2008.

\bibitem{Emelyanov:2011ze}
V.~Emelyanov and F.~R. Klinkhamer.
\newblock {Possible solution to the main cosmological constant problem}.
\newblock {\em Phys. Rev.}, D85:103508, 2012.

\bibitem{Emelyanov:2011wn}
V.~Emelyanov and F.~R. Klinkhamer.
\newblock {Reconsidering a higher-spin-field solution to the main cosmological
  constant problem}.
\newblock {\em Phys. Rev.}, D85:063522, 2012.

\bibitem{Emelyanov:2011kn}
V.~Emelyanov and F.~R. Klinkhamer.
\newblock {Vector-field model with compensated cosmological constant and
  radiation-dominated FRW phase}.
\newblock {\em Int. J. Mod. Phys.}, D21:1250025, 2012.

\bibitem{Kouwn:2015cdw}
S.~Kouwn, P.~Oh, and C.-G. Park.
\newblock {Massive Photon and Dark Energy}.
\newblock {\em Phys. Rev.}, D93(8):083012, 2016.

\bibitem{Wetterich:1994bg}
C.~Wetterich.
\newblock {The Cosmon model for an asymptotically vanishing time dependent
  cosmological 'constant'}.
\newblock {\em Astron.Astrophys.}, 301:321--328, 1995.

\bibitem{Amendola:1999er}
L.~Amendola.
\newblock {Coupled quintessence}.
\newblock {\em Phys.Rev.}, D62:043511, 2000.

\bibitem{Zimdahl:2001ar}
W.~Zimdahl and D.~Pavon.
\newblock {Interacting quintessence}.
\newblock {\em Phys.Lett.}, B521:133--138, 2001.

\bibitem{Chimento:2003iea}
L.~P. Chimento, A.~S. Jakubi, D.~Pavon, and W.~Zimdahl.
\newblock {Interacting quintessence solution to the coincidence problem}.
\newblock {\em Phys.Rev.}, D67:083513, 2003.

\bibitem{Guo:2004vg}
Z.-K. Guo and Y.-Z. Zhang.
\newblock {Interacting phantom energy}.
\newblock {\em Phys. Rev. D.}, 71:023501, 2005.

\bibitem{Cai:2004dk}
R.-G. Cai and A.~Wang.
\newblock {Cosmology with interaction between phantom dark energy and dark
  matter and the coincidence problem}.
\newblock {\em JCAP}, 0503:002, 2005.

\bibitem{Guo:2004xx}
Z.-K. Guo, R.-G. Cai, and Y.-Z. Zhang.
\newblock {Cosmological evolution of interacting phantom energy with dark
  matter}.
\newblock {\em JCAP}, 0505:002, 2005.

\bibitem{Bi:2004ns}
X.-J. Bi, B.~Feng, H.~Li, and X.~Zhang.
\newblock {Cosmological evolution of interacting dark energy models with mass
  varying neutrinos}.
\newblock {\em Phys. Rev. D.}, 72:123523, 2005.

\bibitem{Gumjudpai:2005ry}
B.~Gumjudpai, T.~Naskar, M.~Sami, and S.~Tsujikawa.
\newblock {Coupled dark energy: Towards a general description of the dynamics}.
\newblock {\em JCAP}, 0506:007, 2005.

\bibitem{Wang:2005jx}
B.~Wang, Y.-G. Gong, and E.~Abdalla.
\newblock {Transition of the dark energy equation of state in an interacting
  holographic dark energy model}.
\newblock {\em Phys. Lett.}, B624:141--146, 2005.

\bibitem{Wang:2005pk}
B.~Wang, Y.~Gong, and E.~Abdalla.
\newblock {Thermodynamics of an accelerated expanding universe}.
\newblock {\em Phys. Rev.}, D74:083520, 2006.

\bibitem{Wang:2005ph}
B.~Wang, C.-Y. Lin, and E.~Abdalla.
\newblock {Constraints on the interacting holographic dark energy model}.
\newblock {\em Phys. Lett.}, B637:357--361, 2006.

\bibitem{Wang:2007ak}
B.~Wang, C.-Y. Lin, D.~Pavon, and E.~Abdalla.
\newblock {Thermodynamical description of the interaction between dark energy
  and dark matter}.
\newblock {\em Phys. Lett.}, B662:1--6, 2008.

\bibitem{Costa:2013sva}
A.~A. Costa, X.-D. Xu, B.~Wang, E.~G.~M. Ferreira, and E.~Abdalla.
\newblock {Testing the Interaction between Dark Energy and Dark Matter with
  Planck Data}.
\newblock {\em Phys. Rev.}, D89(10):103531, 2014.

\bibitem{Abdalla:2014cla}
E.~Abdalla, E.~G.~M. Ferreira, J.~Quintin, and B.~Wang.
\newblock {New evidence for interacting dark energy from BOSS}.
\newblock 2014.

\bibitem{Costa:2014pba}
A.~A. Costa, L.~C. Olivari, and E.~Abdalla.
\newblock {Quintessence with Yukawa Interaction}.
\newblock {\em Phys. Rev.}, D92(10):103501, 2015.

\bibitem{Costa:2016tpb}
A.~A. Costa, X.-D. Xu, B.~Wang, and E.~Abdalla.
\newblock {Constraints on interacting dark energy models from Planck 2015 and
  redshift-space distortion data}.
\newblock {\em JCAP}, 1701(01):028, 2017.

\bibitem{Marcondes:2016reb}
R.~J.~F. Marcondes, R.~C.~G. Landim, A.~A. Costa, B.~Wang, and E.~Abdalla.
\newblock {Analytic study of the effect of dark energy-dark matter interaction
  on the growth of structures}.
\newblock {\em JCAP}, 2016(12):009, 2016.

\bibitem{Farrar:2003uw}
G.~R. Farrar and P.~J.~E. Peebles.
\newblock {Interacting dark matter and dark energy}.
\newblock {\em Astrophys. J.}, 604:1--11, 2004.

\bibitem{Abdalla:2012ug}
E.~Abdalla, L.~L. Graef, and B.~Wang.
\newblock {A Model for Dark Energy decay}.
\newblock {\em Phys. Lett.}, B726:786--790, 2013.

\bibitem{D'Amico:2016kqm}
G.~D'Amico, T.~Hamill, and N.~Kaloper.
\newblock {Quantum Field Theory of Interacting Dark Matter/Dark Energy: Dark
  Monodromies}.
\newblock {\em Phys. Rev.}, D94(10):103526, 2016.

\bibitem{micheletti2009}
S.~Micheletti, E.~Abdalla, and B.~Wang.
\newblock {A Field Theory Model for Dark Matter and Dark Energy in
  Interaction}.
\newblock {\em Phys.Rev.}, D79:123506, 2009.

\bibitem{copeland1998}
E.~J. Copeland, A.~R. Liddle, and D.~Wands.
\newblock {Exponential potentials and cosmological scaling solutions}.
\newblock {\em Phys.Rev.}, D57:4686--4690, 1998.

\bibitem{ng2001}
S.~C.~C. Ng, N.~J. Nunes, and F.~Rosati.
\newblock {Applications of scalar attractor solutions to cosmology}.
\newblock {\em Phys.Rev.}, D64:083510, 2001.

\bibitem{Copeland:2004hq}
E.~J. Copeland, M.~R. Garousi, M.~Sami, and S.~Tsujikawa.
\newblock {What is needed of a tachyon if it is to be the dark energy?}
\newblock {\em Phys.Rev.}, D71:043003, 2005.

\bibitem{Zhai2005}
X.-H. Zhai and Y.-B. Zhao.
\newblock {A cosmological model with complex scalar field}.
\newblock {\em Nuovo Cim.}, B120:1007--1016, 2005.

\bibitem{DeSantiago:2012nk}
J.~De-Santiago, J.~L. Cervantes-Cota, and D.~Wands.
\newblock {Cosmological phase space analysis of the F(X) - V($\phi$) scalar
  field and bouncing solutions}.
\newblock {\em Phys. Rev.}, D87(2):023502, 2013.

\bibitem{Dutta:2016bbs}
J.~Dutta, W.~Khyllep, and N.~Tamanini.
\newblock {Cosmological dynamics of scalar fields with kinetic corrections:
  Beyond the exponential potential}.
\newblock {\em Phys. Rev.}, D93(6):063004, 2016.

\bibitem{TsujikawaGeneral}
S.~Tsujikawa.
\newblock {General analytic formulae for attractor solutions of scalar-field
  dark energy models and their multi-field generalizations}.
\newblock {\em Phys.Rev.}, D73:103504, 2006.

\bibitem{amendola2006challenges}
L.~Amendola, M.~Quartin, S.~Tsujikawa, and I.~Waga.
\newblock {Challenges for scaling cosmologies}.
\newblock {\em Phys.Rev.}, D74:023525, 2006.

\bibitem{ChenPhantom}
X.-M. Chen, Y.-G. Gong, and E.~N. Saridakis.
\newblock {Phase-space analysis of interacting phantom cosmology}.
\newblock {\em JCAP}, 0904:001, 2009.

\bibitem{Landim:2015poa}
R.~C.~G. Landim.
\newblock {Coupled tachyonic dark energy: a dynamical analysis}.
\newblock {\em Int. J. Mod. Phys.}, D24:1550085, 2015.

\bibitem{Landim:2015uda}
R.~C.~G. Landim.
\newblock {Coupled dark energy: a dynamical analysis with complex scalar
  field}.
\newblock {\em Eur. Phys. J.}, C76(1):31, 2016.

\bibitem{Mahata:2015lja}
N.~Mahata and S.~Chakraborty.
\newblock {Dynamical system analysis for DBI dark energy interacting with dark
  matter}.
\newblock {\em Mod. Phys. Lett .A}, 30(02):1550009, 2015.

\bibitem{Stojkovic:2007dw}
D.~Stojkovic, G.~D. Starkman, and R.~Matsuo.
\newblock {Dark energy, the colored anti-de Sitter vacuum, and LHC
  phenomenology}.
\newblock {\em Phys. Rev.}, D77:063006, 2008.

\bibitem{Greenwood:2008qp}
E.~Greenwood, E.~Halstead, R.~Poltis, and D.~Stojkovic.
\newblock {Dark energy, the electroweak vacua and collider phenomenology}.
\newblock {\em Phys. Rev.}, D79:103003, 2009.

\bibitem{Shafieloo:2016bpk}
A.~Shafieloo, D.~K. Hazra, V.~Sahni, and A.~A. Starobinsky.
\newblock {Metastable Dark Energy with Radioactive-like Decay}.
\newblock 2016.

\bibitem{Aulakh:1998nn}
C.~S. Aulakh, A.~Melfo, and G.~Senjanovic.
\newblock {Minimal supersymmetric left-right model}.
\newblock {\em Phys. Rev.}, D57:4174--4178, 1998.

\bibitem{Duka:1999uc}
P.~Duka, J.~Gluza, and M.~Zralek.
\newblock {Quantization and renormalization of the manifest left-right
  symmetric model of electroweak interactions}.
\newblock {\em Annals Phys.}, 280:336--408, 2000.

\bibitem{Dobrescu:2015qna}
B.~A. Dobrescu and Z.~Liu.
\newblock {$W'$ Boson near 2 TeV: Predictions for Run 2 of the LHC}.
\newblock {\em Phys. Rev. Lett.}, 115(21):211802, 2015.

\bibitem{Dobrescu:2015jvn}
B.~A. Dobrescu and P.~J. Fox.
\newblock {Signals of a 2 TeV $W'$ boson and a heavier $Z'$ boson}.
\newblock {\em JHEP}, 05:047, 2016.

\bibitem{Ko:2015uma}
P.~Ko and T.~Nomura.
\newblock {SU(2)$_L\times$SU(2)$_R$ minimal dark matter with 2 TeV $W'$}.
\newblock {\em Phys. Lett.}, B753:612--618, 2016.

\bibitem{Bezrukov:2009th}
F.~Bezrukov, H.~Hettmansperger, and M.~Lindner.
\newblock {keV sterile neutrino Dark Matter in gauge extensions of the Standard
  Model}.
\newblock {\em Phys. Rev.}, D81:085032, 2010.

\bibitem{Esteves:2011gk}
J.~N. Esteves, J.~C. Romao, M.~Hirsch, W.~Porod, F.~Staub, and A.~Vicente.
\newblock {Dark matter and LHC phenomenology in a left-right supersymmetric
  model}.
\newblock {\em JHEP}, 01:095, 2012.

\bibitem{An:2011uq}
H.~An, P.~S.~B. Dev, Y.~Cai, and R.~N. Mohapatra.
\newblock {Sneutrino Dark Matter in Gauged Inverse Seesaw Models for
  Neutrinos}.
\newblock {\em Phys. Rev. Lett.}, 108:081806, 2012.

\bibitem{Nemevsek:2012cd}
M.~Nemevsek, G.~Senjanovic, and Y.~Zhang.
\newblock {Warm Dark Matter in Low Scale Left-Right Theory}.
\newblock {\em JCAP}, 1207:006, 2012.

\bibitem{Bhattacharya:2013nya}
S.~Bhattacharya, E.~Ma, and D.~Wegman.
\newblock {Supersymmetric left-right model with radiative neutrino mass and
  multipartite dark matter}.
\newblock {\em Eur. Phys. J.}, C74:2902, 2014.

\bibitem{Heeck:2015qra}
J.~Heeck and S.~Patra.
\newblock {Minimal Left-Right Symmetric Dark Matter}.
\newblock {\em Phys. Rev. Lett.}, 115(12):121804, 2015.

\bibitem{Garcia-Cely:2015quu}
C.~Garcia-Cely and J.~Heeck.
\newblock {Phenomenology of left-right symmetric dark matter}.
\newblock {\em JCAP}, 1603:021, 2015.

\bibitem{Berlin:2016eem}
A.~Berlin, P.~J. Fox, D.~Hooper, and G.~Mohlabeng.
\newblock {Mixed Dark Matter in Left-Right Symmetric Models}.
\newblock {\em JCAP}, 1606(06):016, 2016.

\bibitem{Landim:2015upa}
R.~C.~G. Landim.
\newblock {Cosmological tracking solution and the Super-Higgs mechanism}.
\newblock {\em Eur. Phys. J.}, C76(8):430, 2016.

\bibitem{Landim:2016dxh}
R.~C.~G. Landim.
\newblock {Dynamical analysis for a vector-like dark energy}.
\newblock {\em Eur. Phys. J.}, C76:480, 2016.

\bibitem{Landim:2015hqa}
R.~C.~G. Landim.
\newblock {Holographic dark energy from minimal supergravity}.
\newblock {\em Int. J. Mod. Phys.}, D25(4):1650050, 2016.

\bibitem{Landim:2016isc}
R.~G. Landim and E.~Abdalla.
\newblock {Metastable dark energy}.
\newblock {\em Phys. Lett. B.}, 764:271, 2017.

\bibitem{Landim:2016gpz}
Ricardo~G. Landim and Fabrizio~F. Bernardi.
\newblock {Coupled quintessence and the impossibility of some interactions: a
  dynamical analysis study}.
\newblock 2016.

\bibitem{Marcondes:2016zte}
R.~J.~F. Marcondes.
\newblock {\em {Interacting dark energy models in Cosmology and large-scale
  structure observational tests}}.
\newblock PhD thesis, University of S\~ao Paulo, 2016.

\bibitem{weinberg2008cosmology}
S.~Weinberg.
\newblock {\em Cosmology}.
\newblock OUP Oxford, 2008.

\bibitem{Dodelson-Cosmology-2003}
S.~Dodelson.
\newblock {\em Modern Cosmology}.
\newblock Academic Press. Academic Press, 2003.
\newblock ISBN: 9780122191411.

\bibitem{Kolb:1990vq}
E.~W. Kolb and M.~S. Turner.
\newblock {\em The Early Universe}.
\newblock Addison-Wesley, 1990.
\newblock Frontiers in Physics, 69.

\bibitem{Weinberg:100595}
S.~Weinberg.
\newblock {\em {Gravitation and Cosmology: Principles and Applications of the
  General Theory of Relativity}}.
\newblock Wiley, New York, NY, 1972.

\bibitem{friedmann1}
A.~Friedman.
\newblock {Über die Krümmung des Raumes}.
\newblock {\em Z. Phys.}, 10:377, 1922.

\bibitem{friedmann2}
A.~Friedmann.
\newblock {Über die Möglichkeit einer Welt mit konstanter negativer Krümmung
  des Raumes}.
\newblock {\em Z. Phys.}, 21:326, 1924.

\bibitem{Lemaitre}
G.~Lemaître.
\newblock {\em Ann. Soc. Sci. Brux.}, A47:49, 1927.

\bibitem{robertson1}
H.~P. Robertson.
\newblock {Kinematics and World-Structure}.
\newblock {\em Astrophys. J.}, 83:187, 257, 1936.

\bibitem{robertson2}
H.~P. Robertson.
\newblock {Kinematics and World-Structure III}.
\newblock {\em Astrophys. J.}, 82:284, 1935.

\bibitem{walker}
A.~G. Walker.
\newblock {On Milne's Theory of World-Structure}.
\newblock {\em Proc. London Math. Soc.}, s2-42:90, 1937.

\bibitem{Ade:2015xua}
P.~A.~R. Ade et~al.
\newblock {Planck 2015 results. XIII. Cosmological parameters}.
\newblock {\em Astron. Astrophys.}, 594:A13, 2016.

\bibitem{Armendariz-Picon:2013jej}
C.~Armendariz-Picon and J.~T. Neelakanta.
\newblock {How Cold is Cold Dark Matter?}
\newblock {\em JCAP}, 1403:049, 2014.

\bibitem{bean2001}
R.~Bean, S.~H. Hansen, and A.~Melchiorri.
\newblock {Early universe constraints on a primordial scaling field}.
\newblock {\em Phys.Rev.}, D64:103508, 2001.

\bibitem{Boehmer:2011tp}
C.~G. Boehmer, N.~Chan, and R.~Lazkoz.
\newblock {Dynamics of dark energy models and centre manifolds}.
\newblock {\em Phys. Lett.}, B714:11--17, 2012.

\bibitem{Amendola}
L.~Amendola and S.~Tsujikawa.
\newblock {\em Dark Energy: Theory and Observations}.
\newblock Cambridge University Press, 2010.

\bibitem{amendola2001}
L.~Amendola and D.~Tocchini-Valentini.
\newblock {Stationary dark energy: The present universe as a global attractor}.
\newblock {\em Phys.Rev.}, D64:043509, 2001.

\bibitem{Hebecker2000}
A.~Hebecker and C.~Wetterich.
\newblock {Quintessential adjustment of the cosmological constant}.
\newblock {\em Phys.Rev.Lett.}, 85:3339--3342, 2000.

\bibitem{Sen:2004nf}
A.~Sen.
\newblock {Tachyon dynamics in open string theory}.
\newblock {\em Int.J.Mod.Phys.}, A20:5513--5656, 2005.

\bibitem{Sen:1998sm}
A.~Sen.
\newblock {Tachyon condensation on the brane anti-brane system}.
\newblock {\em JHEP}, 9808:012, 1998.

\bibitem{Sen:1999xm}
A.~Sen.
\newblock {Universality of the tachyon potential}.
\newblock {\em JHEP}, 9912:027, 1999.

\bibitem{Sen:2002in}
A.~Sen.
\newblock {Tachyon matter}.
\newblock {\em JHEP}, 0207:065, 2002.

\bibitem{Padmanabhan:2002cp}
T.~Padmanabhan.
\newblock {Accelerated expansion of the universe driven by tachyonic matter}.
\newblock {\em Phys.Rev.}, D66:021301, 2002.

\bibitem{Bagla:2002yn}
J.~S. Bagla, H.~K. Jassal, and T.~Padmanabhan.
\newblock {Cosmology with tachyon field as dark energy}.
\newblock {\em Phys.Rev.}, D67:063504, 2003.

\bibitem{Abramo:2003cp}
L.~R.~W. Abramo and F.~Finelli.
\newblock {Cosmological dynamics of the tachyon with an inverse power-law
  potential}.
\newblock {\em Phys.Lett.}, B575:165--171, 2003.

\bibitem{Aguirregabiria:2004xd}
J.~M. Aguirregabiria and R.~Lazkoz.
\newblock {Tracking solutions in tachyon cosmology}.
\newblock {\em Phys.Rev.}, D69:123502, 2004.

\bibitem{wess1992}
J.~Wess and J.~Bagger.
\newblock {\em Supersymmetry and Supergravity}.
\newblock Princeton University Press, 1992.

\bibitem{Freedman2012}
D.~Z. Freedman and A.~Van Proeyen.
\newblock {\em Supergravity}.
\newblock Cambridge University Press, 2012.

\bibitem{MoroiThesis}
T.~Moroi.
\newblock {\em Effects of the {G}ravitino on the {I}nflationary {U}niverse}.
\newblock {Ph.D.} thesis, Tohoku University, 1995.

\bibitem{weinberg2000}
S.~Weinberg.
\newblock {\em The Quantum Theory of Fields: Supersymmetry}.
\newblock Cambridge University Press, 2000.

\bibitem{kallosh2002supergravity}
R.~Kallosh, A.~D. Linde, S.~Prokushkin, and M.~Shmakova.
\newblock {Supergravity, dark energy and the fate of the universe}.
\newblock {\em Phys. Rev.}, D66:123503, 2002.

\bibitem{kallosh2002gauged}
R.~Kallosh, A.~D. Linde, S.~Prokushkin, and M.~Shmakova.
\newblock {Gauged supergravities, de Sitter space and cosmology}.
\newblock {\em Phys. Rev.}, D65:105016, 2002.

\bibitem{Bergshoeff:2015tra}
E.~A. Bergshoeff, D.~Z. Freedman, R.~Kallosh, and A.~Van~Proeyen.
\newblock {Pure de Sitter Supergravity}.
\newblock {\em Phys. Rev.}, D92(8):085040, 2015.

\bibitem{Pagels1982}
H.~Pagels and J.~R. Primack.
\newblock {Supersymmetry, Cosmology, and New Physics at Teraelectronvolt
  Energies}.
\newblock {\em Phys. Rev. Lett.}, 48:223--226, 1982.

\bibitem{Takayama2000}
F.~Takayama and M.~Yamaguchi.
\newblock {Gravitino Dark Matter without R-parity}.
\newblock {\em Phys. Lett. B}, 485:388--392, 2000.

\bibitem{Buchmuller:2007ui}
W.~Buchmuller, L.~Covi, K.~Hamaguchi, A.~Ibarra, and T.~Yanagida.
\newblock Gravitino dark matter in r-parity breaking vacua.
\newblock {\em JHEP}, 0703:037, 2007.

\bibitem{Weinberg1982}
S.~Weinberg.
\newblock {Cosmological Constraints on the Scale of Supersymmetry Breaking}.
\newblock {\em Phys. Rev. Lett.}, 48:1303--1306, 1982.

\bibitem{Kawasaki:1994af}
M.~Kawasaki and T.~Moroi.
\newblock {Gravitino Production in the Inflationary Universe and the Effects on
  Big-Bang Nucleosynthesis}.
\newblock {\em Prog.Theor.Phys.}, 93:879--900, 1995.

\bibitem{Kawasaki:1994bs}
M.~Kawasaki and T.~Moroi.
\newblock {Gravitino decay into a neutrino and a sneutrino in the inflationary
  universe}.
\newblock {\em Phys.Lett.}, B346:27--34, 1995.

\bibitem{nakamura2006}
S.~Nakamura and M.~Yamaguchi.
\newblock { Gravitino production from heavy moduli decay and cosmological
  moduli problem revived}.
\newblock {\em Phys. Lett. B}, 638:389--395, 2006.

\bibitem{Davidson2002}
S.~Davidson and A.~Ibarra.
\newblock {A lower bound on the right-handed neutrino mass from leptogenesis}.
\newblock {\em Phys. Lett. B}, 535:25--32, 2002.

\bibitem{Buchmuller2002}
W.~Buchmuller, P.~Di Bari, and M.~Plumacher.
\newblock {Cosmic Microwave Background, Matter-Antimatter Asymmetry and
  Neutrino Masses}.
\newblock {\em Nucl. Phys. B}, 643:367--390, 2002.

\bibitem{Coughlan1983}
G.~D. Coughlan, W.~Fischler, E.~W. Kolb, S.~Raby, and G.~G. Ross.
\newblock {Cosmological problems for the polonyi potential}.
\newblock {\em Phys.Lett.}, B131:59, 1983.

\bibitem{Cohen:1998zx}
A.~G. Cohen, D.~B. Kaplan, and A.~E. Nelson.
\newblock {Effective field theory, black holes, and the cosmological constant}.
\newblock {\em Phys. Rev. Lett.}, 82:4971--4974, 1999.

\bibitem{Bekenstein:1973ur}
J.~D. Bekenstein.
\newblock {Black holes and entropy}.
\newblock {\em Phys. Rev.}, D7:2333--2346, 1973.

\bibitem{Bekenstein:1974ax}
J.~D. Bekenstein.
\newblock {Generalized second law of thermodynamics in black hole physics}.
\newblock {\em Phys. Rev.}, D9:3292--3300, 1974.

\bibitem{Bekenstein:1980jp}
J.~D. Bekenstein.
\newblock {A Universal Upper Bound on the Entropy to Energy Ratio for Bounded
  Systems}.
\newblock {\em Phys. Rev.}, D23:287, 1981.

\bibitem{Hawking:1974sw}
S.~W. Hawking.
\newblock {Particle Creation by Black Holes}.
\newblock {\em Commun. Math. Phys.}, 43:199--220, 1975.
\newblock [,167(1975)].

\bibitem{Bekenstein:1993dz}
J.~D. Bekenstein.
\newblock {Entropy bounds and black hole remnants}.
\newblock {\em Phys. Rev.}, D49:1912--1921, 1994.

\bibitem{'tHooft:1993gx}
G.~'t~Hooft.
\newblock {Dimensional reduction in quantum gravity}.
\newblock In {\em {Salamfest 1993:0284-296}}, pages 0284--296, 1993.

\bibitem{Susskind:1994vu}
L.~Susskind.
\newblock {The World as a hologram}.
\newblock {\em J. Math. Phys.}, 36:6377--6396, 1995.

\bibitem{Hsu:2004ri}
S.~D.~H. Hsu.
\newblock {Entropy bounds and dark energy}.
\newblock {\em Phys. Lett.}, B594:13--16, 2004.

\bibitem{Li:2004rb}
M.~Li.
\newblock {A model of holographic dark energy}.
\newblock {\em Phys. Lett.}, B603:1, 2004.

\bibitem{Pavon:2005yx}
D.~Pavon and W.~Zimdahl.
\newblock {Holographic dark energy and cosmic coincidence}.
\newblock {\em Phys. Lett.}, B628:206--210, 2005.

\bibitem{Maldacena:1997re}
J.~M. Maldacena.
\newblock {The Large N limit of superconformal field theories and
  supergravity}.
\newblock {\em Int. J. Theor. Phys.}, 38:1113--1133, 1999.
\newblock [Adv. Theor. Math. Phys.2,231(1998)].

\bibitem{Sheykhi:2011cn}
A.~Sheykhi.
\newblock {Holographic Scalar Fields Models of Dark Energy}.
\newblock {\em Phys. Rev.}, D84:107302, 2011.

\bibitem{Nastase:2015pua}
H.~Nastase.
\newblock {General $f(R)$ and conformal inflation from minimal supergravity
  plus matter}.
\newblock {\em Nucl. Phys.}, B903:118--131, 2016.

\bibitem{Nastase:2016sji}
H.~Nastase.
\newblock {Quantum gravity and the holographic dark energy cosmology}.
\newblock {\em JHEP}, 04:149, 2016.

\bibitem{Coleman:1977py}
S.~R. Coleman.
\newblock {The Fate of the False Vacuum. 1. Semiclassical Theory}.
\newblock {\em Phys. Rev.}, D15:2929--2936, 1977.
\newblock [Erratum: Phys. Rev.D16,1248(1977)].

\bibitem{Weinberg:1996kr}
S.~Weinberg.
\newblock {\em {The quantum theory of fields. Vol. 2: Modern applications}}.
\newblock Cambridge University Press, 2013.

\bibitem{griffiths2005introduction}
D.~J. Griffiths.
\newblock {\em Introduction to quantum mechanics}.
\newblock Pearson Education, 2nd edition, 2005.

\bibitem{Coleman:1977th}
S.~R. Coleman, V.~Glaser, and A.~Martin.
\newblock {Action Minima Among Solutions to a Class of Euclidean Scalar Field
  Equations}.
\newblock {\em Commun. Math. Phys.}, 58:211, 1978.

\bibitem{Coleman:1980aw}
S.~R. Coleman and F.~De~Luccia.
\newblock {Gravitational Effects on and of Vacuum Decay}.
\newblock {\em Phys. Rev.}, D21:3305, 1980.

\bibitem{Agashe:2014kda}
K.~A. Olive et~al.
\newblock {Review of Particle Physics}.
\newblock {\em Chin. Phys.}, C38:090001, 2014.

\bibitem{Wang:2016lxa}
B.~Wang, E.~Abdalla, F.~Atrio-Barandela, and D.~Pavon.
\newblock {Dark Matter and Dark Energy Interactions: Theoretical Challenges,
  Cosmological Implications and Observational Signatures}.
\newblock {\em Rep. Prog. Phys.}, 79(9):096901, 2016.

\end{thebibliography}
